\documentclass{aa}

\usepackage{graphicx} 
\usepackage{txfonts}
\usepackage[colorlinks,linkcolor=blue,citecolor=blue]{hyperref}
\usepackage{natbib}

\begin{document}

\title{The 700 ks {\sl Chandra} Spiderweb Field II: Evidence 
for inverse-Compton and thermal diffuse emission in the
Spiderweb galaxy}
\author{
P. Tozzi\inst{1}, 
R. Gilli\inst{2},
A. Liu\inst{3},
S. Borgani\inst{4,5,6,7}, 
M. Lepore\inst{1}, 
L. Di Mascolo\inst{4}, 
A. Saro\inst{4,5,6,7}, 
L. Pentericci\inst{8},  
C. Carilli\inst{9}, 
G. Miley\inst{10},
T. Mroczkowski\inst{11}, 
M. Pannella\inst{4}, 
E. Rasia\inst{5,6},
P. Rosati\inst{12}, 
C. S. Anderson\inst{13}, 
A. Calabrò\inst{8}, 
E. Churazov\inst{3},
H. Dannerbauer\inst{14,15}, 
C. Feruglio\inst{5}, 
F. Fiore\inst{5}, 
R. Gobat\inst{16},  
S. Jin\inst{17,18}, 
M. Nonino\inst{5},
C. Norman\inst{19,20}, 
H.J.A. R\"ottgering\inst{9} 
% V. Strazzullo\inst{5} 
}

\institute{
INAF - Osservatorio Astrofisico di Arcetri, Largo E. Fermi, I-50122 Firenze, Italy 
\email{paolo.tozzi@inaf.it} 
\and
INAF - Osservatorio di Astrofisica e Scienza dello Spazio, via Piero Gobetti 93/3, 40129 Bologna, Italy
\and Max Planck Institute for Extraterrestrial Physics, Giessenbachstrasse
1, 85748 Garching, Germany
\and
Astronomy Unit, Department of Physics, University of Trieste, via Tiepolo 11, I-34131 Trieste, Italy
\and
INAF-Osservatorio Astronomico di Trieste, via G. B. Tiepolo 11, I-34143 Trieste, Italy
\and
IFPU - Institute for Fundamental Physics of the Universe, Via Beirut 2, 34014 Trieste, Italy
\and
INFN–Sezione di Trieste, Trieste, Italy
\and
INAF - Osservatorio Astronomico di Roma, Via Frascati 33, I-00040 Monteporzio (RM), Italy 
\and
National Radio Astronomy Observatory, P. O. Box 0, Socorro, NM 87801, USA
\and
Leiden Observatory, PO Box 9513, 2300 RA Leiden, The Netherlands
\and
European Southern Observatory (ESO), Karl-Schwarzschild-Str. 2, D-85748 Garching, Germany
\and
Dipartimento di Fisica e Scienze della Terra, Universit\`a degli Studi 
di Ferrara, via Saragat 1, I-44122 Ferrara, Italy
\and
Jansky Fellow of the National Radio Astronomy Observatory, P. O. Box 0, Socorro, NM 87801, USA
\and 
Instituto de Astrofísica de Canarias (IAC), 38205 La Laguna, Tenerife, Spain
\and
Universidad de La Laguna, Dpto. Astrofísica, 38206 La Laguna, Tenerife, Spain
\and 
Instituto de Física, Pontificia Universidad Católica de Valparaíso, Casilla 4059, Valparaíso, Chile
\and
Cosmic Dawn Center, Rådmandsgade 62, 2200 København N, Danimarca
\and
DTU-Space, Technical University of Denmark, Elektrovej 327, DK-2800 Kgs. Lyngby, Denmark 
\and
Space Telescope Science Institute, 3700 San Martin Dr., Baltimore, MD 21210, USA
\and
Johns Hopkins University, 3400 N. Charles Street, Baltimore, MD 21218, USA
}

\titlerunning{{\sl Chandra} Spiderweb Field II}
\authorrunning{Tozzi et al.}

\abstract
   {}
{We present the X-ray imaging and spectral analysis of the diffuse emission around the 
radio galaxy J1140-2629 (the Spiderweb galaxy) at $z=2.16$ and of 
its nuclear emission, based on a deep (700 ks) {\sl Chandra} observation.}
{We obtained a robust characterization of the unresolved nuclear emission, and carefully 
computed the contamination in the surrounding regions due to the wings of the 
instrument point spread function.  
Then, we quantified the extended emission within a radius of 12 arcsec.  
We used the {\sl Jansky} Very Large Array radio image to identify the regions overlapping 
the jets, and performed X-ray spectral analysis separately 
in the jet regions and in the complementary area. }
{We find that the Spiderweb galaxy hosts a mildly absorbed quasar, showing 
a modest yet significant spectral 
% (a factor 1.7 in the intrinsic absorption) 
and flux variability on a timescale of $\sim 1$ year.
We find that the emission in the jet regions is well described by a power law
with a spectral index of $\Gamma \sim 2-2.5$, and it is consistent with inverse-Compton 
upscattering of the cosmic microwave background  photons by the relativistic electrons. 
We also find a roughly symmetric, diffuse 
emission within a radius of $\sim 100$ kpc centered on the Spiderweb galaxy.
This emission, which is not associated with the jets, is significantly softer 
and consistent with thermal bremsstrahlung from a hot intracluster medium 
(ICM)  with a temperature of 
$kT = 2.0_{-0.4}^{+0.7} $ keV, and a metallicity of $Z<1.6 \, Z_\odot$
at 1 $\sigma$ c.l. 
The average electron density within 100 kpc is $n_e=(1.51 \pm 0.24\pm 0.14)\times 10^{-2}$ cm$^{-3}$,
corresponding to an upper limit for the total ICM mass of $\leq (1.76\pm 0.30 \pm 0.17) \times 10^{12}M_\odot$ (where
error bars are $1\sigma$ statistical and systematic, respectively).
The rest-frame luminosity $L_{0.5-10 {\rm keV}}=(2.0 \pm 0.5) \times 10^{44}$ erg s$^{-1}$
is about a factor of 2 higher than the  
extrapolated $L-T$ relation for massive clusters, but still consistent within the scatter. 
If we apply hydrostatic equilibrium to the ICM, we measure a total gravitational mass 
$M(<100~{\rm kpc}) = (1.5^{+0.5}_{-0.3})\times 10^{13}\, M_\odot$ and, extrapolating 
at larger radii, we estimate a total mass 
$M_{500} = (3.2^{+1.1}_{-0.6})\times 10^{13}\, M_\odot$ 
within a radius of $r_{500} = (220\pm 30)$ kpc. }
{We conclude that the Spiderweb protocluster 
shows significant diffuse emission within a radius of 12 arcsec, whose major contribution 
is provided by inverse-Compton scattering associated with the radio jets.  Outside the
jet regions, we also identified thermal emission within a radius of $\sim 100$ kpc, 
revealing the presence of hot, diffuse baryons that may represent the embryonic virialized halo 
of the forming cluster.  
}

\keywords{galaxies: clusters: general -- galaxies: clusters: intracluster medium -- 
X-rays: AGN -- X-rays: galaxies: clusters}

\maketitle

\section{Introduction}

% PROTOCLUSTERS: WHAT THEY ARE, WHY THEY ARE IMPORTANT
Protoclusters are defined as overdense regions in the high-$z$ Universe
which are predicted to evolve into massive, virialized clusters of galaxies by 
$z=0$ \citep[see][for a review]{2016Overzier}. Identifying 
and studying their properties is key to studying the formation and evolution of the 
large-scale structure of the Universe.  In particular, in recent years the scientific community
has mostly focused on the identification of the first virialized, dark-matter-dominated halos, 
on the origin and evolution of the hot, diffuse baryons permeating the potential wells 
-- the intracluster medium (ICM) -- and on the transformational processes that affect star 
formation and nuclear activity in the member galaxies. 
From the observational point of view, a protocluster is usually identified as 
a high-z region that is overdense in galaxy counts compared to the field.  However, 
the dynamical state of the structure and the virialized mass that its halo
ultimately achieves by $z=0$ is usually highly uncertain \citep[see, e.g.,][]{2015Muldrew}. 
At present, there are no standard methods to search for protoclusters, and this is the 
reason why biased-tracer techniques are often used to identify promising candidates.  
For example, several protoclusters have been found around high redshift, powerful radio 
galaxies.  Spectroscopic follow-up has provided evidence that a significant fraction 
% (roughly 75\%) 
of powerful, high-z radio galaxies reside in a protocluster or overdense 
regions \citep[see][]{2007Venemans,2012Galametz,2013Wylezalek}.  
In addition to massive spectroscopic surveys, intensive multiwavelength campaigns in the 
radio and X-ray bands are necessary to trace the many processes occurring in the protocluster
member galaxies during their rapid evolution, such as nuclear activity, feedback into the
surrounding medium, star formation, chemical enrichment, among others.

% THE ROLE OF X-RAY OBSERVATIONS
In this paper we focus on the role of deep X-ray observations.  X-ray data are unique  to use when studying
the unresolved emission from active galactic nuclei (AGN), due to the accretion
onto nuclear supermassive black holes (SMBHs), the thermal diffuse emission from the ICM, 
and the inverse-Compton (IC) emission from relativistic plasma. To a lesser extent, X-ray observations 
can also be used to trace the less intense emission from 
strongly star-forming galaxies, but this signal is too weak to be detected 
beyond $z\sim 1.5$ \citep[see, e.g.,][]{2016Lehmer}. 
%  A QUICK REVIEW OF PROTOCLUSTER STUDIES WITH X-RAY
Until now, only a few deep ($t_{\rm exp}\geq 200$ ks) 
observations of protoclusters have been carried out with {\sl Chandra} or XMM-Newton. 
While the unresolved emission from AGN is relatively easy to detect, the faint, extended emission 
associated with the expected proto-ICM has been very hard to detect, with a few  
promising candidates \citep[][]{2011Gobat,2016Valentino,2016Wang}, and 
some other cases where it is not possible to determine the thermal or relativistic nature 
of the diffuse emission unambiguously \citep{2019Gilli,2021Champagne}.  
This is not surprising since the expected thermal emission from $z\gtrsim 2$ protoclusters 
is predicted to be very faint from semianalytic models and numerical simulations 
\citep[see][]{2009Saro}.  In addition, the presence of nonthermal diffuse emission associated with the
relativistic electrons in the radio jets, coupled to the bright, unresolved nuclear emission
from protocluster members, which are both often present, may easily overwhelm the thermal emission 
from the proto-ICM.  This implies that only high-sensitivity and  
angular-resolution ($\sim 1$ arcsec) observations can be used to search and characterize
the diffuse, thermal emission in high-z protocluster and, at the same time, 
the unresolved emission from AGN members and the diffuse, 
nonthermal emission due to IC from radio jets.

A key piece of information in protoclusters is
the identification and characterization of thermal emission from surrounding diffuse baryons, 
which directly traces the thermodynamics of the largest baryonic component and, 
at the same time, the virialization status of the halo.
As of today, there have been no observations that can track the origin of the
ICM from the birth of the protocluster, to the formation of an evolved, 
virialized clusters.  In this short yet intense phase, which is expected to occur
in the redshift range $2<z<3$, the heating of the proto-ICM can be
due to several mechanisms, namely the following: accretion shocks; heating from stellar winds 
associated with strong star formation episodes in the protocluster galaxies; and 
mechanical heating and turbulence caused by radio jets from the central massive galaxy. 
Therefore, a characterization of the proto-ICM, including its morphology, 
thermodynamical, and chemical properties, can provide an independent and direct measure
of the heating efficiency of the feedback and of the gas accretion and circulation
into the protocluster potential well.  In addition, 
feedback processes not only affect the ICM, but can also couple directly
with the member galaxies, as suggested by the recent X-ray observation
of a $z\sim 1.7$ protocluster \citep{2019Gilli}.

%  ON THE NEED OF X-RAY OBSERVATIONS IN PROTOCLUSTERS
% WHY WE ARE HERE: THE DEEP CHANDRA DATA ON THE SPIDERWEB
In this paper we investigate the diffuse emission (both thermal and nonthermal) and 
the central nuclear emission of the Spiderweb galaxy (J1140-2629) 
protocluster at $z=2.16$, using a deep {\sl Chandra} ACIS-S observation ($\sim 700 $ ks).
The central, powerful radio galaxy is embedded in a giant Ly$\alpha$ 
halo \citep{1997Pentericci,2006Miley}, and surrounded by a $\gtrsim 2$ Mpc--sized 
overdensity of star-forming galaxies 
(Ly$\alpha$ and H$\alpha$ emitters), dusty starbursts, and galaxies that are
rapidly migrating toward a  nascent red sequence \citep{2008Zirm}.  
The archetypal Spiderweb protocluster is expected to evolve 
into a massive cluster in less than 1.5 Gyr, with the radio galaxy itself showing the 
properties of a cD progenitor \citep{2006Miley}. 
% (Miley et al.  2009). 
% PAPER I - SUMMARY
In a companion paper \citep[][hereafter Paper I]{2022Tozzi}, we investigated 
the nuclear activity in the protocluster members, finding a high fraction of
bright AGN with $L_{0.5-10 keV}>3\times 10^{43}$ erg s$^{-1}$ in 
spectroscopically confirmed members with
${\rm log}(M_*/M_\odot)>10.5$, equal to $ 25.5 \pm 4.5 $\%, and estimating an
enhancing factor of $6.0^{+9.0}_{-3.0}$ for the nuclear activity with 
respect to the COSMOS field at comparable redshifts and stellar mass range.
Here we focus on the very central regions of the Spiderweb protocluster, which is 
a rapidly evolving, high-density region where complex and intricate phenomena 
concur to shape the properties of the nascent brightest cluster galaxy 
(BCG) and possibly an embryonic ICM halo. 
Such diffuse X-ray emission is  also investigated in connection with 
the radio (JVLA) and submillimeter (ALMA) spatially resolved signal in other papers of the
collaboration \cite[][Di Mascolo et al. in preparation]{2022Carilli,2022Anderson}.
Finally, in a forthcoming paper, the structure and thermodynamics of the ICM will be
investigated by combining X-ray and submillimeter data (Lepore et al. in preparation). 

% PAPER ORGANIZATION
The paper is organized as follows. In Section 2 we summarize the properties of the Spiderweb 
galaxy and its immediate surroundings on the basis of previous multiwavelength 
observations. In Section 3 we briefly
recall the X-ray data acquisition and reduction, which has already been detailed
in Paper I.  In 
Section 4 we present the ray-tracing simulations we used to accurately model the 
unresolved nuclear emission, a step needed to estimate the contamination of the 
AGN to the extended emission and, therefore, exploit the imaging capability of 
{\sl Chandra} when studying the morphology of such emission.  
In Section 5 we present a detailed analysis of the
nuclear emission, including its variability.  In Section 6 we 
finally investigate the nature of the diffuse X-ray emission surrounding the Spiderweb galaxy, 
spatially separating the thermal and nonthermal contribution.  
In Section 7 we focus on the nature of the thermal diffuse component, and 
discuss the implication of the presence of a virialized halo.  
Finally, our conclusions are summarized in Section \ref{conclusions}.
Throughout this paper, we adopt the seven-year Wilkinson Microwave Anisotropy Probe cosmology 
with $\Omega_{\Lambda} =0.73 $, $\Omega_m =0.27$, and $H_0 = 70.4$ km s$^{-1}$ 
Mpc$^{-1}$ \citep{2011Komatsu}. In this cosmology, at $z=2.156$, 1 arcsec 
corresponds to 8.473 kpc, the Universe is 3.13 Gyr old, 
and the lookback time is 77\% of the age of the Universe.  
Therefore, the size of a {\sl Chandra} ACIS pixel (0.492 arcsec) 
corresponds to $4.237$ kpc. These values are especially relevant in the following 
section where we interchangeably report distances in pixels, arcsecs, or kiloparsecs.
Quoted error bars correspond to a 1 $\sigma$ confidence level, unless noted otherwise.

%%%%%%%%%%%%%%%%%%%%%%%%%%%%%%%%%%%%%%%%%%%%%%%%%%%%%%%%%%%%
% REVIEW OF LITERATURE ON THE SPIDERWEB
% A SHORT SUMMARY OF OBSERVATIONAL PAPERS ON THE SPIDERWEB
% CLASSIFIED ACCORDING TO SCIENCE TOPICS
%%%%%%%%%%%%%%%%%%%%%%%%%%%%%%%%%%%%%%%%%%%%%%%%%%%%%%%%%%%%

\section{The Spiderweb galaxy: Previous observational campaign and main results}

The Spiderweb Galaxy (J1140-2629) and its environment have been 
extensively targeted in the last 25 years, with 54 papers published 
about the Spiderweb field including observations in radio, sub-mm, infra red (IR), 
optical and X--ray wavelengths.  In this section we summarize the main 
results that have been obtained on the Spiderweb Galaxy 
focusing on its immediate environment, while a review on the properties
of the protocluster galaxy population has been presented in 
Paper I.  In the last subsection we point out
the open science cases that are addressed by this work.

\subsection{Discovery and peculiarities}

The Spiderweb Galaxy was identified as a 
high-z radio galaxy (HzRG) while targeting ultra-steep spectrum radio 
sources \citep{1994Roettgering} in an ESO Key Program.  The peculiarity of this
object was immediately noticed thanks to the unusually clumpy and bent radio morphology, 
and the exceptionally high rotation measure ($\sim 6200$ rad/m), 
showing that the radio synchrotron jet is severely affected 
by a surrounding, dense ($10^{-1}- 10^{-2}$cm$^{-3}$) external medium 
\citep{1997Carilli,1997Pentericci,1998Athreya}.  From the optical point of view, 
the Spiderweb Galaxy has been classified as a narrow emission line galaxy at 
$z=2.16$ \citep[see][]{1997Rottgering}. 
A clumpy morphology has been observed also in the optical band, 
and HST ACS observations revealed that this is due to $\sim 10$ star-forming 
satellite galaxies moving with peculiar velocities of several hundred km s$^{-1}$, 
with a distribution resembling that of flies trapped in a spiderweb \citep{2006Miley}, 
and probably in the process of merging with the central radio galaxy within a few
hundreds of Myr.  Nevertheless, the 
most striking feature is constituted by a spectacular 200 kpc Ly$\alpha$ halo 
with a luminosity of $42.5 \times 10^{44}$ erg s$^{-1}$
elongated along the direction of the jets.  The total stellar mass inferred
from the K-band luminosity is $10^{12} M_\odot$ \citep{1997Pentericci}, 
and it has been confirmed by Spitzer data \citep{2007Seymour}.  
The large majority of 
the stellar mass appears to be already in place well before $z=2.16$, while the 
satellite galaxies have measured stellar masses in the range $10^8<M_*<10^{10} M_\odot$, 
contributing only $\sim 1/10$ of the total mass of the central galaxy.

The optical and NIR spectrum of the nucleus 
of J1140-2629 shows a broad, spatially unresolved H$\alpha$ line
(blended with [NII]) with a FWHM of $\sim 14900$ km s$^{-1}$ \citep{2011Nesvadba}.
The presence of broad nuclear lines is rare for optical
counterparts of high redshift radio sources, which typically 
show narrow emission lines and are classified as TypeII AGN. 
Other emission lines which have been detected include [OII], [NeIII], 
H$_\beta$ narrow, [OIII], [OI], [NII], [SII], and a strong CIV and HeII 
\citep[see][]{2008Humphrey}. Such an optical spectrum is consistent 
with photoionization from the AGN, 
powered by a SMBH with estimated mass $M_{BH}=2\times 10^{10}M_\odot$, 
putting J1140-2629 a factor of $\sim 2$ above the local $M-\sigma$ relation 
\citep{2011Nesvadba}. 

Taken together, the observed characteristics suggest that the 
Spiderweb is indeed a massive radio galaxy forming at the center of a dynamically evolving 
protocluster region, that is likely to evolve into a BCG
\citep[see][]{1997Pentericci}.
These findings sparkled a wide interest in this object and its surroundings, 
as one of the most promising targets where the evolutionary transition from a 
collapsing large scale structure to a gravitationally bound halo heavily 
affects the entire galaxy population.

\subsection{Star formation activity in and around J1140-2629}

The Spiderweb Galaxy shows a star formation rate of $15.6\pm 0.7 M_\odot$/yr 
from dust-uncorrected optical/UV emission, 
as opposed to a total of $\sim 50 M_\odot$/yr
contributed from all the "flies" \citep{2009Hatch}.  
IR observations with Spitzer \citep{2010DeBreuck}, Herschel and LABOCA \citep{2012Seymour} 
provide, however, a different picture, finding that the contribution to the 
$8-1000\mu\, $m luminosity is $1.17\pm 0.27 \times 10^{13} L_\odot$ and 
$0.79\pm0.09\times 10^{13}L_\odot$ for an AGN and a starburst component, respectively. 
This implies a 20\% Eddington accretion rate onto the SMBH
and, most important, a star formation rate (SFR) equal to $1390\pm 150 M_\odot$/yr, 
dramatically higher than 
the value derived from the UV/optical rest-frame\footnote{Despite the 
source appears as unresolved in the 8$\, \mu$m to 250$\, \mu$m Herschel bands, 
the IR flux includes also the surrounding star-forming galaxies (the so called {\sl flies}),
and the possible diffuse star formation, 
and, in the 350, 500, and 870$\, \mu$m bands, also additional member galaxies 
out to 30 arcsec from the protocluster center.  This value, therefore, should 
be considered an upper limit \citep{2012Seymour}.}.  Such a large value
is confirmed by other works based on spectroscopic analysis of 
IR data \citep{2012Ogle,2013Rawlings}, and it cannot be reconciled with the 
dust-corrected estimate based on the rest-frame UV and the simplistic assumption 
of a uniform dust screen by \citet{2008Hatch}, implying that most of the 
star formation in the Spiderweb must be highly obscured at rest-UV wavelengths.

The diffuse UV-continuum light surrounding the Spiderweb Galaxy shows 
a distribution similar to the $Ly\alpha$ halo, and it is likely due to 
a young stellar population \citep{2008Hatch} originated by a diffuse, {\sl in situ}  
star formation of $57\pm 8 \, M_\odot$/yr, or  $\sim 140\, M_\odot$/yr 
if corrected by dust extinction.  
This interpretation is not free from complexities, though.  The presence of 
CIV and HeII emission lines suggest an ionization source harder than 
stellar continuum, implying a lower limit of only $\sim 7$\% of the $Ly\alpha$ emission 
associated with star formation \citep{2008Hatch}. 

A key point here is the presence of a large gas reservoir that feeds both the 
central and the diffuse star formation.
The molecular gas mass inferred from observation of the redshifted 
CO(1-0) line with ATCA \citep{2013Emonts} is $M_{H_2} = 6\times 10^{10} 
M_\odot$ (for $M_{H2}/L^{'}_{CO}=0.8$).  This reservoir is able to sustain a 
SFR of $\sim 1400 M_\odot/$yr for only $\sim  40$ Myr, while the doubling-mass time 
would be as long as 0.8 Gyr.  
A deeper ATCA observation \citep{2016Emonts}, coupled with 
JVLA, was able to identify an even larger amount of molecular gas 
$M_{H_2} = 1.5\pm 0.4 \times 10^{11} M_\odot$ 
(assuming $\alpha_{CO}=4 \, M_\odot$ (K km s$^{-1}$ pc$^2$)$^{-1}$), associated with the
$Ly\alpha$ halo, showing that
there is enough molecular gas to fuel the star formation within the IGM up to $z\sim 1.6$.

Summing the $3\times 10^9 M_\odot$ in ionized gas found by \citet{2006Nesvadba}, 
the gas masses of various components of the ISM are about  an 
order of magnitude lower than the stellar mass, 
suggesting that this is already quite an evolved galaxy with a gas fraction in line for
a main sequence galaxy at $z\sim 2$.  
Again, this suggests that we might be witnessing a transition from a 
very active starburst phase triggered by a merging event to a radio-feedback 
dominated phase, leading to the more quiescent existence of a BCG.  

How this transition will be happening is still very much debated and unclear. 
On the one hand we have the expectation that this transition should be short 
lived, and that strong nuclear activity is indeed able to rapidly quench star 
formation, to be consistent with the lack of bright IR sources
at X-ray luminosities comparable to the Spiderweb ($L_X>10^{45}$ erg s$^{-1}$) 
in the GOODS-North deep field \citep{2012Page}. 
On the other hand, results obtained with improved statistics \citep{2012Harrison}, 
and actually extending a trend well established at lower X-ray luminosities, 
seem to suggest that high-accretion, X-ray luminous, nuclear activity phases 
do not show any obvious impact on the star formation rate and stellar mass 
growth of the host galaxies and that super massive black hole and galaxy growth 
rather coexist over cosmic times.  In this perspective, it is key to investigate
a large sample of massive, high-z transitional galaxies similar to the Spiderweb
to assess the relative contribution of radiative and mechanical feedback.

\subsection{Diffuse gas and feedback effects in the halo of J1140-2629}

The X-ray band is a promising spectral window to explore the
Spiderweb, particularly when considering the expected X-ray
brightening of high-z radio galaxies driven by the cosmic microwave background 
(CMB) photons \citep[see][]{2021Hodges-Kluck}.
The discovery of extended X-ray emission, that was not possible to constrain 
in an early ROSAT HRI image \citep{1998Carilli}, was measured to reach a 
luminosity of $\sim 3\times 10^{44}$ erg s$^{-1}$ in the 2-10 keV band 
aligned along the direction of the jet on a scale of $\sim 20$ arcsec, 
thanks to the first shallow {\sl Chandra} exposure of about $\sim 30$ ks 
\citep{2002Carilli}.  From these preliminary data it was not possible to
spectrally verify whether the dominant emission mechanism is thermal 
bremsstrahlung from hot gas or Inverse Compton due to a relativistic 
electron population. 

The nature of the diffuse X-ray emission is clearly a key ingredient
in any picture aimed at constraining the presence of a virialized halo and the 
feedback processes in the 
diffuse medium around the Spiderweb Galaxy.  Nevertheless, signatures 
of preheating have been searched for at other wavelengths.  
Integral field unit spectroscopy with SPIFFI (SINFONI) at VLT, found evidences for strong
outflows with velocities on the order of $\sim 2000$ km s$^{-1}$, 
and total kinetic power in the range $2\times 10^{46}$-$3\times 10^{47}$erg s$^{-1}$
\citep[][]{2006Nesvadba,2011Nesvadba}. The 
high ratio of kinetic over radiative power may indicate that J1140-2629 is about 
to complete the transition to the radio-mode feedback regime \citep[see][]{2012Fabian}.
This picture is reinforced by the gas kinematics derived from the 
[OIII]$\lambda 5007$ line, that highlights a structure reminiscent of 
"bubbles" intended as expanding spheres or conical outflows \citep{2008Humphrey},
implying a relatively efficient interaction between the AGN and the 
interstellar medium. 
With a sufficient coupling, the jet has enough energy to 
totally remove the ISM of the host within $\sim 5 \times 10^8$ yr.

\subsection{Dynamical state of the Spiderweb Complex}

A first dynamical estimate of the total mass from the member galaxy velocity
dispersion provided a value of $\sim 2.3 \times 10^{14} M_\odot$, 
contributed by two subgroups of $1.7$ and $0.6 \times 10^{14} M_\odot$ 
and virial radii of 1.1 and 0.8 Mpc, respectively, assuming they are 
both virialized \citep{2004aKurk}.
However, the entire system, as well as the two subgroups separately, 
seem far from being virialized, not only for the lack of evident 
extended, thermal X-ray emission, but also for the lack of 
a well-formed red-sequence in the J-K vs K color magnitude diagram
(despite this does not necessarily follow or precede virialization). 
Thanks to near-infrared spectroscopy with SINFONI at VLT, 
\citet{2011Kuiper} found that the velocity distribution within 60 
kpc of the Spiderweb Galaxy is consistent with bimodality, 
with no global peak resembling that of a virialized halo.
They described the system as the merger of two halos 
that would predict, in case of virialization of both subhalos, a rest-frame 2-10 keV 
luminosity of $2\times 10^{43}$ erg s$^{-1}$.
Finally \citet{2014Shimakawa} identified a region possibly virialized with a
$R_{200}\sim 0.53 $ Mpc and $\sigma_{cl}\sim 680$ km s$^{-1}$, while the inclusion of all the 
spectroscopic members would give $\sigma_{cl}\sim 880$ km s$^{-1}$, reflecting
some velocity structures particularly in the outer regions. 
The inferred mass is consistent with that of a
progenitor of present-day most massive class of galaxy clusters.

\citet{2009Saro} analyzed at $z=2.2$ cosmological hydrodynamical simulations of two protocluster 
regions, which form by $z=0$ two clusters with virial masses of $\sim 10^{14}$ and 
$\sim 10^{15}M_\odot$.  The line-of-sight velocity dispersion of galaxies
in the largest halo is roughly in agreement with the total observed dispersion
in the Spiderweb Complex, thus suggesting that it
is consistent with being the progenitor of a massive cluster by $z=0$ and that a diffuse 
atmosphere of hot gas already in equilibrium should be already present.
Their prediction for the 0.5-2 keV flux from the ICM is $\sim 10^{-15}$ and 
$\sim 10^{-14}$ erg s$^{-1}$ cm$^{-2}$ for subhalo temperatures of 2 and 5 keV, respectively.  
These findings underline the importance of confirming the thermal nature of at least part of 
the diffuse X-ray emission around the Spiderweb Galaxy.

\subsection{Open issues and role of the new X-ray observations}

Previous works on the Spiderweb Complex found a clear overdensity 
in discrete source counts in all the 
observed bands, from sub-mm to the X-ray, that ranges from a factor of $\sim 2-6$ to $\sim 100$, 
depending on the source selection and on the specific region considered within the complex.
In Paper I we explored the galaxy population of the Spiderweb Protocluster, finding an 
X-ray AGN fraction enhanced by a factor of $6.0^{+9.0}_{-3.0}$ with 
respect to the COSMOS field at comparable redshifts and stellar mass range. 

In this work we focus on other key aspects concerning the Spiderweb Galaxy and 
its immediate surroundings: i) the nature of the diffuse emission; ii) the presence 
of diffuse thermal emission possibly associated with the virialization of a central halo;
iii) the nuclear activity in the Spiderweb Galaxy.
Exploring these aspects is mandatory to explore more general aspects 
concerning galaxy evolution, such as: the main process
(gravitational or feedback) responsible for the heating of the proto-ICM; 
the compresence of hot and cold gas within the halo as a typical 
phase of the formation of BCG progenitors; the eventual evolution of 
the Spiderweb protocluster into a virialized structure by $z=0$.
The new set of data obtained in the last couple of years in the Spiderweb field, 
and in particular the deep {\sl Chandra} observation presented in this work, 
allows us to better address all the science goals listed here.  
The presence of hot, diffuse baryons around the 
Spiderweb Galaxy and its relation to its nuclear activity 
are discussed in other papers based on the combination of {\sl Chandra} X-ray 
and JVLA Radio data \citep{2022Carilli,2022Anderson}.
Moreover, the ICM of the Spiderweb Galaxy is also investigated with ALMA SZ data
(Di Mascolo et al. in preparation).  
The main results and the data products can be found on the 
project webpage.\footnote{\url{http://www.arcetri.inaf.it/spiderweb/ .}}

\section{X-ray data reduction}

The Spiderweb Galaxy was observed with a Chandra Large Program for 700 ks with ACIS-S 
granted in Cycle 20 (PI P. Tozzi). The observations were completed in the 
period November 2019 - August 2020, splitted in 21 separate Obsid.  To this data 
set we add the first X-ray observation 
with ACIS-S, dating back to June 2000 for a total of 39.5 ks.  
All the Obsid are listed in Table 1 of Tozzi et al. (2021), where 
we also describe the followed standard data reduction procedure.  We used the 
latest release of the Chandra Calibration Database at the time of writing {\tt (CALDB 4.9.3)}
We briefly recall that we run the task 
{\tt acis\_process\_events} with the parameter {\tt apply\_cti=yes} 
to flag background events, most likely associated with cosmic 
rays, and, by rejecting them, obtain a significant reduction of 
the background, thanks to the VFAINT mode of data acquisition. 

This step is particularly relevant in this work 
since the background may significantly affect the detection of the low surface-brightness, diffuse
emission surrounding the Spiderweb Galaxy.
The small price to pay is the removal of some of photons from bright sources.   
We find that slightly more than 100 net counts in the total band are removed at the position of 
the nucleus of J1140-2629 across the 22 exposures.  This indicates that 
the source, which is by far the brightest in the field, suffers a small amount of pile-up. However, 
we have verified {\sl a posteriori} that this effect does not impact the spectral analysis of the
nucleus and therefore it is not worth to give up the VFAINT cleaning to recover less 
than 1\% of the flux in the nucleus. 

\begin{figure*}
\begin{center}
\includegraphics[width=0.49\textwidth, trim=0 55 55 140, clip]{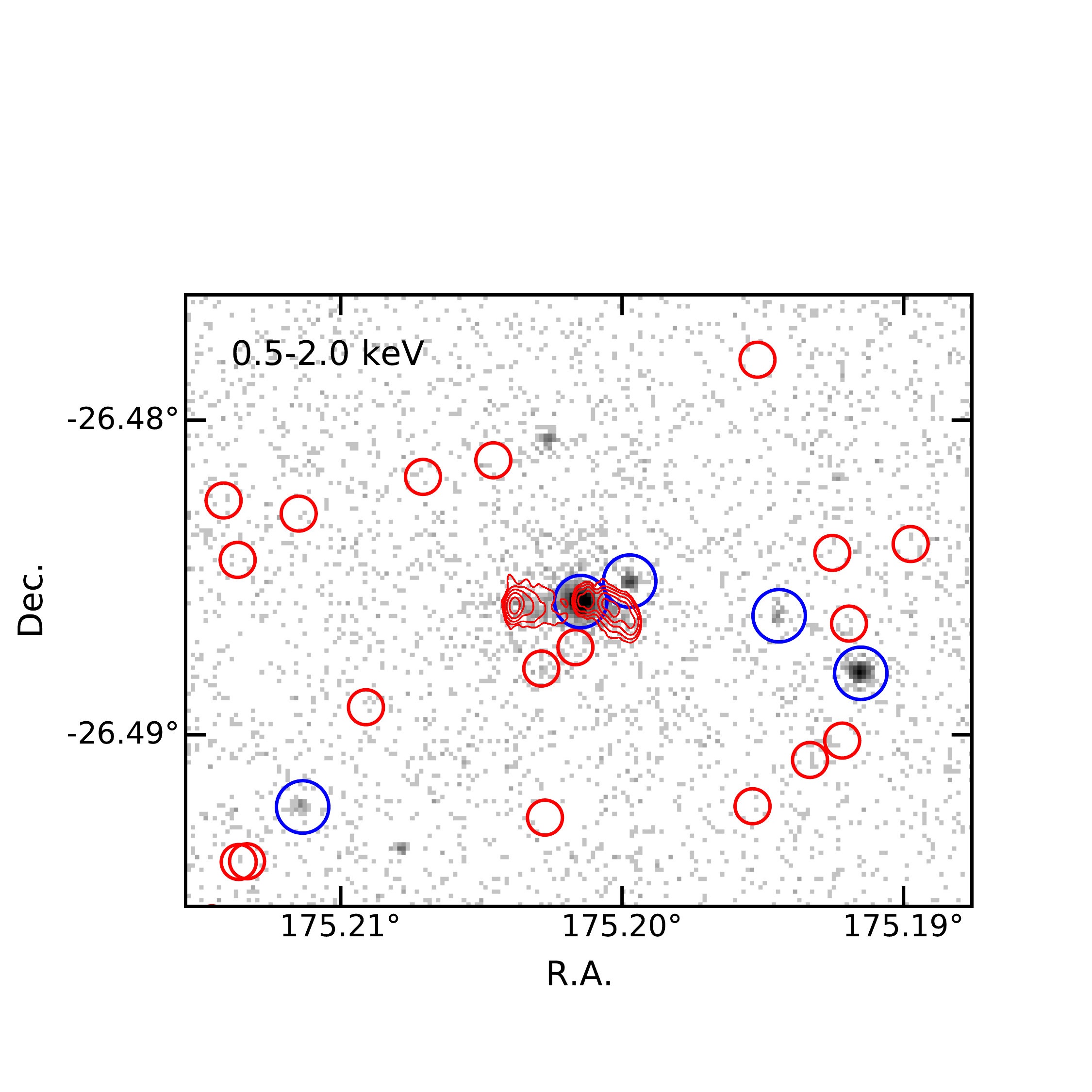}
\includegraphics[width=0.49\textwidth, trim=0 55 55 140, clip]{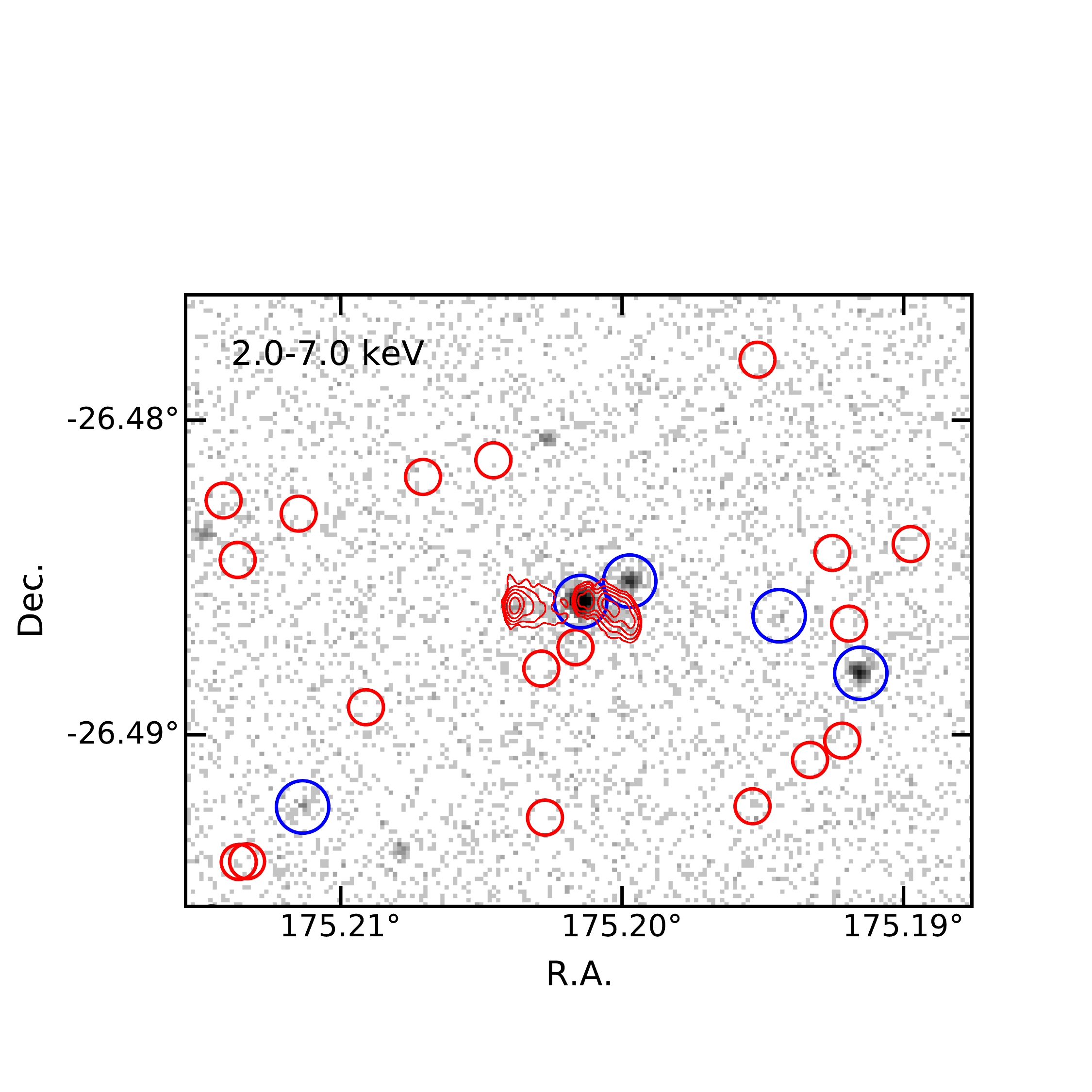}
\caption{Close-up of the Spiderweb Galaxy ($90\times 70$ arcsec by side)
at full angular resolution, in the soft (0.5-2 keV, left) 
and hard (2-7 keV, right) band.  Extended emission is clearly visible 
in both bands.  Red contours show radio emission observed in the 10 GHz band with the JVLA
\citep[see][]{2022Carilli} at levels of $0.03$, $0.2$, $2$ and
$20$ mJy/beam.  It is possible to appreciate the larger background in the hard band, 
and the enhancement of the diffuse emission along the radio jets.  
Also, the extended emission appear to be stronger and more isotropic in the soft band. 
Blue circles correspond to X-ray emitting, spectroscopically confirmed members, while
red circles to spectroscopically confirmed members with no X-ray 
emission \citep[see][]{2022Tozzi}.
}
\label{softhard}
\end{center}
\end{figure*}

The final total exposure time after data reduction and 
excluding the dead-time correction amounts to 715 ks (corresponding to the 
LIVETIME keyword in the header of {\sl Chandra} fits files), including the
first observation. 
The 22 level-2 event files are merged together with the tool {\tt reproject\_obs}, 
using the reference coordinates of Obsid 21483.  
The soft (0.5-2 keV) and hard (2-7 keV) band images at full resolution in the 
immediate surroundings ($90\times 70$ arcsec$^2$) 
of the Spiderweb Galaxy are shown in Figure \ref{softhard}.  
It is possible to notice the larger background in the hard band, and to appreciate 
the enhancement of the diffuse emission along the radio emission associated with the jets, 
shown with red contours \citep[after][]{2022Carilli}.  Fairly isotropic extended mission 
is also present far from the jets in the soft band.  Clearly, a meaningful analysis of the
extended emission must take into account all the complexity arising from the 
overlapping of four components: emission from the jets, isotropic extended emission, 
unresolved nucleus, and background.  The choice of the 
extraction regions corresponding to different components is based on 
a detailed imaging analysis as described in the next Section.
Finally, to perform spectral analysis, we extracted the spectra and 
computed the ancillary response file (ARF) and redistribution matrix file (RMF) 
for each observation separately with the command {\tt mkarf} and {\tt mkacisrmf}. 
Our default spectral analysis uses a local background that, given the small 
extent of the sources, is directly extracted from a source-free region on the 
same CCD-7 chip.

%%%%%%%%%%%%%%%%%%%%%%%%%%%%%%%%%%%%%%%%%%%%%%%%%%%%%%%%%%%%%%%%%%%%%%%%%%%%%
%%%%%%%%%%%%%%%%%%%%%%%%%%%%%%%%%%%%%%%%%%%%%%%%%%%%%%%%%%%%%%%%%%%%%%%%%%%%%
%%%%%%%%%%%%%%%%%%%%%%%%%%%%%%%%%%%%%%%%%%%%%%%%%%%%%%%%%%%%%%%%%%%%%%%%%%%%%
%
% SPIDERWEB GALAXY
%
%%%%%%%%%%%%%%%%%%%%%%%%%%%%%%%%%%%%%%%%%%%%%%%%%%%%%%%%%%%%%%%%%%%%%%%%%%%%%
%%%%%%%%%%%%%%%%%%%%%%%%%%%%%%%%%%%%%%%%%%%%%%%%%%%%%%%%%%%%%%%%%%%%%%%%%%%%%
%%%%%%%%%%%%%%%%%%%%%%%%%%%%%%%%%%%%%%%%%%%%%%%%%%%%%%%%%%%%%%%%%%%%%%%%%%%%%

\section{X-ray properties of the Spiderweb Galaxy: Imaging analysis}

The X-ray images of the Spiderweb Galaxy show a prominent, dominant unresolved component
due to the central AGN, and a clear diffuse emission limited to a radius 
of $\sim 12$ arcsec.  Due to the brightness of the unresolved emission, the 
small extension and the composite nature of the diffuse emission, 
a detailed imaging and morphological analysis of the image is 
required to account for multiple components.  The major contribution is due 
to the unresolved nucleus, that, due to its intensity, contaminates the 
surrounding regions up to radii of several arcsec.  Therefore, as a first step, 
we perform ray-tracing simulations to obtain an accurate modalization 
of the unresolved AGN, assuming the same instrumental setup of the real data.

\begin{figure*}
\begin{center}
\includegraphics[width=0.49\textwidth, trim=85 80 55 140, clip]{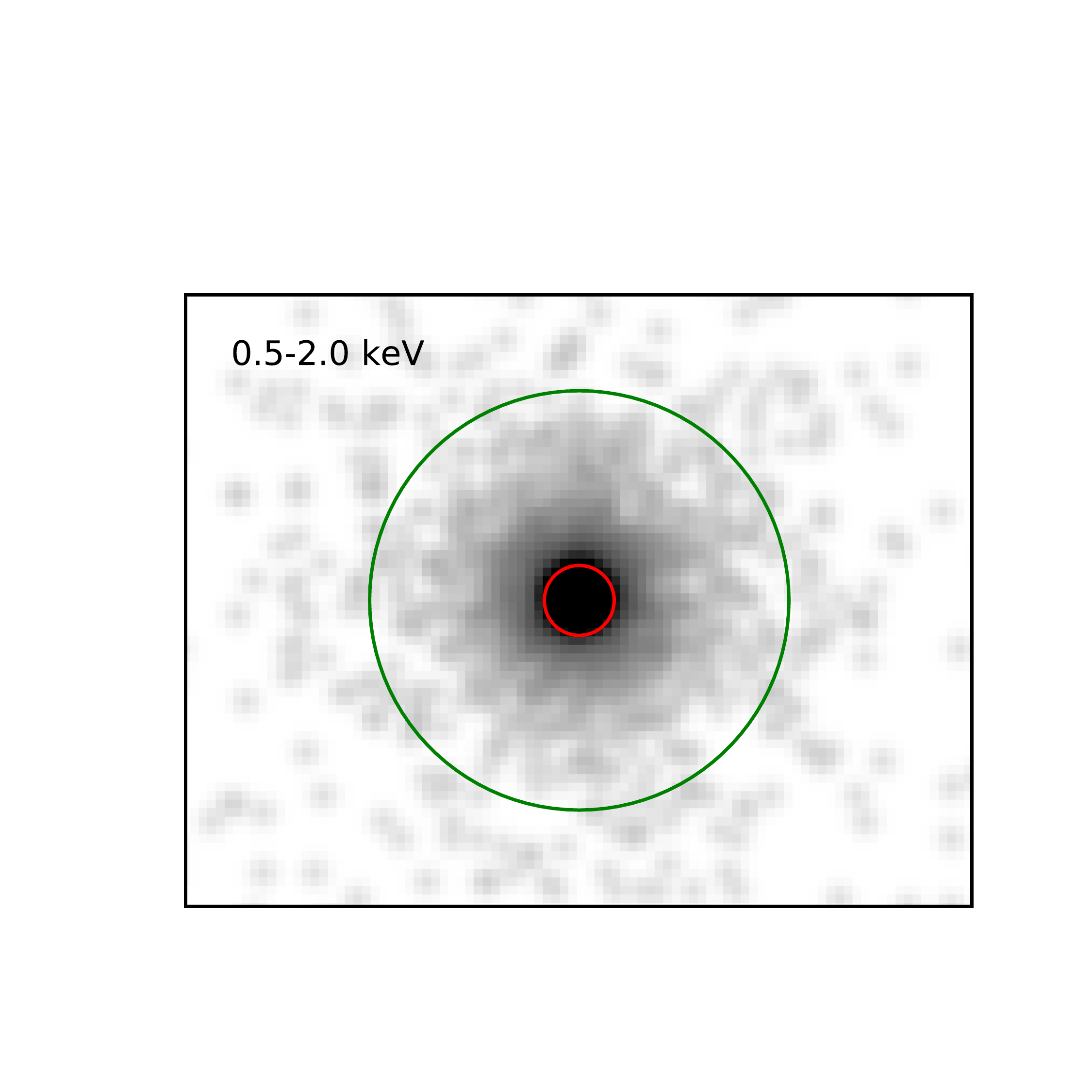}
\includegraphics[width=0.49\textwidth, trim=85 80 55 140, clip]{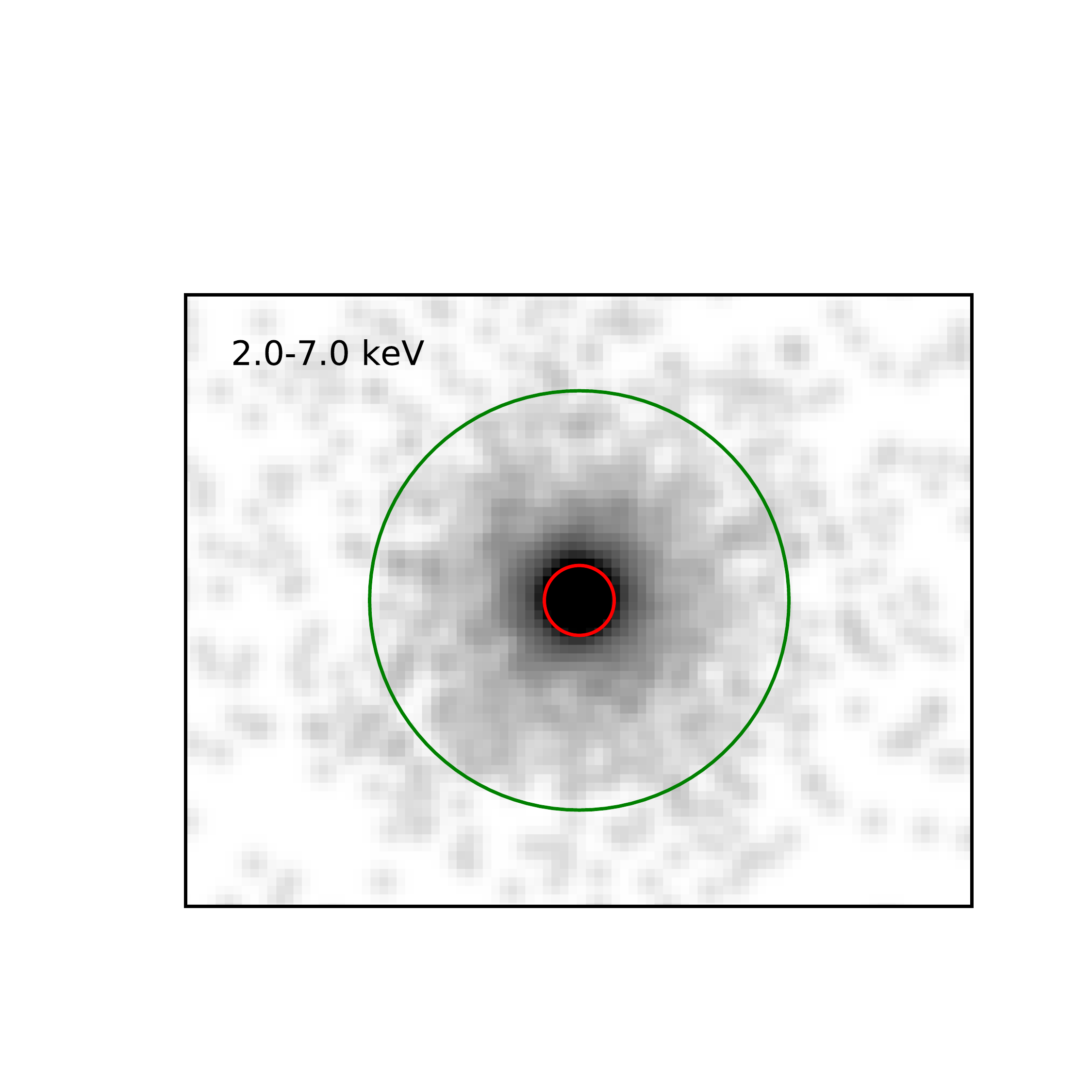}
\caption{Simulated images of an unresolved source with the same
observational characteristics (position, roll angle, exposure time and 
observation date) of the merged image of the Spiderweb Galaxy, in the 
0.5-2 keV (left) and 2-7 keV (right) band.  The inner red circle (2 arcsec radius) 
is used to extract the AGN spectrum, while the outer green circle  (12 arcsec radius)
encompasses the diffuse emission.   The color scale is the same on both bands. }
\label{chart}
\end{center}
\end{figure*}

\subsection{Ray-tracing simulations}

To perform accurate ray-tracing simulations, we used the Chandra Ray Tracer ({\tt Chart}) 
tool\footnote{\url{https://cxc.cfa.harvard.edu/ciao/PSFs/chart2/.}}.
First, we created spectral files with {\tt Sherpa} to reproduce the spectrum 
of the Spiderweb nucleus.  The spectral fits were performed with 
{\tt Xspec 12.11.1} \citep{1996Arnaud} for each Obsid in order to 
track possible fluctuations in the flux and spectral shape as well.  
To model the nuclear emission,  we considered the emission in the 
extraction region within a radius of 2 arcsec, which is clearly dominated by 
the AGN, but it still may include some minor, unknown contribution from diffuse 
emission. Despite this, the background 
subtraction in this first step was computed simply rescaling the emission in an annulus 
with inner and outer radii of 3$^{\prime\prime}$ and 5$^{\prime\prime}$, 
respectively, centered on the nucleus.  
This background subtraction can be considered as an 
amenable approximation to the non-AGN component within 2 arcsec, and
it is sufficient to obtain spectra accurate enough to compute the 
expected image of the point spread function (PSF), which 
has a mild dependence on the photon energy\footnote{See, for example,
\url{https://cxc.cfa.harvard.edu/ciao/PSFs/psf_central.html} and references therein.}.  
The spectral fits of the AGN obtained in this
first step are not discussed, and the detailed description of the spectral 
analysis, with all the components properly included, are presented in Section 5.
Spectral analysis of the diffuse emission, both thermal and nonthermal, 
is presented in Section 6.

The best-fit spectral parameters of a power law emission with intrinsic absorption 
obtained for each Obsid are given as input for the image simulations.  
We perform 10 ray tracing simulations for each one of the 22 Obsid.  We pay attention 
to normalize the exposure time of the simulations to the actual effective
exposure time of each Obsid after data reduction.
Eventually, we create the evt2 file corresponding to a given simulation with 
{\tt Marx}\footnote{\url{https://space.mit.edu/cxc/marx/} .}.  We merge the ten files of 
each Obsid to obtain an event file corresponding to an exposure $10\times$ larger 
than in the real data. Finally, we merge all the simulated Obsid by reprojecting 
the evt2 files onto the reference frame of Obsid 21483 as in the reduction of real data. 
We obtain the soft and hard band images of the simulated source directly 
from the reprojected and merged file, dividing the outcome by a factor of 10
to obtain an image with the same nominal exposure time, but a Poisson noise a factor 
of $\sim 3$ lower.  The images of the merged PSF in the soft and hard bands are 
shown in Figure \ref{chart}, where the extraction region  
is shown as a red circle (with a radius of 2 arcsec), 
while the approximate extent of the diffuse emission is 
shown as a green circle (with a radius of 12 arcsec).  The color scale is the same 
in both bands.  The large majority of the signal is within a radius of 2 arcsec, however
the wings of the PSF contribute a significant fraction of the total intensity in the
region between 2 and 12 arcsec, with the PSF having slightly more extended wings in the hard band. 
From the simulated image we can robustly estimate that in the soft band 94.8\% of the 
source emission is within 2 arcsec, and 4.0\% between 2 and 12 arcsec, while only 1.2\%
is beyond 12 arcsec.  
% 1.5\% WITHIN 2.6 AND 4.6 arcsec in the soft band.
In the hard band these values are 91.7\%, 5.2\% and 3.1\%, respectively. 
% 1.9\% WITHIN 2.6 AND 4.6 arcsec on the hard band
From a preliminary estimate, the counts contributed in our data (full exposure) by the 
AGN in the 2-12 arcsec region, where the diffuse emission dominates, amounts to 
$\sim 300$ net photons in the 0.5-7 keV band. Since we measure a total of 
$\sim 900$ net counts in the 0.5-7 keV band in the 2-12 arcsec region, 
this implies that a non-negligible fraction ($\sim 0.33$) of the 
diffuse signal is due to the wings of the AGN emission.   

\subsection{Reconstruction of the diffuse emission}

The next step is to accurately normalize the simulated image to the actual nuclear emission. 
We note, in fact, that the normalization of the spectra used for the ray-tracing simulations, 
were based on an approximate background subtraction. Therefore, while we trust the 
distribution of the emission according to the combined PSF, the overall normalization can 
be offset by a few percent.  While this translates in an almost
negligible correction to the normalization of the AGN emission, 
it is very important to estimate the relation between the AGN normalization and the 
diffuse component expected in the inner circle of 
2 arcsec (corresponding to $\sim 17$ kpc), 
for example, to infer the degree of "cool coreness".
In both cases, we note that in this step we consider both thermal and nonthermal 
components in the extended emission, since it is impossible, without spectral analysis, 
to disentangle the two components. 

Therefore, we proceed as follows. We measure the emission in an annulus 
with inner and outer radius of 3 and 5 arcsec, respectively, that would
correspond to the standard choice in case of an isolated, unresolved sources
\citep[see][]{2001Tozzib}. Within this annulus we find 1.5\% and 1.9\% of the total emission 
from the nucleus in the soft and hard bands, respectively.  In addition, we also estimate
the instrumental (plus unresolved extragalactic X-ray background) from an annulus with 
inner and outer radius of 16 and 29.5 arcsec, respectively. This last region (from where two 
unresolved sources have been previously identified and removed) has negligible contribution 
from the extended emission and the AGN emission, if any. Therefore, we estimate the 
instrumental background per pixel as $b_i=B_i/A_B$, where $B_i$ is the photometry in 
the 16-29.5 arcsec annulus and $A_B$ is the area of the annulus.  
The surface brightness of the diffuse emission in the range 3-5 arcsec can 
be written as $d=[D-(A_d\times b_i) - p_d\times S_{AGN}]/A_d$, where $D$ 
is the total photometry in the 3$^{\prime\prime}$-5$^{\prime\prime}$ annulus, 
$A_d$ the area of the annulus, 
and $p_d$ is the fraction of the actual total signal from the AGN falling in the annulus, 
(computed from the simulated PSF images).  Finally, the total photometry of the 
nuclear emission is $S_{AGN} = [C_{AGN}-A_{AGN}\times b_i - n_d\times A_{AGN}\times d]/p_{AGN}$, 
where $C_{AGN}$ is the photometry in the 2 arcsec extraction region, $A_{AGN}$ is the
area of the extraction region, and $p_{AGN}$ is the fraction of the AGN emission 
falling into the extraction region (computed again from the simulated PSF images).  
This system can be easily solved, apart from the unknown, free parameter $n_d$ that 
is defined as the ratio of the average surface brightness within a radius of 
2 arcsec with respect to the average surface brightness in the range 3-5 arcsec. 
We use the parameter $n_d$ to quantify our ignorance on the surface brightness of
the diffuse emission in the 2 arcsec extraction region 
in terms of multiple of the average surface brightness observed in 
the 3.5$^{\prime\prime}$-5$^{\prime\prime}$ annulus.

Since it is likely to have an increase 
in the central surface brightness with respect to the outer annulus, 
we expect $n_d\geq 1$.  In the case of a strong, small-scale cooling 
flow the spectral shape may change on the scale of $\sim 1$ arcsec (corresponding to 
$\sim 8.5$ kpc), and, therefore, the parameter $n_d$ may be different in the soft and in the hard band.  
If we consider, as a reference, the strong cool-core cluster CL1415 at $z\sim 1$ 
\citep{2012Santos}, and the same physical
regions we are using here, we find that the surface brightness in the inner region is 3.3 
and 2.7 times larger than that in the annulus in the soft and hard bands, respectively. 
Despite the fact that
CL1415 is a mature, massive cluster observed at much lower redshift, 
we adopt values in the range 1-6 for the parameter $n_d$, where $n_d=1$
correspond to a constant surface brightness, $n_d\sim 3.5$ correspond to a fully 
developed cool core, and $n_d\sim 6$ to an exceptionally peaked cool core.

We compute the total photometry of the AGN in the soft and hard bands as a function of the 
enhancement factor $n_d$, and use these values to normalize the simulated images and compare them
with the real data. The elative normalization of the diffuse emission within a radius of 2 arcsec 
(the $n_d$ parameter), nevertheless, has a marginal impact on the AGN emission.
In the soft band the uncertainty
on the AGN normalization is at most 100 net counts for a strong cool core with $n_d=3$
and 200 net counts for an extreme cool core with $n_d=6$ (corresponding to a relative uncertainty of 
1.8\% and 3.5\%, respectively). In the hard band the uncertainty is much less, due to the 
lower surface brightness measured on the 3$^{\prime\prime}$-5$^{\prime\prime}$ annulus. 
The hard-band net counts expected from the 
diffuse emission within 2 arcsec are in the range 8-15, corresponding to an 
uncertainty of 0.2\%-0.4\% in the normalization of the nuclear hard emission.  

If we focus on the estimated flux from diffuse emission within 2 arcsec, 
we find that the energy flux associated with diffuse
emission within a radius of 2 arcsec is $(1.17\pm 0.09)\times 10^{-15}\times (n_d/3)$
erg s$^{-1}$ cm$^{-2}$ and $(2.9\pm 0.4)\times 10^{-16}\times (n_d/3)$
erg s$^{-1}$ cm$^{-2}$ in the soft and hard (2-10 keV) bands, respectively.  
These values have been obtained using the conversion factors at the aimpoint
listed in Table 3 of Paper I.  The lower value in the hard band
is due to the low hardness ratio of the emission measured in the 
3$^{\prime\prime}$-5$^{\prime\prime}$ annulus, 
which is measured to be $HR=-0.85$, and it provides a first, preliminary hint  
that the diffuse emission may be dominated by thermal
bremsstrahlung at least in this region.

\begin{figure*}
\begin{center}
\includegraphics[width=0.32\textwidth, trim=85 120 55 170, clip]{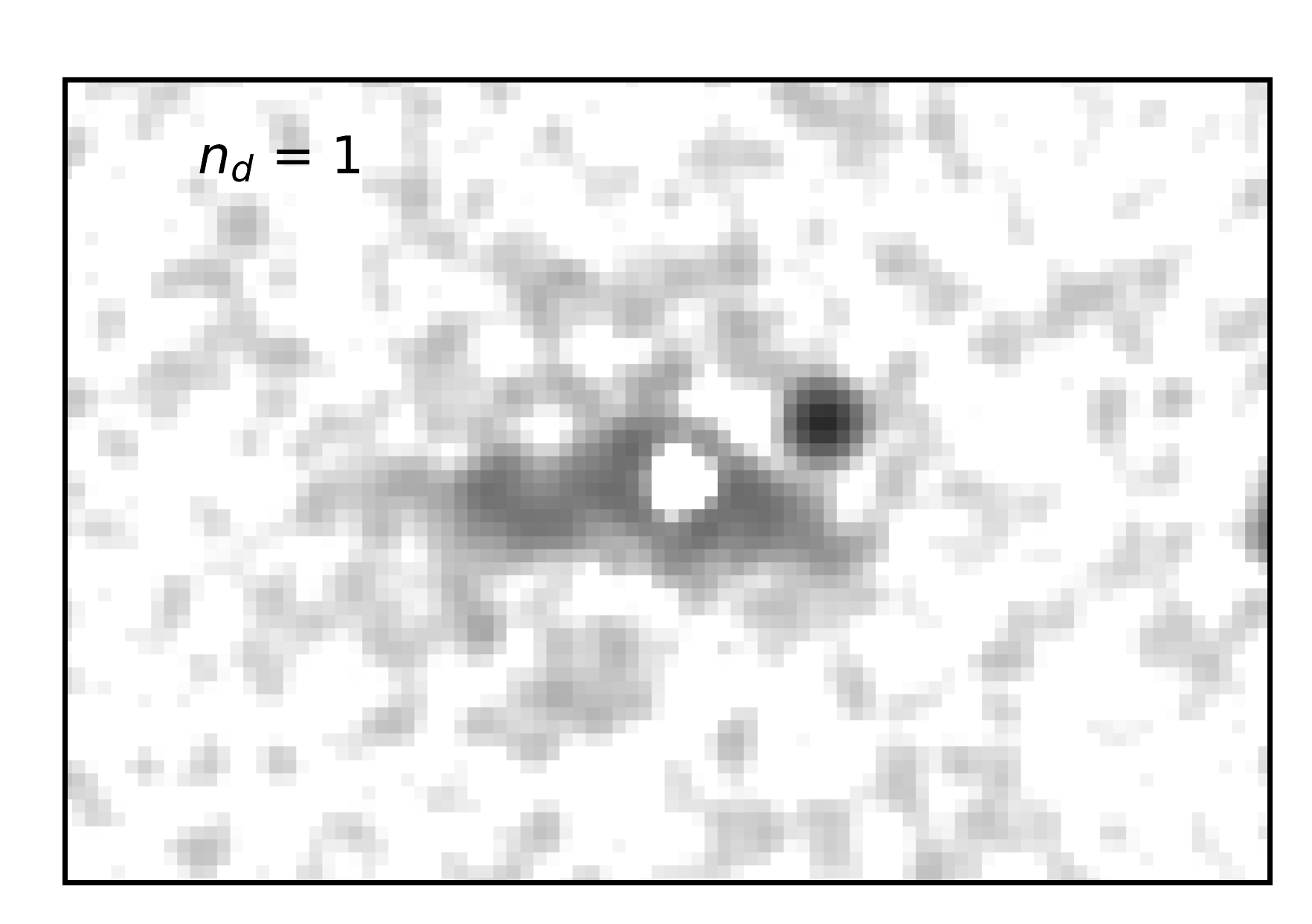}
\includegraphics[width=0.32\textwidth, trim=85 120 55 170, clip]{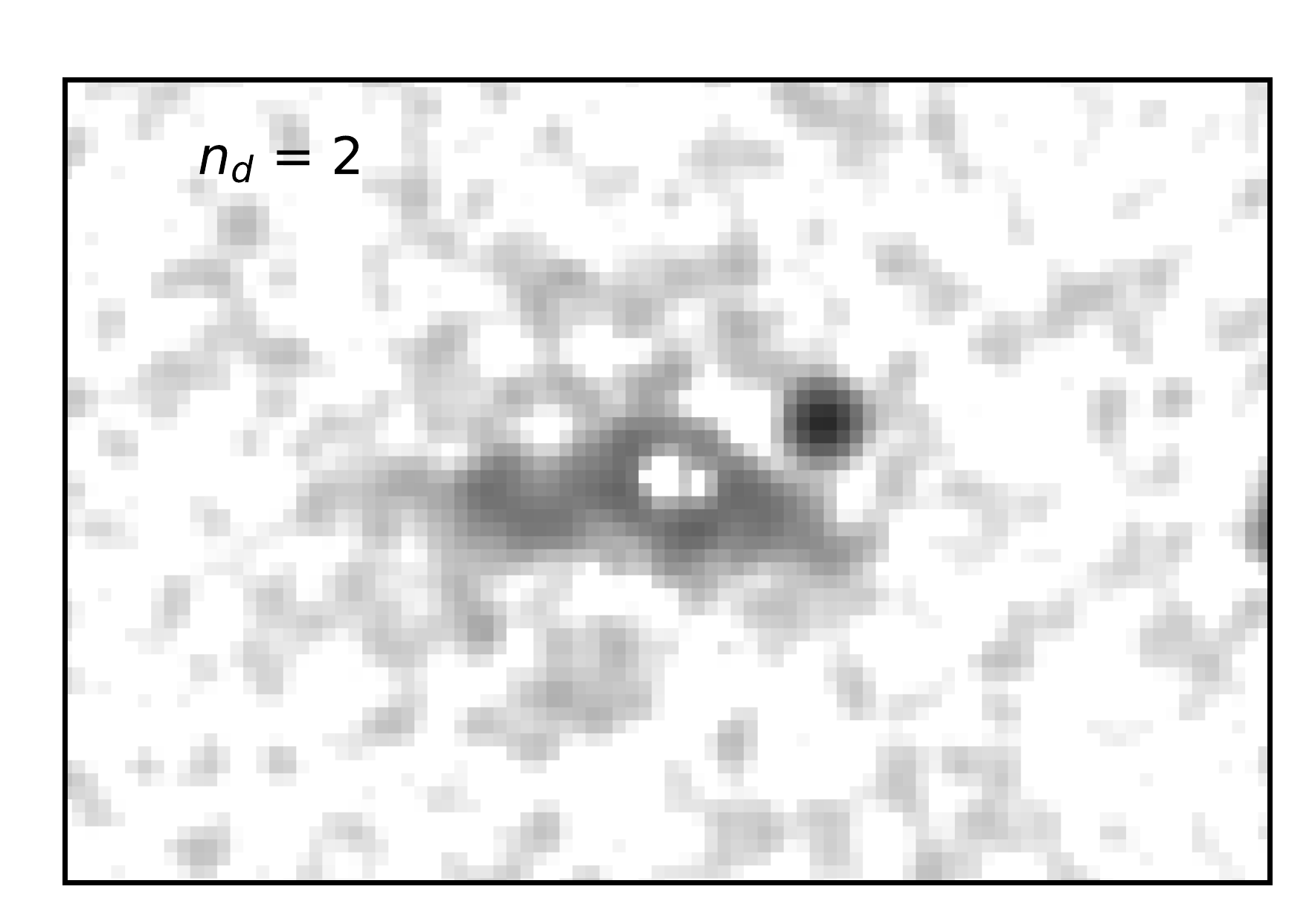}
\includegraphics[width=0.32\textwidth, trim=85 120 55 170, clip]{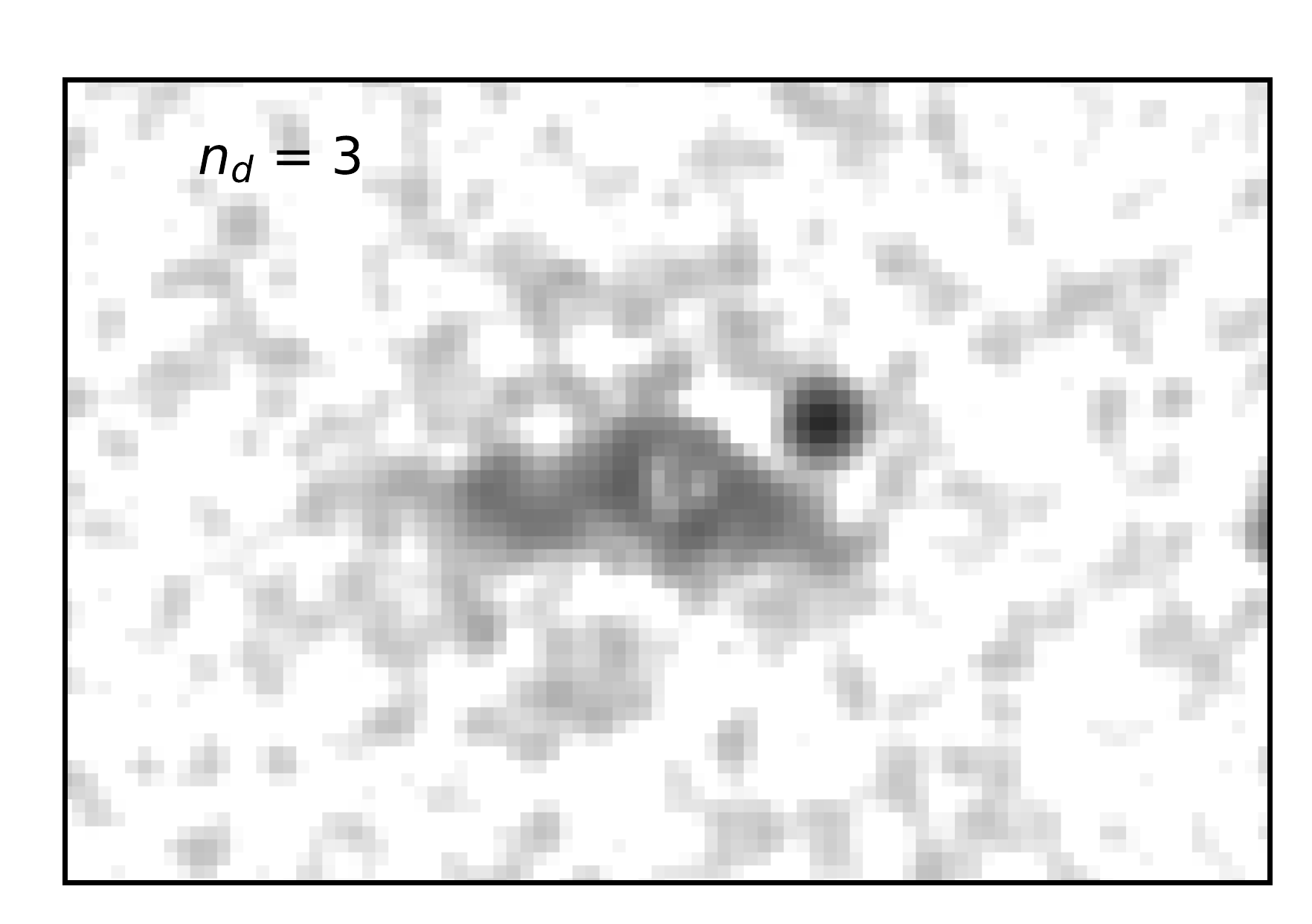}
\includegraphics[width=0.32\textwidth, trim=85 120 55 170, clip]{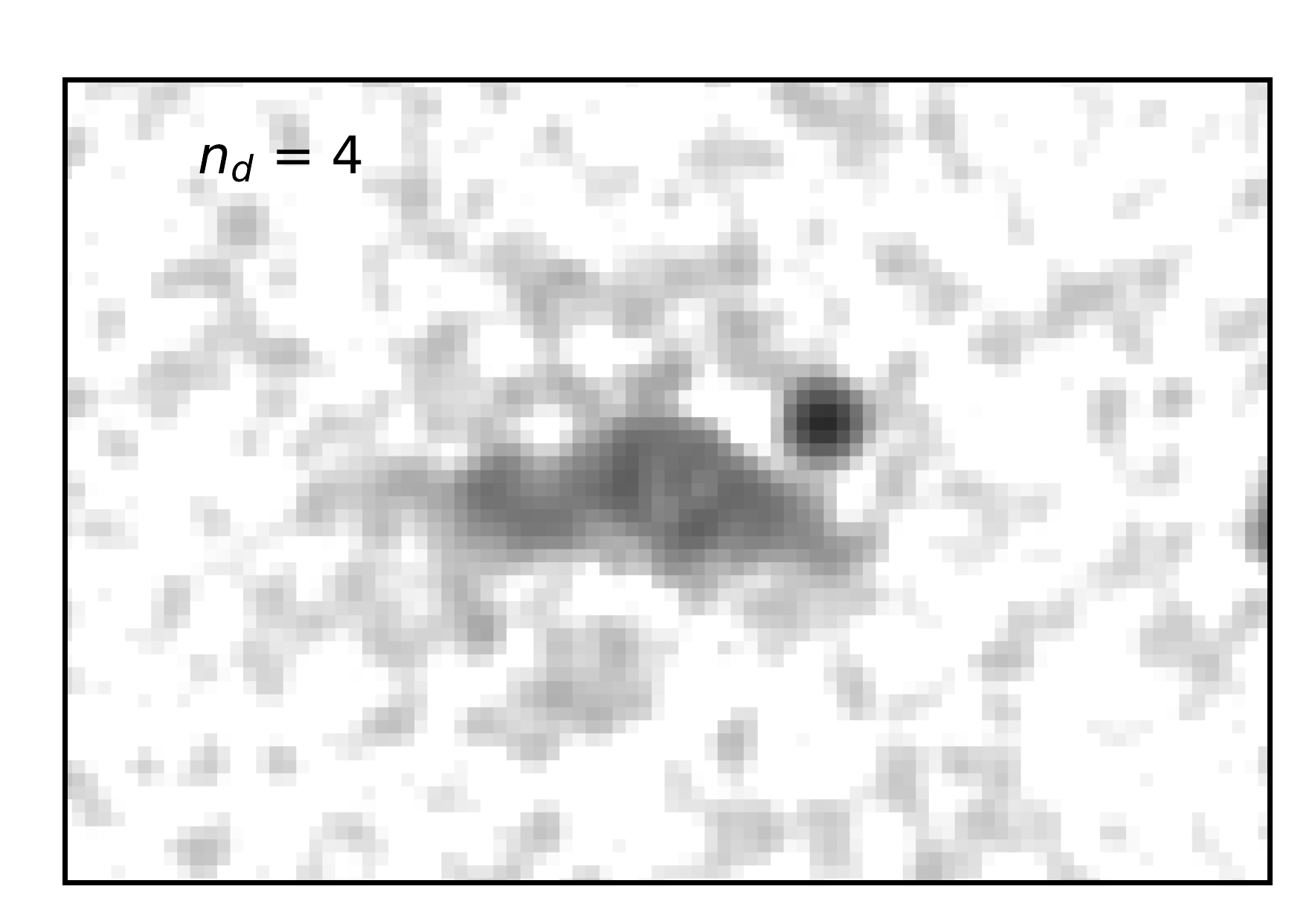}
\includegraphics[width=0.32\textwidth, trim=85 120 55 170, clip]{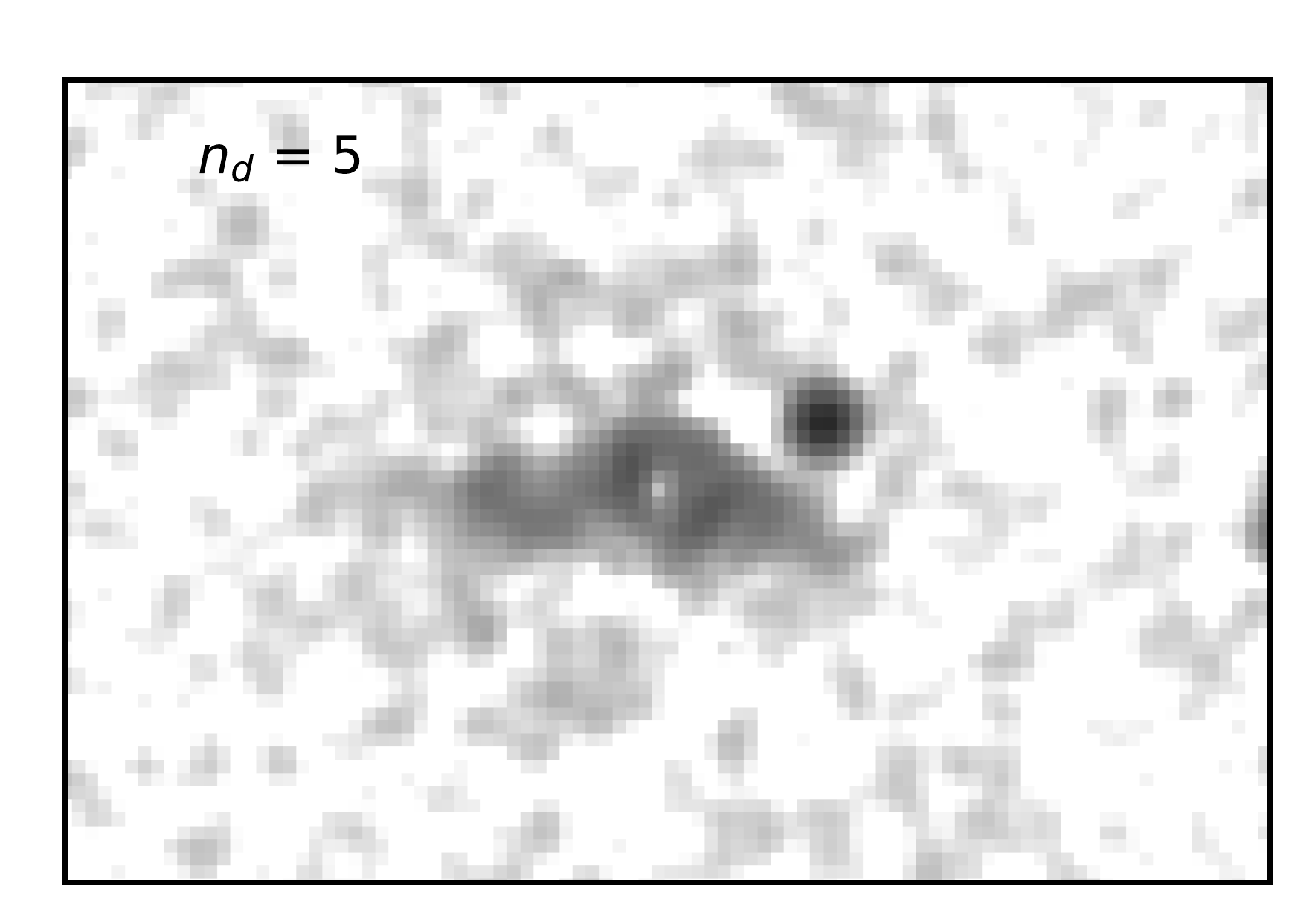}
\includegraphics[width=0.32\textwidth, trim=85 120 55 170, clip]{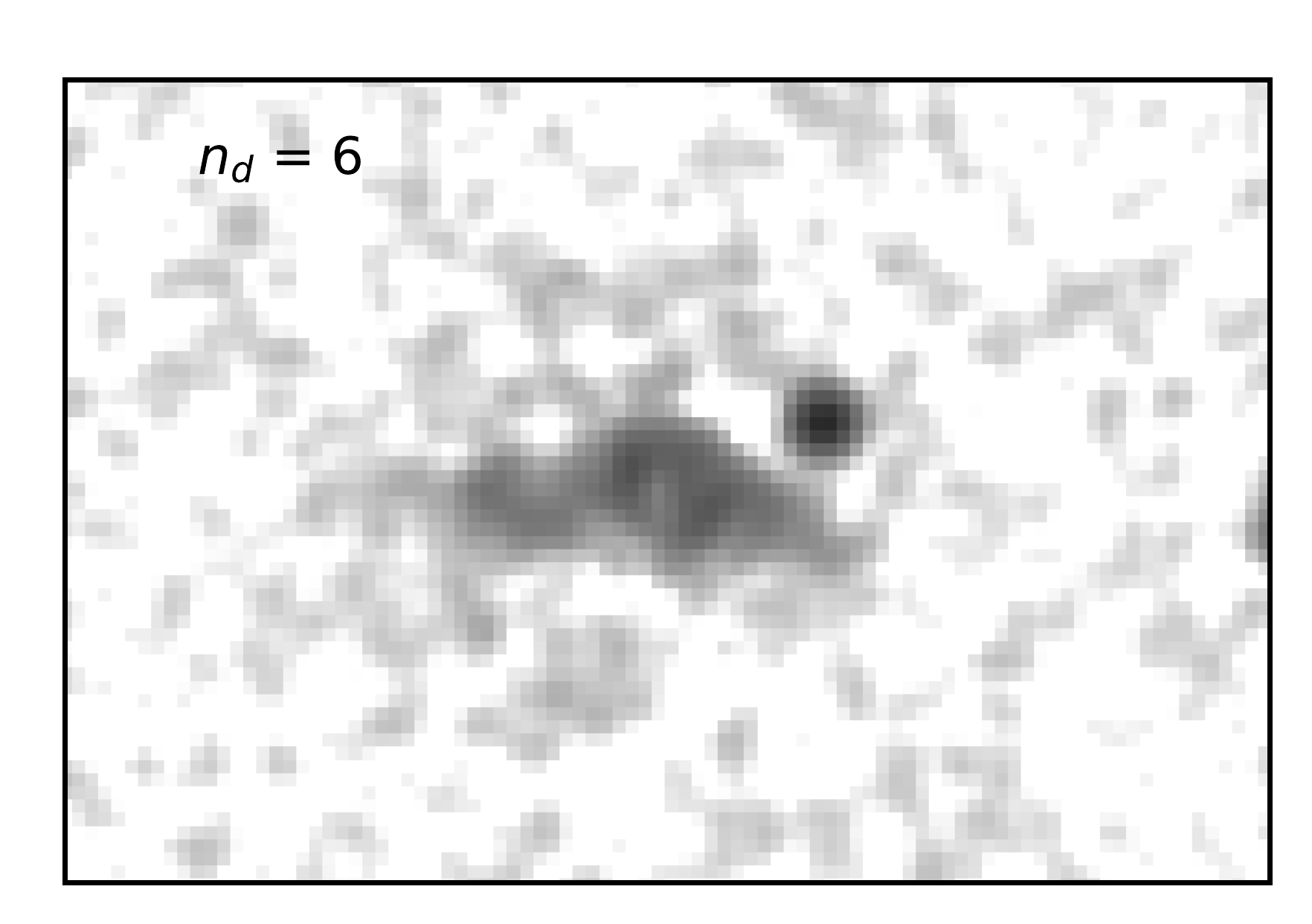}
\caption{Background-subtracted soft band images of the Spiderweb Galaxy after removing the
simulated image of the AGN.  The image has been smoothed with a 
Gaussian kernel with a sigma of 1 pixel, slightly degrading the effective
resolution.  The normalization of the surface brightness of the diffuse emission 
within a radius of 2 arcsec is free and it is parametrized with $n_d$ ranging from 
1 (upper left image) to 6 (lower-right image).  There are a few slightly negative pixels
in the image, so that the average pixel value far from the Spiderweb Galaxy is zero. The unresolved
source close to the center is an AGN that has not been subtracted from the image.}
\label{sb_soft_nd}
\end{center}
\end{figure*}

Assuming that the diffuse emission in the hard band is negligible, we focus on the soft
band and attempt to reconstruct an image of the Spiderweb Galaxy removing the 
dominant AGN emission.  This can be obtained simply by subtracting
the normalized PSF image from the real data.  Clearly, the slope of the 
PSF image on the scale of 1 arcsec is so steep that an uncertainty on the order 
of 1 pixel on the relative astrometry of the real data and of
the ray-tracing simulation does not allow an accurate subtraction, and 
may result in a distorted distribution of net counts. 
After the subtraction, in
fact, the resulting image shows a small number of pixels with negative values 
(about ten), all very close
to the center.  Such negative values can be eliminated with a simple redistribution, 
consisting in averaging iteratively the maximum and minimum pixel values within the 
inner 2 arcsec, untill no negative pixels are left (about 10 iterations). This procedure 
preserves the photometry, but the final distribution of the pixel 
values is not accurate on the scale of $\sim 1$ arcsec.  
Finally, we smooth the images with a 2D Gaussian kernel with a size of 1 pixel, to obtain 
a less noisy image without loss of information, and to subtract the average
instrumental plus sky background as estimated from 
the 16$^{\prime\prime}$-29.5$^{\prime\prime}$ annulus.   

The background-subtracted, soft-band images of the Spiderweb Galaxy after the removal of the 
AGN emission are shown in Figure \ref{sb_soft_nd} for $n_d$ ranging from 1 to 6.  
As previously mentioned, the presence of a hole in the center 
for $n_d<3$ is an artifact of the method used
to reconstruct the image, and reflects our ignorance on the 
distribution of surface brightness on scales on the order of 1 arcsec in the 
regions dominated by the AGN emission. 
As a consequence, we focus only on the global photometry 
within a radius of 2 arcsec, ignoring anomalies in the surface brightness 
distribution. The images shown in Figure \ref{sb_soft_nd}, however,
still allow us to appreciate that the choice $n_d\geq 4$ corresponds to
an almost uniform, prominent core in the innermost
17 kpc, nicely matching the surface brightness level robustly measured 
outside 2 arcsec (where the effects of a sub-pixel mismatch are much 
smaller due to the rapidly flattening shape of the {\sl Chandra} PSF).  
As previously discussed, the choice 
$n_d=4$ corresponds to a prominent, but still plausible, cool core. Thus, we 
adopt $n_d=4$ as a reference value to estimate the central ICM density, 
stressing that the exact value has a mild impact on all the
conclusions derived in this work.

\begin{figure*}
\begin{center}
\includegraphics[width=0.49\textwidth]{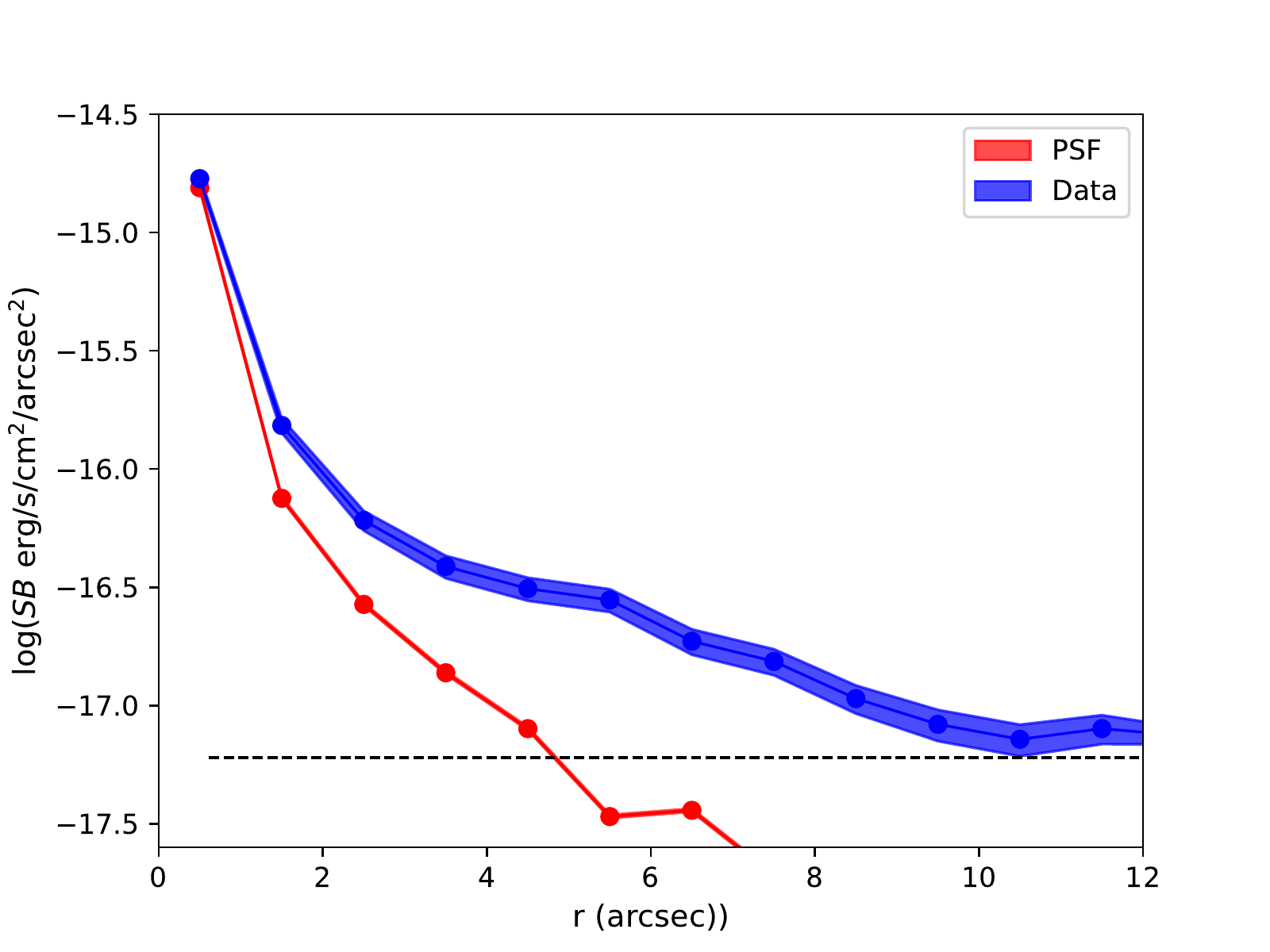}
\includegraphics[width=0.49\textwidth]{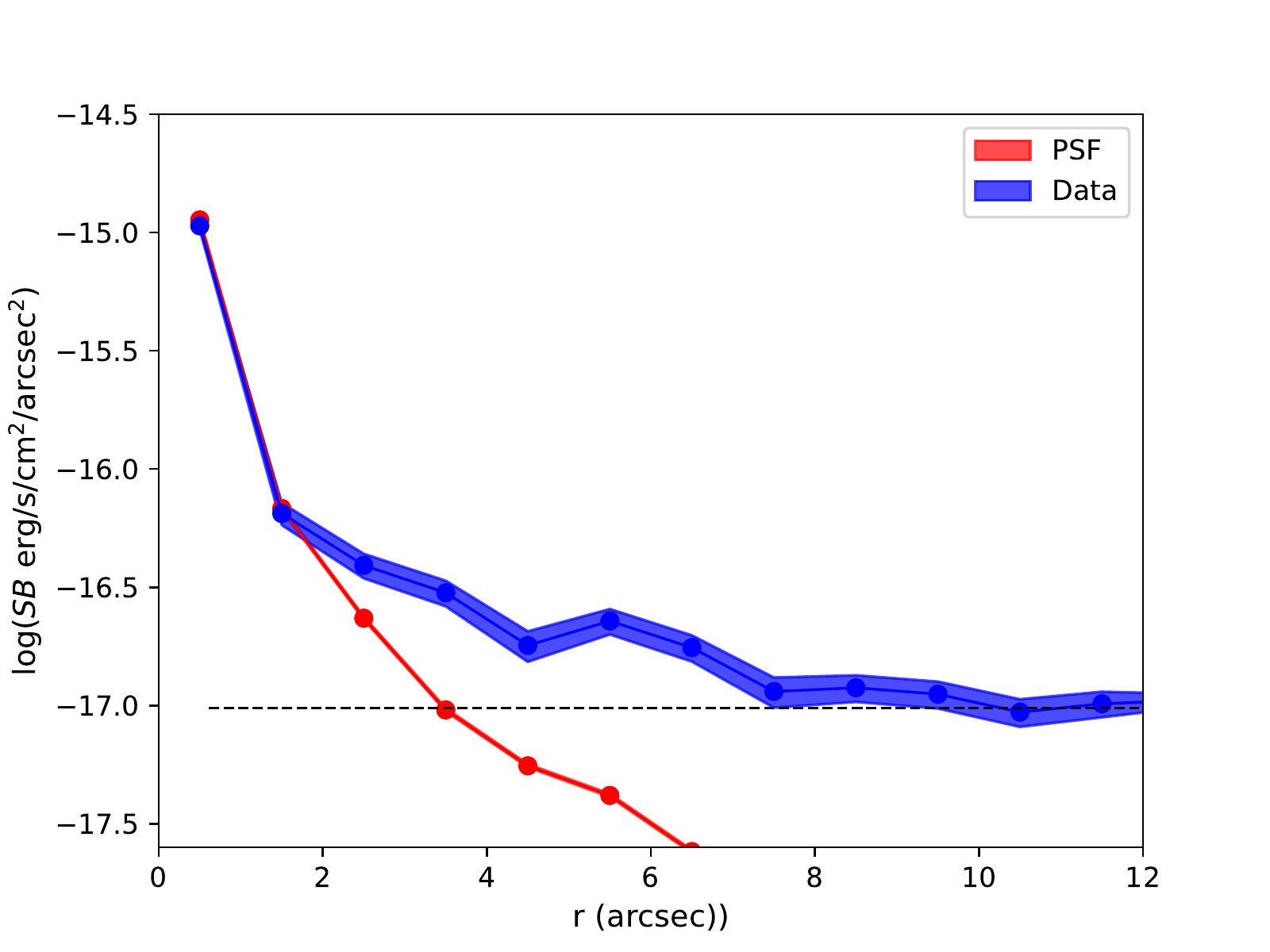}
\caption{Left: Azymuthally averaged profile of the AGN simulated 
image (normalized to $n_d=4$, red line) compared to the soft-band Spiderweb profile (blue line). 
The horizontal line shows the instrumental plus unresolved X-ray 
background level. Right: same in the hard (2-7 keV) band. 
}
\label{psf_image}
\end{center}
\end{figure*}

\begin{figure}
\begin{center}
\includegraphics[width=0.49\textwidth]{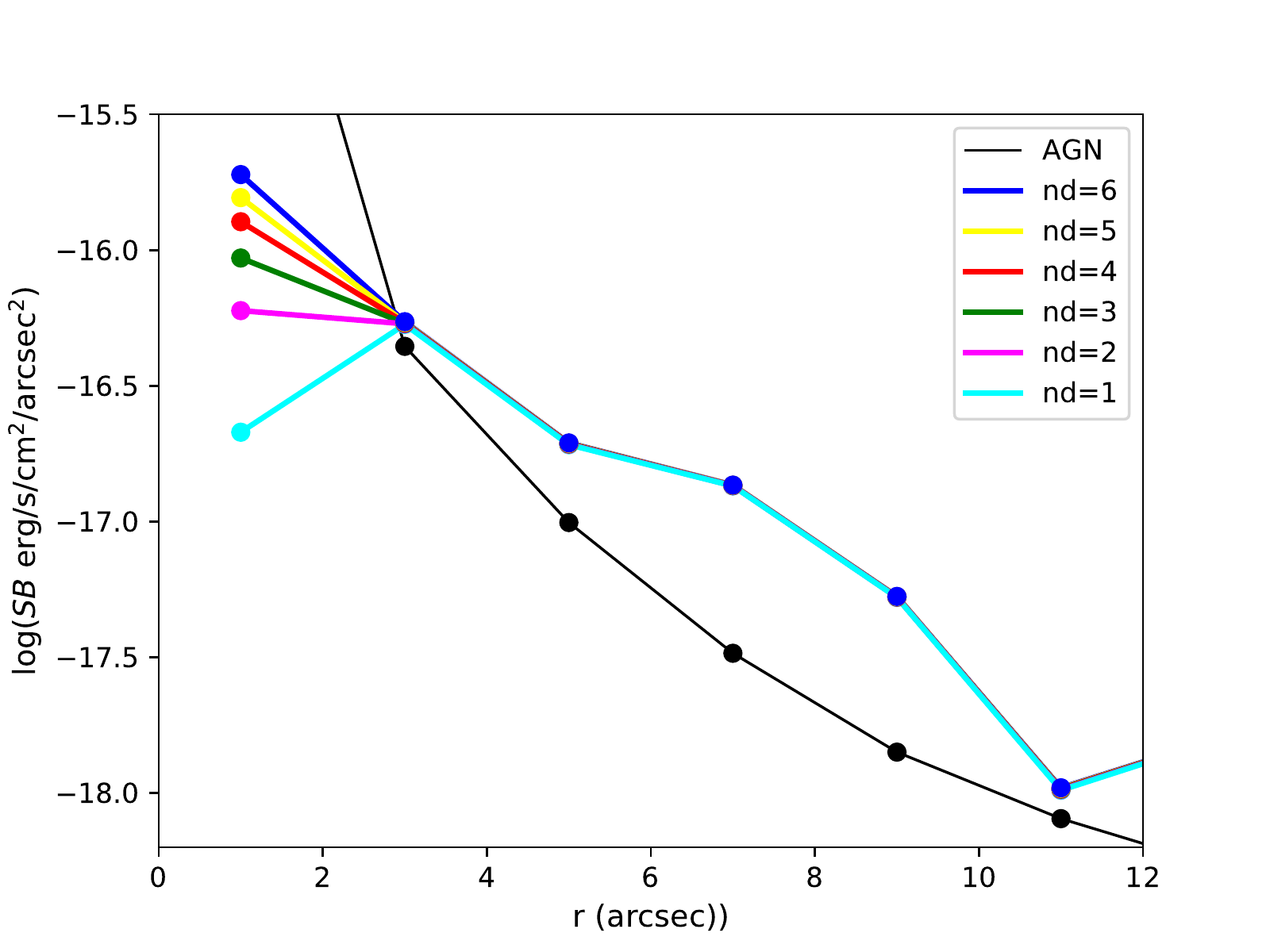}
\caption{Azymuthally averaged profile of the diffuse emission
(after subtraction of the AGN emission and of the background)
at different $n_d$ values.  As expected, only the innermost bin is affected
by the choice of $n_d$, while the indirect effects on the normalization 
of the AGN contamination, thus on the overall normalization of
the diffuse emission, is negligible.  The black line is the simulated AGN emission for
comparison. 
}
\label{ext_em}
\end{center}
\end{figure}

To further support this choice, we look directly at the surface brightness profile of the
simulated AGN and the real data, in both bands.  We note that the effect of the normalization
of the AGN image corresponding to different values of $n_d$ is affecting only the 
first bin of the profile ($r<1$ arcsec), while all the other bins 
(at $r>1$ arcsec) are practically unaffected.  In the left panel of 
Figure \ref{psf_image} we show the azymuthally averaged profile of the AGN 
simulated image (for the specific choice $n_d=4$)
compared to the Spiderweb profile in the soft band.   We notice that 
the data show signal in excess of the unresolved nuclear source at $r\geq 1.5$ arcsec, 
while in the hard band (shown in the right panel), the excess starts to appear only 
at $r\geq 2.5$ arcsec.  
Since the measurement of the diffuse emission at $r\geq 1.5$ arcsec 
is practically independent of the $n_d$ value, we note that an approximately constant increase
of the surface brightness toward the center is obtained for $3<n_d<4$. 
This is shown in Figure \ref{ext_em}, where we plot the azymuthally averaged 
surface brightness profile of the extended emission in the soft band
(that is, the difference of the data and the PSF model shown in the left panel of
Figure \ref{psf_image}).  
Unfortunately, at the moment we do not have compelling argument to strongly 
constrain the value of $n_d$.  The implications of values different from
$n_d=4$, as adopted in this work, will be investigated in Lepore et al. (in preparation).

From the right panel of Figure \ref{psf_image}, 
we also notice that the hard band emission is less extended 
than the soft emission, that is detected well above the
background up to 12 arcsec ($\sim 100$ kpc) from the nucleus of the Spiderweb Galaxy.
Instead, in the hard band there is no excess at 
$r\leq 2^{\prime\prime}$, and the diffuse emission
is less significant and limited to the range $3^{\prime\prime}<r<7^{\prime\prime}$. 
This suggests that the diffuse emission is contributed
by two distinct components.  The one with  larger hardness ratio is less extended and
is most likely associated with the IC from the 
relativistic population of electrons in the jet.  The  
soft component is more extended up, to a distance of $\sim 100$ kpc 
from the center, and is possibly associated with hot gas.
In the following section we proceed with the spectral analysis of the 
AGN-dominated emission included within a radius of 17 kpc, while 
in Section 6 we investigate the diffuse emission at radii $>2^{\prime\prime}$, 
after accounting for the background and AGN contamination.

\section{X-ray properties of the Spiderweb Galaxy: spectral analysis of the central AGN}

\subsection{Global nuclear emission}

Compared to the shallower X-ray observation presented in \citet{2002Carilli}, 
our new, deep exposure allows us to obtain a more detailed spectral 
analysis of the nuclear emission, and, in particular, to investigate the 
contribution of the diffuse component, on the basis of the accurate imaging 
analysis presented in the previous Section.  In the following, we 
perform our reference analysis of spectra extracted separately from each Obsid, 
and for which we consider the respective Obsid ARF and RMF calibration files. 
Ultimately, this info are considered all together during the fitting procedure.
This approach allows us to keep track of the different 
effective area corresponding to each Obsid.  Eventually, we also analyze 
the spectrum obtained by merging the single Obsid spectra, 
using cumulative ARF and RMF files obtained by weighting single
Obsid files by the corresponding exposure times.
We find that this approach gives results that are in agreement
with our reference spectral analysis.  Given the brightness of the source, 
each Obsid has at least 300 net counts within the extraction region (a 2$^{\prime\prime}$ radius circle).  
Therefore, we compute the ARF and RMF files for each one of the 22 Obsid. 

We fit the nuclear emission with {\tt Xspec 12.11.1} \citep{1996Arnaud}
over the  0.5-9.0 keV energy range, adopting the simplest possible
model, as done for the analysis of the protocluster members in Paper I.  
The AGN emission is, therefore, 
described by an intrinsically absorbed power law, using the model components {\tt zwabs}
and {\tt powerlaw}. The galactic absorption is described with the model {\tt tbabs}\footnote{See \url{https://heasarc.gsfc.nasa.gov/xanadu/xspec/manual/node268.html} .}, 
and its value is fixed to $3.18\times$ 10$^{20}$ cm$^{-2}$ according to the HI map of
the Milky Way \citep{2016HI4PI}.  

As previously discussed, the contribution from diffuse emission within 2 arcsec 
cannot be constrained accurately but it must be parametrized by the enhancement 
factor $n_d$, that represents the ratio of the surface brightness within 2 
arcsec over that measured in the 3-5 arcsec annulus.  
As a preliminary test, we perform a straightforward spectral analysis assuming 
two background files: the first including only the instrumental plus unresolved 
background, sampled in the annular region between 16 and 24 arcsec; the second 
including instrumental background and extended emission contribution directly 
sampled in the 3-5 arcsec annulus. The last choice corresponds to the 
assumption that the surface brightness of the diffuse emission is constant within 
5 arcsec with no central enhancement ($n_d=1$).
We find that the fit obtained in the second case significantly improves, with 
$\Delta C_{stat}\sim 23$.  This simply confirms that we must account for some 
contribution from diffuse emission within 2 arcsec, possibly with $n_d>1$, 
as we already shown from the imaging analysis.  At the same time, the 
best-fit spectral parameters of the AGN emission model are in agreement within 
1 $\sigma$ irrespective of the background treatment, 
confirming that the contribution of the diffuse emission is 
very low compared to the nuclear emission.  
This is expected also from photometry, since we measure 9710 net counts in the 0.5-7 keV band
within 2 arcsec, while, from the imaging analysis, we expect only $95\times (n_d/3)$ 
and $8\times (n_d/3)$ net counts in the soft and hard bands, respectively, from
diffuse emission.  Overall, within a radius of 2 arcsec, 
we expect a fraction of $0.015\times (n_d/3)$ of the total emission to be 
not associated with the AGN in the soft band, and only $0.003\times (n_d/3)$ in the hard band.
Such contribution is so small that $n_d$ cannot be constrained by spectral analysis. 

As a next step, we search for the best 
spectral modelization of the diffuse emission, fitting
the emission within the 3-5 arcsec annulus after instrumental background subtraction and
after accounting for 1.5\% of the nuclear emission expected in that region. We find that
statistically equivalent fits are obtained with a thermal {\tt mekal} model 
or power law, with $kT\sim 3.0$ keV and $\Gamma \sim 2.7$, respectively. On the other 
hand, the fit considerably gets worse if $\Gamma=2$ is assumed (with $\Delta C_{stat}\sim 16$).  
This shows us that the diffuse component in this region is better described by a soft, 
thermal emission than by a nonthermal, $\Gamma\sim 2$ power-law emission.  
Therefore, we describe the weak diffuse component within 2 arcsec 
by using a {\tt mekal} model with $kT = 3$ keV.  

Incidentally, we note that we do not consider here a nuclear 
plus diffuse contribution possibly associated with star formation.  
The total SFR associated with the core of the Spiderweb Galaxy 
plus the {star-forming satellite galaxies (the so-called {\sl flies})}
has been estimated to range from $70 M_\odot$/yr \citep{2009Hatch} 
to $1400 M_\odot$/yr \citep{2012Seymour}, and is expected to be 
significantly obscured. If we maximize the star-formation (SF) 
contribution with an unabsorbed spectrum with $\Gamma\sim 2.0$, 
we find that in our data we expect $\sim 2.9\times ({\rm SFR}/100 M_\odot/yr)$ 
net counts in the 0.5-7.0 keV range.  Considering the maximum value of 
$1400 M_\odot$/yr, we estimate an upper limit of 40 net counts to 
the SF contribution within 2 arcsec, which is about 0.3\% of the AGN component and, 
therefore, can be safely ignored.  The possible contribution from 
SF associated with the Ly$\alpha$ halo, has been estimated to be in the range $60-140 M_\odot$/yr, 
and is also expected to be completely negligible in the 
annular region between 2$^{\prime\prime}$ and 12$^{\prime\prime}$ 
that we consider in the Section 6.

\begin{figure*}
\begin{center}
\includegraphics[width=0.49\textwidth]{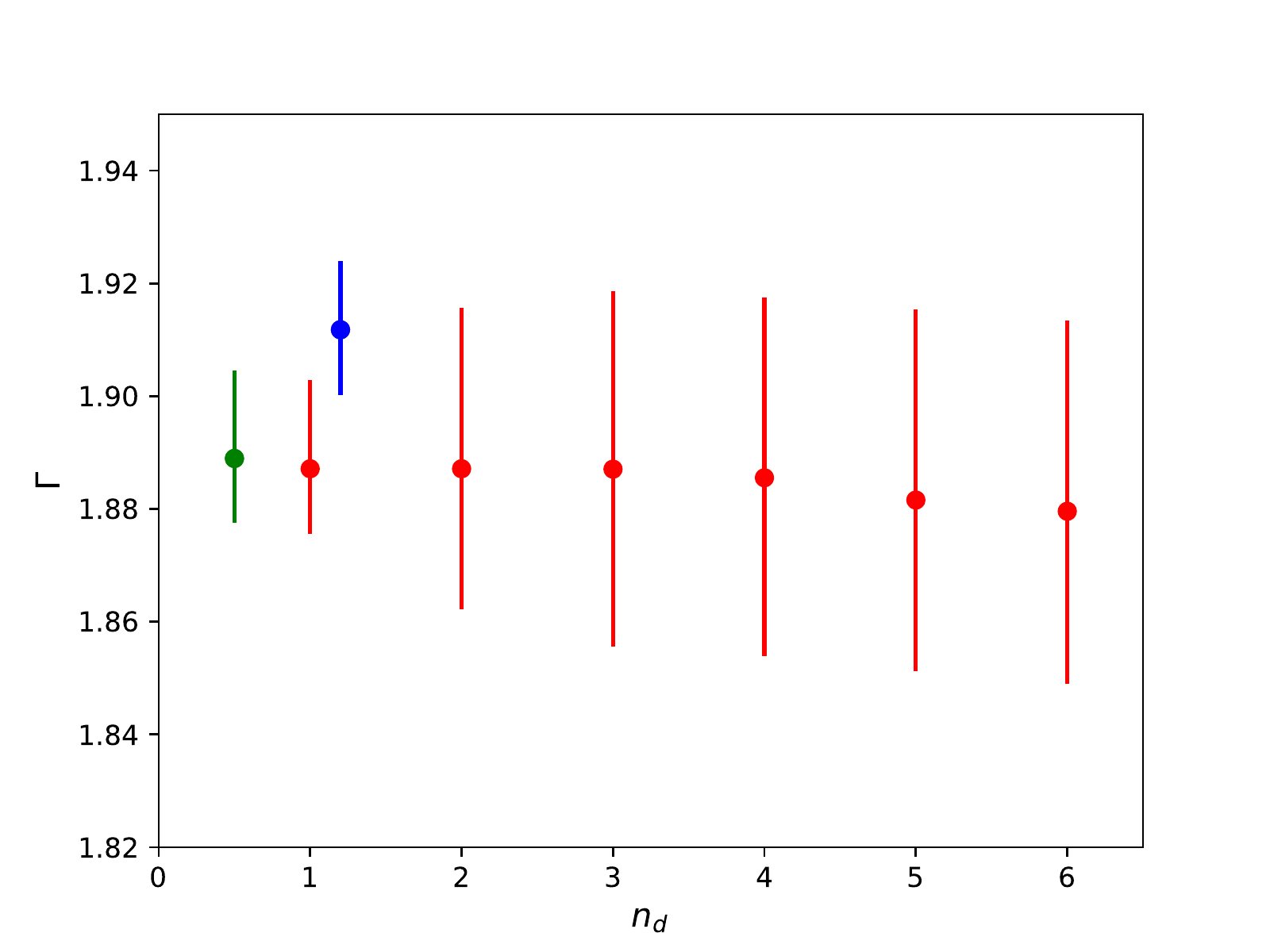}
\includegraphics[width=0.49\textwidth]{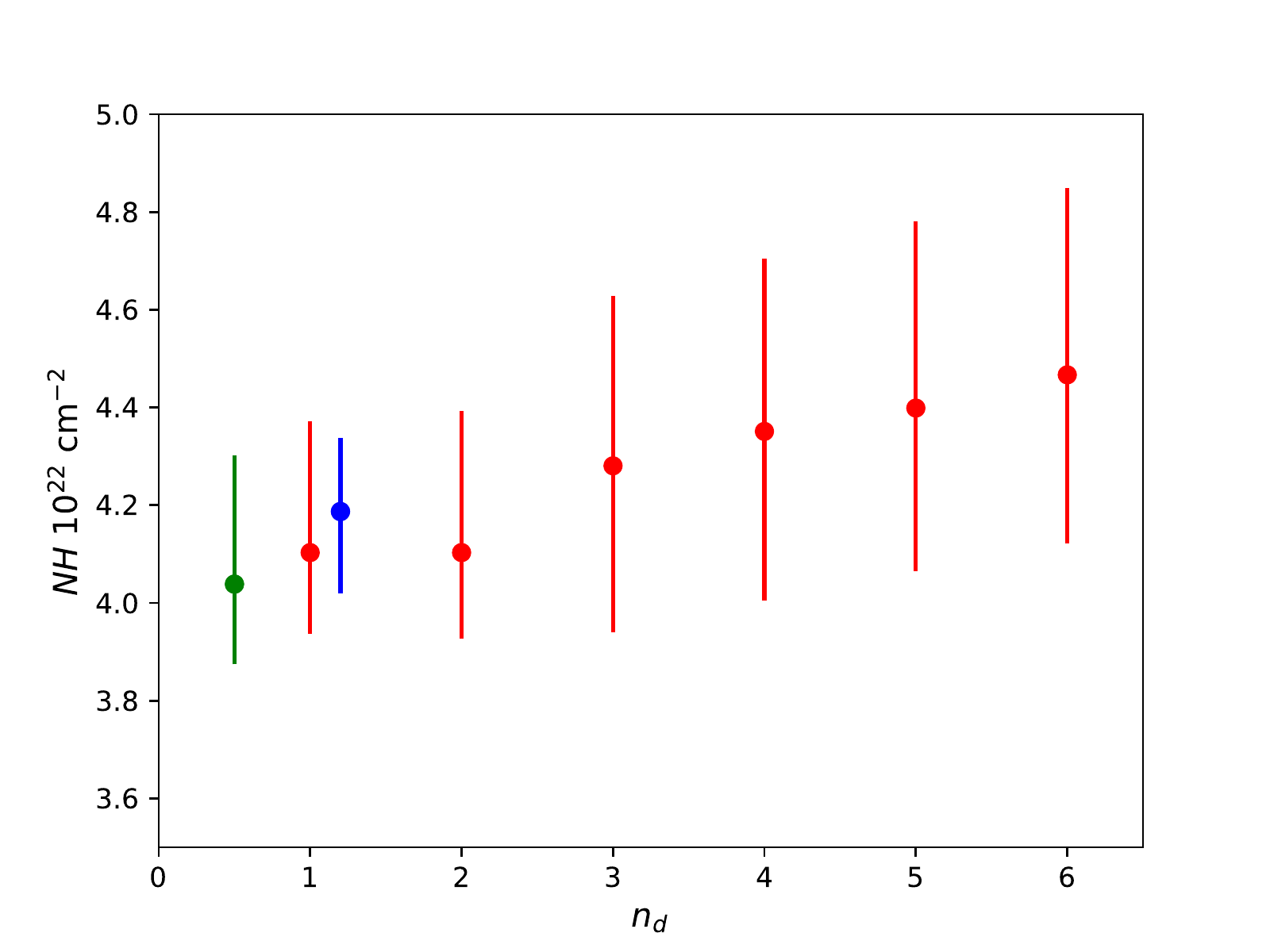}
\caption{Left: Best-fit spectral slope $\Gamma$ of the nuclear emission 
of the Spiderweb as a function of the enhancement factor $n_d$ (red points).  
The best-fit values obtained with direct background subtraction are shown as blue and green 
points, where blue is for the background sampled in the 2.6-4.6 arcsec annulus, and
green in the 16-24 arcsec annulus. Right: Same for the best-fit value of the 
intrinsic absorption $N_H$.
}
\label{bestfit_AGN}
\end{center}
\end{figure*}

Now we focus on the spectral analysis of the nuclear emission.
In Figure \ref{bestfit_AGN} we show the best fit values as a function of the enhancement 
factor $n_d$  that paramterizes the normalization of the diffuse component within 2 arcsec.  
We see that the spectral slope $\Gamma$ is hardly affected, since it is 
robustly constrained by the hard-band signal at energies larger than 2 keV, where the 
contribution of diffuse emission is estimated to be very small.  However, when compared to the 
values obtained with direct background subtraction, with and without the inclusion of the
foreground as estimated in the 3-5 arcsec annulus, we find that the 1 $\sigma$ uncertainties
almost double, even though it is still below 2\%.  This marginal but noticeable effect is 
simply due to the fact that the inclusion of a small thermal contribution makes the model 
more similar to the data, resulting in a milder dependence of $C_{stat}$ on the 
parameters of the fit, hence larger error bars. We also note that the 
inclusion of a thermal component with a fixed $n_d$ does not imply additional 
free parameters (in other words, not only the normalization but also the 
shape, namely temperature and metallicity, are 
kept frozen).  When we turn to the intrinsic absorption $N_H$ (right panel of Figure 
\ref{bestfit_AGN}), we notice that the best fit value increases slightly with $n_d$.  
We do expect this trend since, when the thermal component increases, the nuclear 
emission in the soft band is correspondingly reduced, and this is compensated with 
a slightly larger intrinsic absorption.  Here the typical 1 $\sigma$ error bar is 
constant at the level of 5\%.  As expected, these results do not change if we describe 
the diffuse emission with a $\Gamma\sim 2.7$ power law instead of a $\sim 3$ 
keV {\tt mekal} model.  To summarize, the investigation of the effect of the 
diffuse emission within a radius of 2 arcsec on the spectral analysis of the 
nuclear emission, shows that the properties of the nuclear emission are marginally 
affected by the parameter $n_d$.  According to the discussion in Section 4.2 based 
on our imaging analysis, in the following we assume a value $n_d=4$ for reference, 
bearing in mind that different $n_d$ values would not impact the results presented in 
this work.

\begin{figure}
\begin{center}
\includegraphics[width=0.49\textwidth]{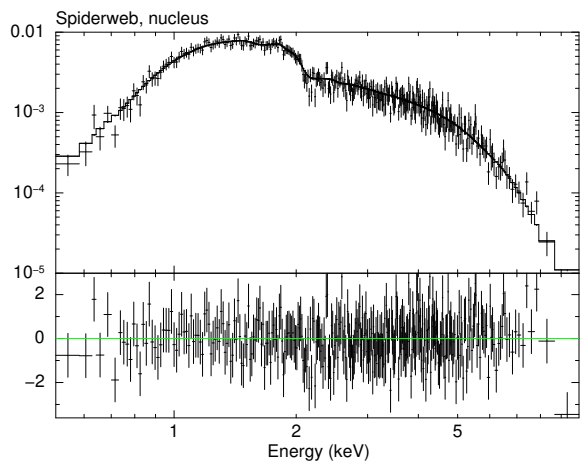}
\caption{Merged spectrum (folded with the instrument spectral response)
of the nucleus of the Spiderweb Galaxy with the 
best fit model (first column of Table 1, with $n_d=4$). 
The residuals are shown in the bottom panel. There is no significant 
excess in the soft band, showing that diffuse emission 
is properly accounted for within the limits of the spectrum quality.
}
\label{AGN_spectrum}
\end{center}
\end{figure}

\begin{table}
\caption{Best-fit values for the reference spectral analysis of the 
nuclear emission of the Spiderweb Galaxy assuming $n_d=4$. Fluxes 
are corrected for Galactic absorption, while luminosities are corrected
both for Galactic and intrinsic absorption}.
\label{bestfit_table}
\begin{center}
\begin{tabular}[width=0.5\textwidth]{lcc}
\hline
parameter   & aperture (2$^{\prime\prime}$)  & PSF correction \\
 \hline
 $\Gamma$             & $1.883_{-0.012}^{+0.016}$                &      $1.85\pm 0.02$ \\
  $N_H$                   & $4.23_{-0.17}^{+0.28}\times 10^{22}$  & $4.2_{-0.2}^{+0.3}\times 10^{22}$         \\
  $F_S$               &  $5.64\pm 0.17 \times 10^{-14}$  &      $5.95\pm 0.18 \times 10^{-14}$    \\
  $F_H$               &  $1.38\pm 0.02\times 10^{-13}$           &      $1.51\pm 0.02\times 10^{-13}$ \\
  $L_{0.5-2 keV}$ &  $3.10\pm 0.09\times 10^{45}$                &       $3.27\pm 0.10\times 10^{45}$\\
  $L_{2-10 keV}$  &     $4.29\pm 0.05\times 10^{45}$     &        $4.68\pm 0.05\times 10^{45}$\\
\hline
\end{tabular}
\end{center}
\end{table}

We also consider aperture correction that introduces a mild spectral distortion 
due to the different PSF in the soft and hard bands. From the AGN ray-tracing simulations, 
we estimate that total soft and hard fluxes and luminosities are larger 
by a factor $1.05$ and $1.09$,
% $1/0.948$ and $1/0.917$, 
respectively, compared to that measured within 2 arcsec.  Fluxes are 
corrected only for the Galactic absorption, while luminosities are corrected
for both Galactic and intrinsic absorption.  
Error on fluxes and luminosities account for
the uncertainty associated with the full range  $n_d=1-6$ 
and are  3\% in the soft and 1\% in the hard band.
The effect of the PSF correction on the 
spectral shape is instead very mild, with the intrinsic slope $\Gamma$ being
harder by less than 2\%.
We verified a posteriori that the fit is not sensitive to the assumed value for 
{the parameter $N_{HGal}$ that describes the } Galactic absorption. 
If left free, $N_{HGal}$ has a best-fit value
of $3.2_{-0.9}^{+1.6}\times 10^{20}$ cm$^{-2}$, right on top of the 
value based on the HI map of the Milky Way \citep{2016HI4PI}.  
Finally, we search for 
the presence of a neutral iron line at 6.4 keV rest-frame, corresponding to 2.03 keV 
in the observing frame, but we found no signal.  The mild obscuration corresponds to 
a source dominated by the transmitted emission, while significant iron line are expected when 
the reflected component is dominant.  
If we perform a blind search for emission lines in the range 1-7
keV, we do not find any significant candidate.  The merged spectrum and
the best-fit model, along with the residuals, are shown in Figure \ref{AGN_spectrum}.
The residuals are very regular with no significant departure from pure noise. In particular, 
we do not find significant residuals in the soft band, showing that the diffuse emission 
is properly accounted for within the limits of the spectrum quality.
Our best-fit model for the nuclear spectrum of the Spiderweb Galaxy is shown in 
Table \ref{bestfit_table}, with and without aperture correction in terms of
normalization and spectral shape.

In conclusion, an X-ray obscuration at the level of 
$4.2 \times 10^{22}$ cm$^{-2}$ can be provided by the surrounding 
galactic medium and circumgalactic medium
as observed in high-z galaxies in the CDFS \citep[][]{2019Circosta,2020Damato}, or 
in the BCG of a $z\sim 1.7$ protocluster \citep[][]{2020bDamato}.
Therefore it is plausible that the obscuration is not 
dominated by a torus, and that the AGN in the Spiderweb Galaxy is an unabsorbed, TypeI quasar, 
as also suggested by the detection of broad lines.  This would be unusual since 
radio galaxies are typically associated with obscured TypeII AGN. However, the peculiar 
phase of the Spiderweb Galaxy may justify the copresence of a strong quasar and 
a strong radio activity.

%#########################
%# SPECTRAL VARIABILITY
%#########################

\subsection{Spectral variability}

We can investigate the variability of the AGN in the Spiderweb Galaxy on a scale of $\sim 9$ 
months, which is the time span over which the observations of the Large Program have been
taken (November 2019 - August 2020). In addition, we have one measurement at a distance of 
$\sim 20$ years (in 2000), and the timescale provided by the single Obsid duration, 
from a minimum of 5 to a maximum of 14 hours.  

\begin{figure}
\begin{center}
\includegraphics[width=0.49\textwidth]{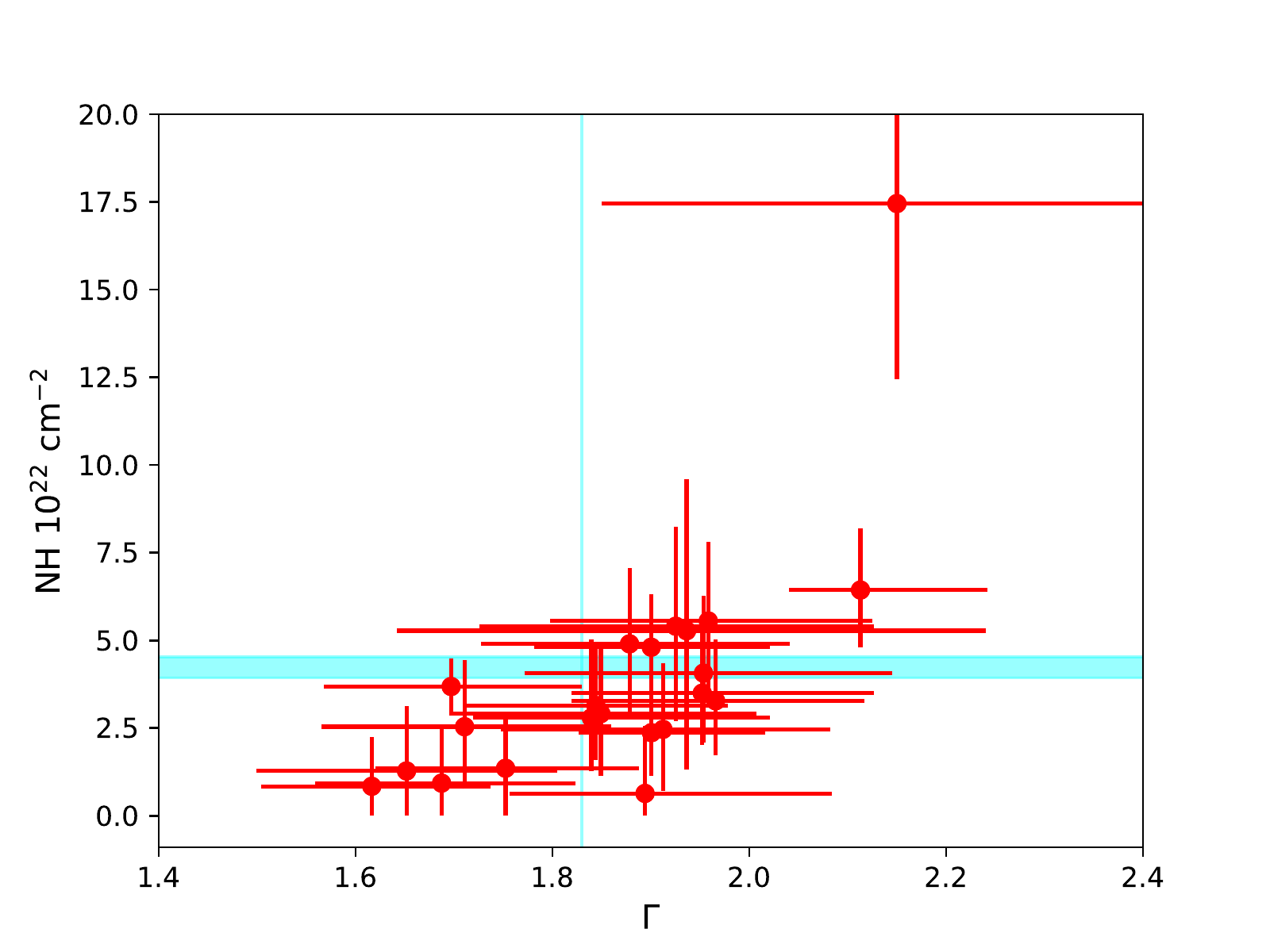}
\caption{Scatter plot of the best fit values of the intrinsic absorption $N_H$ and
the spectral slope $\Gamma$ for each Obsid. Error bars refer to 1 $\sigma$ c.l. on 
a single parameter.
Shaded area show the 1 $\sigma$ interval for the best fit value of $N_H$ from the global fit.
}
\label{nh_gamma}
\end{center}
\end{figure}

First we investigate the spectral variability.  We fit separately each Obsid 
and we plot the distribution of best fit values for $N_H$ and $\Gamma$.  
We note that for the fits of each Obsid we do not apply the small PSF correction
to the spectral shape. 
As we show in Figure \ref{nh_gamma}, the distribution of the best-fit values in the 
$N_H$-$\Gamma$ plane appears to be roughly consistent with the global best-fit, with the
apparent correlation due to the well-known degeneracy of $N_H$ and $\Gamma$. 
If we compute a simple reduced $\chi^2$ with respect to the average value
(after removing the unique outlier with $N_H>10^{23}$ cm$^{-2}$ obtained
for Obsid=22922),
we find $\chi^2\sim 0.98$ and $\chi^2\sim 1.21$ for 21 degrees of freedom, for $\Gamma$ and 
$N_H$, respectively. In both cases we are not able to reject the null hypothesis of a constant
value also in the case of $N_H$, for which we have a probability slightly less than $0.25$ to 
obtain a larger $\chi^2$.  If we do not remove the unique strong outlier (corresponding to 
Obsid=22922), we are able to reject the null hypothesis of a constant $N_H$ on the entire 
observing period with a probability less than $0.05$ to 
obtain a larger $\chi^2$. To summarize, from a simple $\chi^2$ test on the entire data set, 
we get only a weak hint, if any, of spectral variability.

\begin{figure}
\begin{center}
\includegraphics[width=0.49\textwidth]{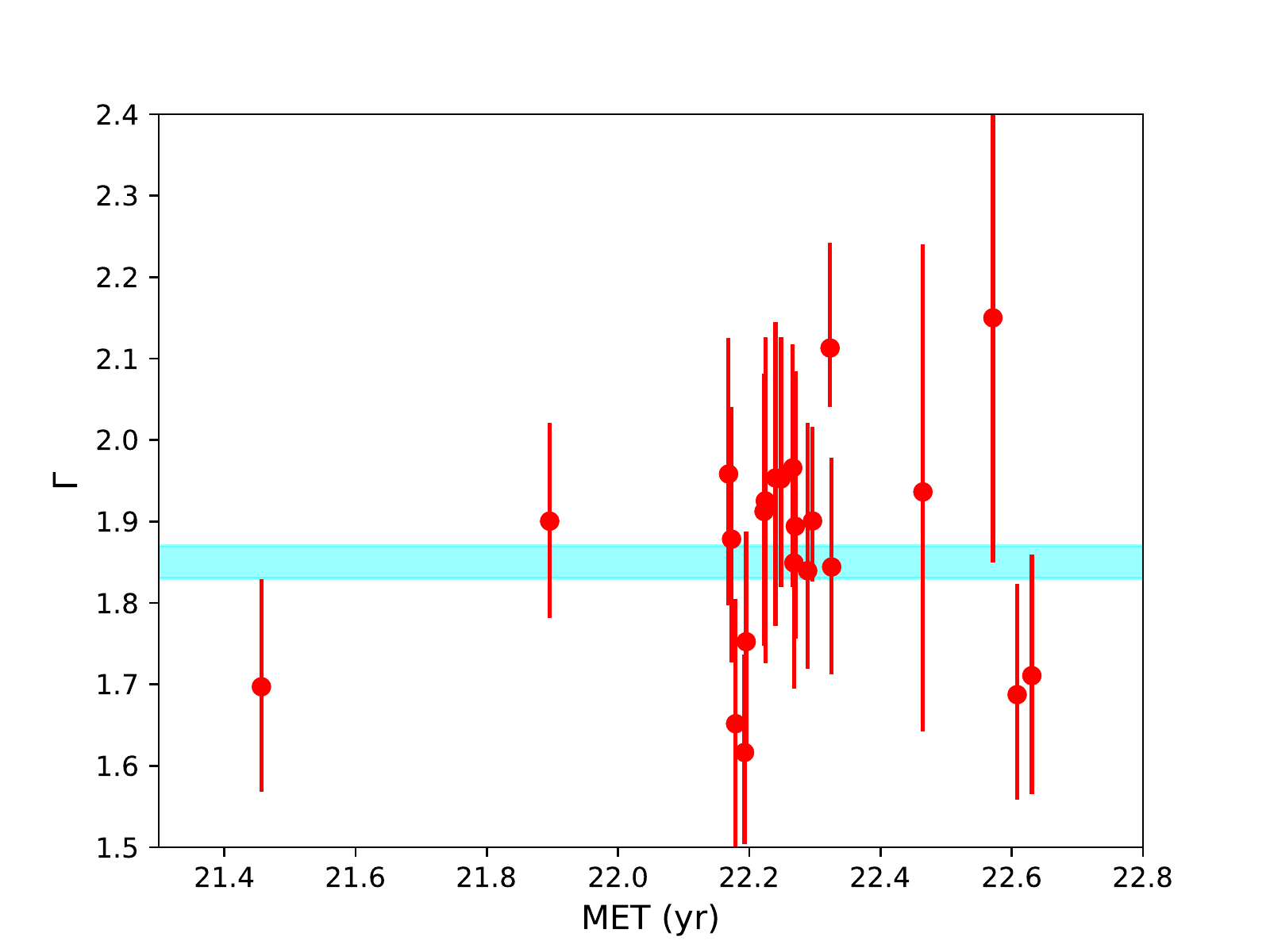}
\caption{Best-fit values of $\Gamma$ for each Obsid as a function of 
the Mission Timeline $MET$ (years). The first point has been moved to $MET=21.5$ to be shown
in the same linear plot. The horizontal shaded line shows the 
best-fit value of the cumulative spectrum with PSF correction, as listed in Table \ref{bestfit_table}.
}
\label{gamma_obsid}
\end{center}
\end{figure}

\begin{figure}
\begin{center}
\includegraphics[width=0.49\textwidth]{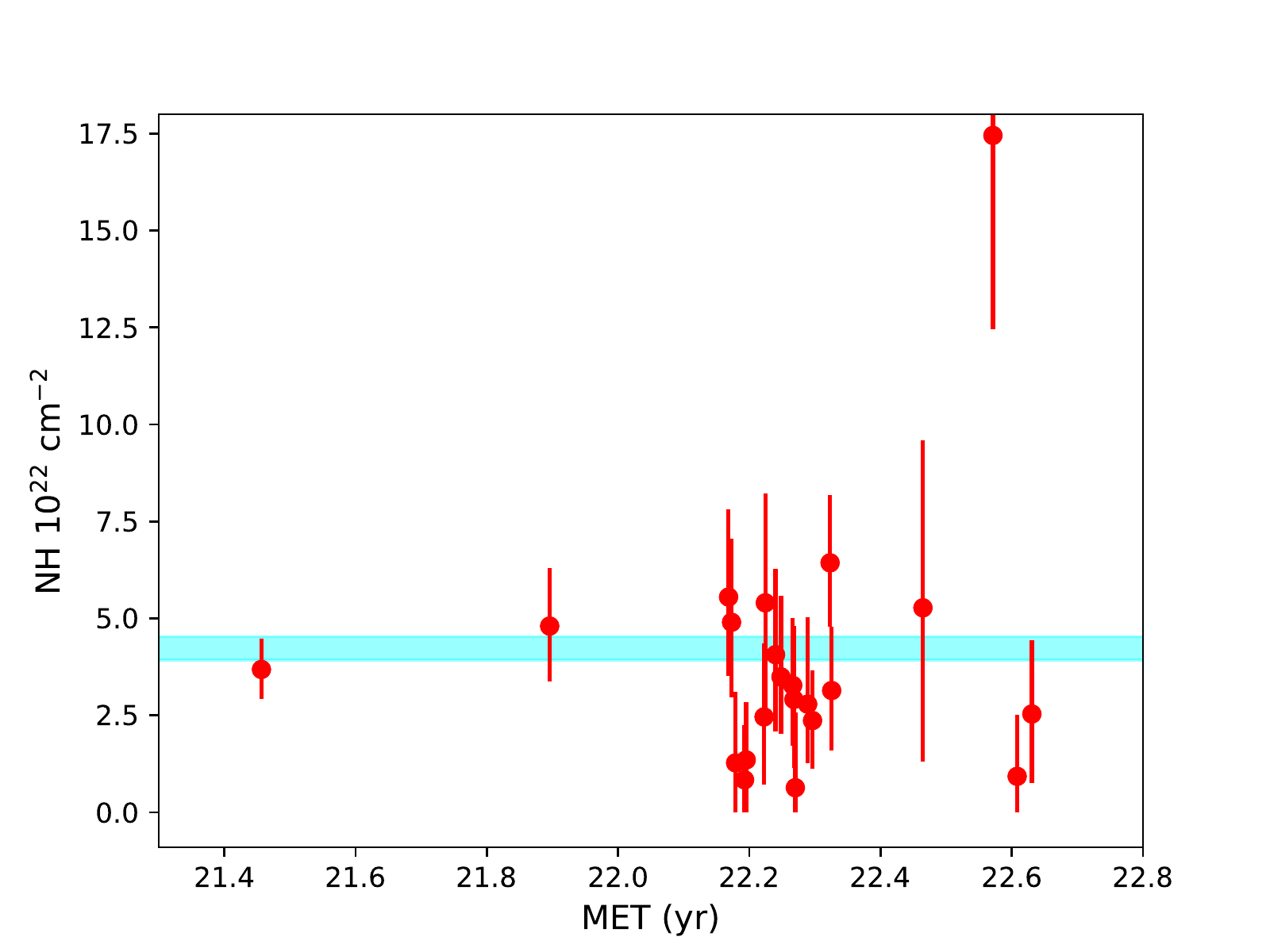}
\includegraphics[width=0.49\textwidth]{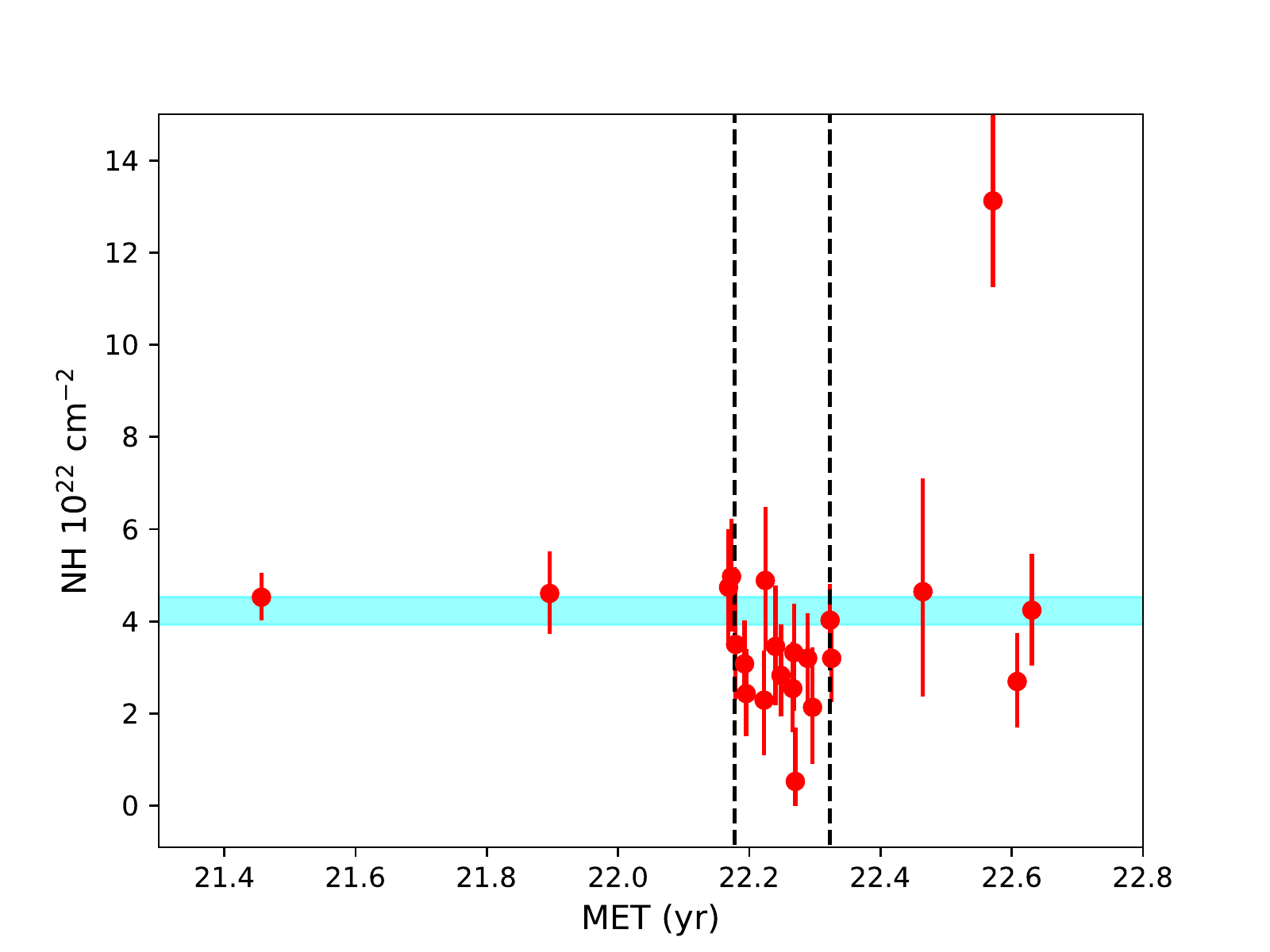}
\caption{Upper panel: Best-fit values of $N_H$ for each Obsid as a function of 
the Mission Timeline $MET$ (years). The first point has been moved to 21.5 to be shown
in the same linear plot. The horizontal shaded line shows the 
best-fit value of the cumulative spectrum with PSF correction, as listed in Table \ref{bestfit_table}.
Lower panel: Same, but with best-fit values of $N_H$ obtained
with $\Gamma$ frozen to 1.88.  The vertical dashed lines bracket the period with low intrinsic
absorption as discussed in the text.
}
\label{nh_obsid}
\end{center}
\end{figure}

Then, in Figure \ref{gamma_obsid} and \ref{nh_obsid}
we plot the best-fit values of $\Gamma$ and $N_H$ as a function of the
Mission Timeline (MET). In Figure \ref{gamma_obsid} we can see that $\Gamma$ 
is consistent with the best-fit value of the spectral analysis of the
cumulative spectrum for each Obsid within 1 $\sigma$, as confirmed 
by a simple $\chi^2$ test. 
On the other hand, in Figure \ref{nh_obsid} we find that $N_H$ shows a higher level
of fluctuations, and that the lowest values of $N_H$ seems
to occur in contiguous periods within a narrow time frame.  
We repeat the plot after freezing the spectral slope to the 
average best-fit value $\Gamma=1.88$. 
This reduces the noise around the average value, removing the degeneracy between 
$N_H$ and $\Gamma$, but the period with low $N_H$ is still visible (see Figure \ref{nh_obsid}, 
lower panel).  Actually, when $\Gamma$ is frozen, 
the reduced $\chi^2$ with respect to a constant $N_H$ value increases
to $\chi^2=1.65$ despite the smaller scatter, due to the smaller error bars, after
the removal of the single outlier (Obsid 22922).
To quantify the spectral variability of $N_H$ we discard the outlier and focus 
on three contiguous periods. One is when a 
low $N_H$ is measured, corresponding to $22.179<{\rm MET}<22.323$, while the rest 
consists in all the observations before ${\rm MET}=22.179$ and after ${\rm MET}=22.323$.  
We obtain $N_{Hhigh} = (5.02\pm 0.32)\times 10^{22}$ cm$^{-2}$ 
and $N_{Hlow} = (2.95\pm 0.24)\times 10^{22}$ cm$^{-2}$, with a difference 
of a factor $\sim 1.7$ of $\sim 5$ $\sigma$\footnote{We note that the statistical errors
of the two measurements refer to the average, central value in the corresponding period, 
and do not reflect the dispersion observed across the measurement in each Obsid shown in 
Figure \ref{nh_obsid}.}.  
To summarize, we conclude that we see a modest but significant spectral variation
corresponding to a change in the intrinsic absorption by a factor $\sim 1.7$ on a timescale of
$\sim 1 $ year. This also implies that part of the intrinsic absorption happens close to the 
SMBH and it is associated with a clumpy obscuring torus, while part of the absorption may 
still be associated with the intergalactic medium, as suggested in Section 5.1.

\subsection{Flux variability}

\begin{figure}
\begin{center}
\includegraphics[width=0.49\textwidth]{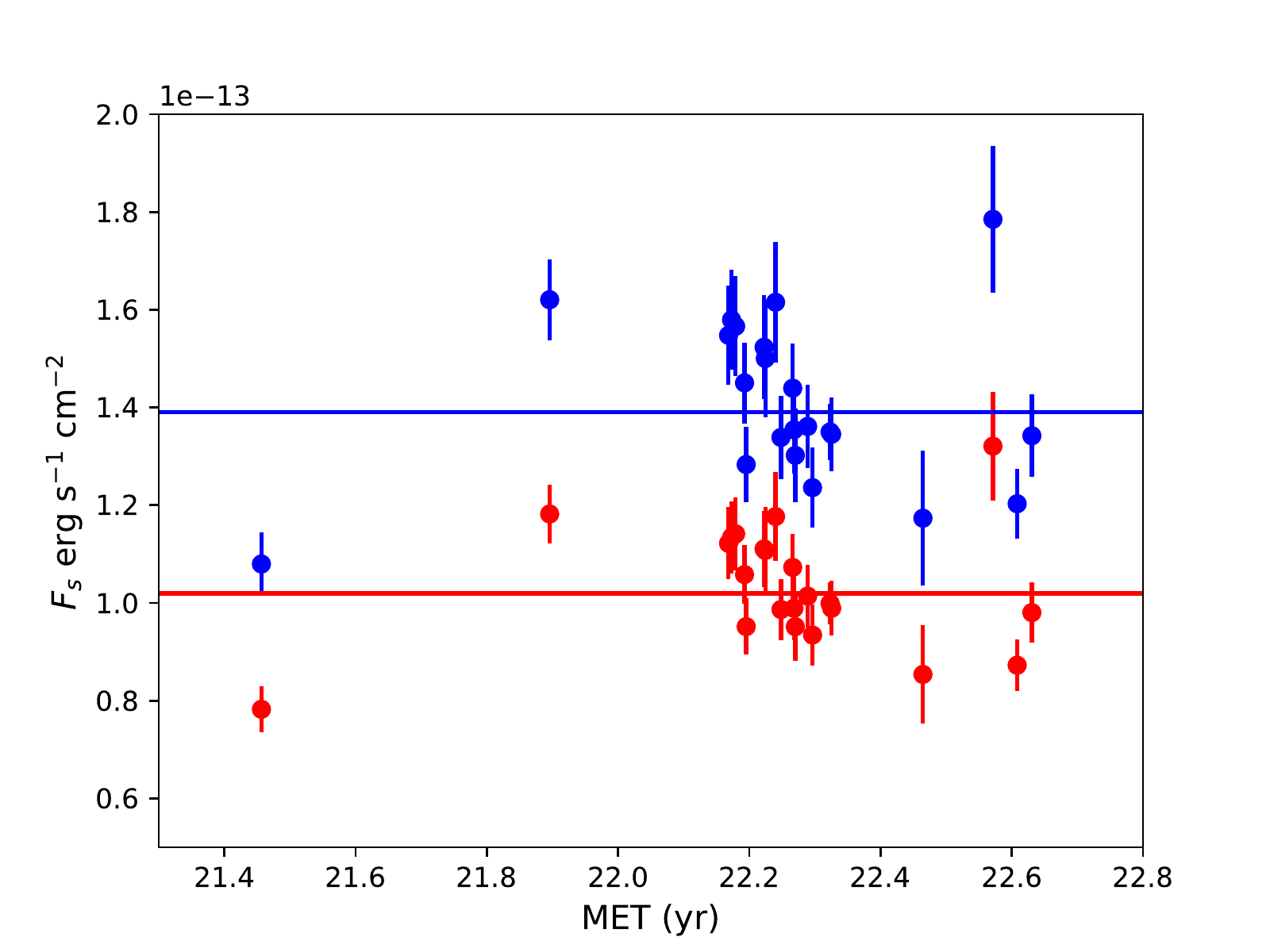}
\caption{Soft (red points) and hard (blue points) flux values measured in each Obsid
within a radius of 2 arcsec
as a function of the Mission Timeline. Fluxes are corrected for the Galactic absorption.
Horizontal lines show the average values. 
}
\label{flux_obsid}
\end{center}
\end{figure}

\begin{figure}
\begin{center}
\includegraphics[width=0.49\textwidth]{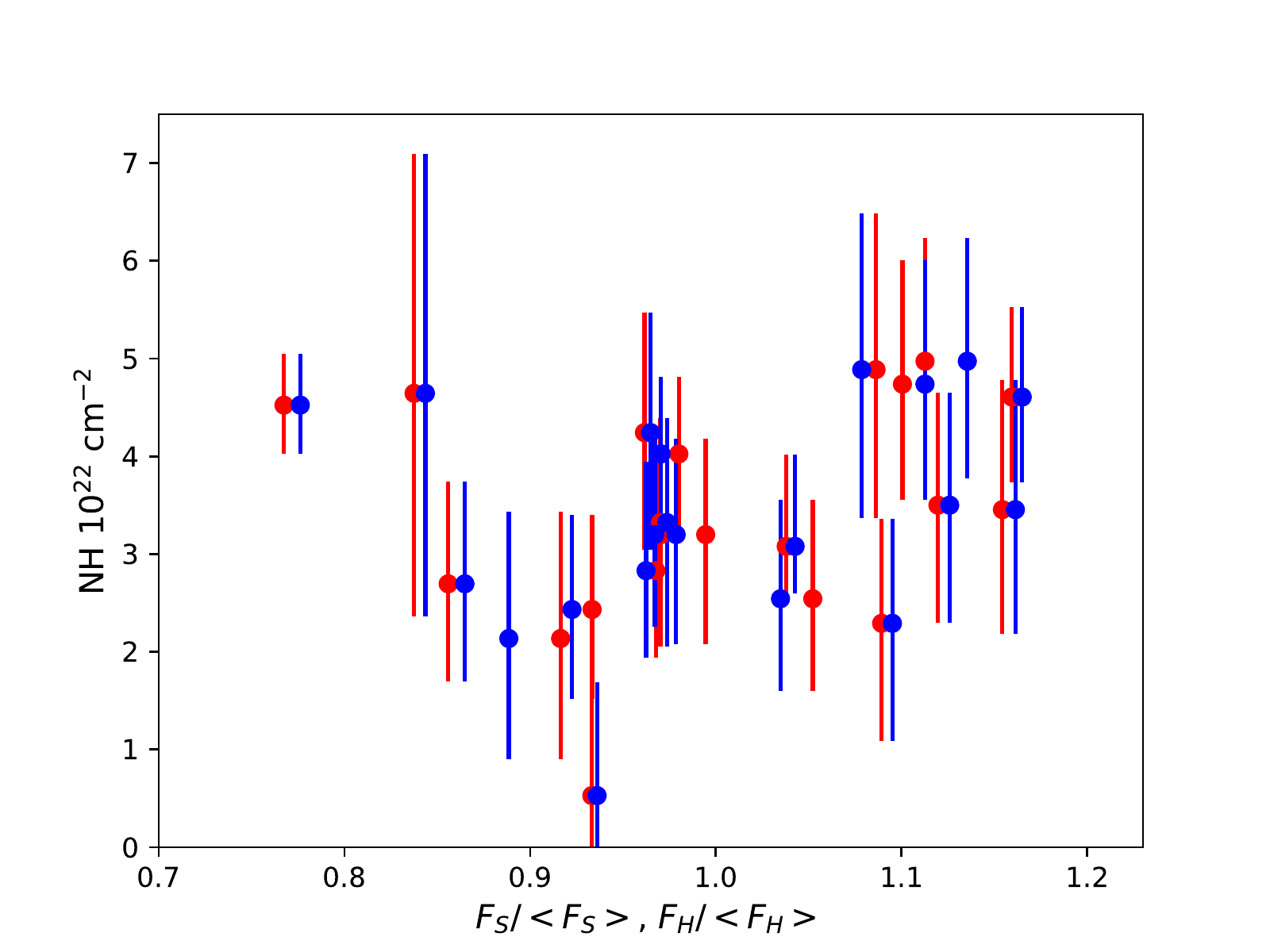}
\caption{Best-fit values of $N_H$ plotted against the ratio of the flux in each Obsid 
over the average flux (red and blue dots correspond to the soft and hard bands, respectively).}
\label{fluxratio}
\end{center}
\end{figure}

We compute the soft and hard band flux values in each Obsid keeping $\Gamma=1.88$ frozen, 
and correcting for the Galactic absorption. The result is shown in Figure \ref{flux_obsid}.
We note that for simplicity here we report the fluxes 
within 2 arcsec, without correcting for the PSF effects
in each band, since our discussion would not be affected by this correction. 
We notice a significant variability on a scale of a few months in the observing frame.  
If we fit the flux as a function of the time with a constant,
the reduced $\chi^2$ is $\sim 3.0$ in both bands
for 21 degrees of freedom (for consistency, we conservatively removed Obsid 22922 that appears to 
be an outlier both for shape and normalization of the spectrum). 
Also if we remove the lowest and highest points, we still 
obtain a reduced $\chi^2\sim 2$ with 19 degrees of freedom. 
The {\sl rms} fluctuation is $12$\% in both bands, therefore very mild, however
it is statistically significant.  
In addition, we note that the flux variability seems to have 
a temporal behavior different from the spectral variation (or change in $N_H$). 
This can be seen by the lack of correlation between the best-fit $N_H$ values 
and the percentage variation of the flux, shown in Figure \ref{fluxratio}.
Therefore, we conclude that we are able to detect variability both in the normalization of the spectrum 
and in the spectral shape, and that there are no signs of correlation between the 
two phenomena. 

\section{X-ray properties of the Spiderweb Galaxy: Diffuse emission}

\subsection{Imaging analysis}

\begin{figure*}
\begin{center}
\includegraphics[width=0.49\textwidth, trim=0 55 55 140, clip]{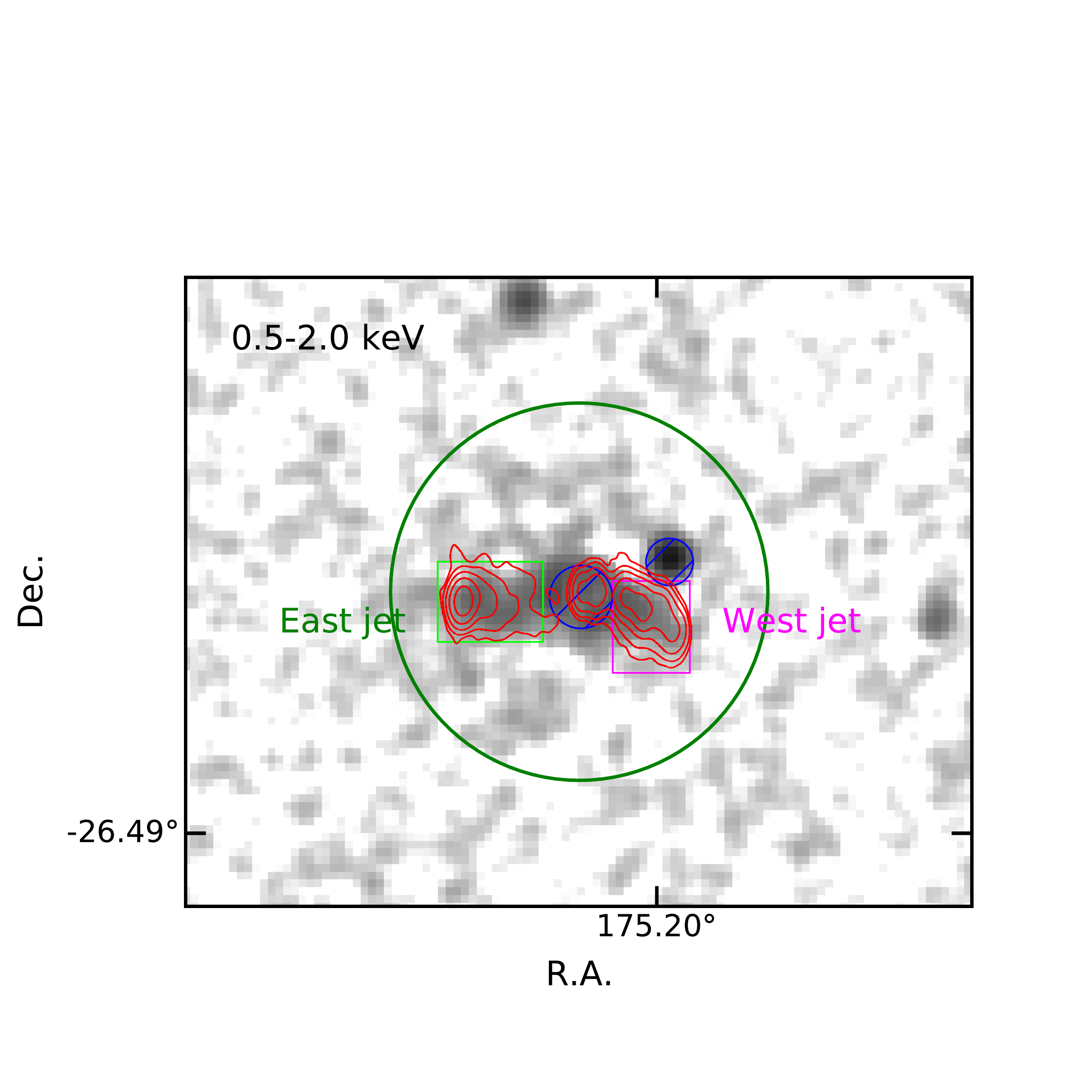}
\includegraphics[width=0.49\textwidth, trim=0 55 55 140, clip]{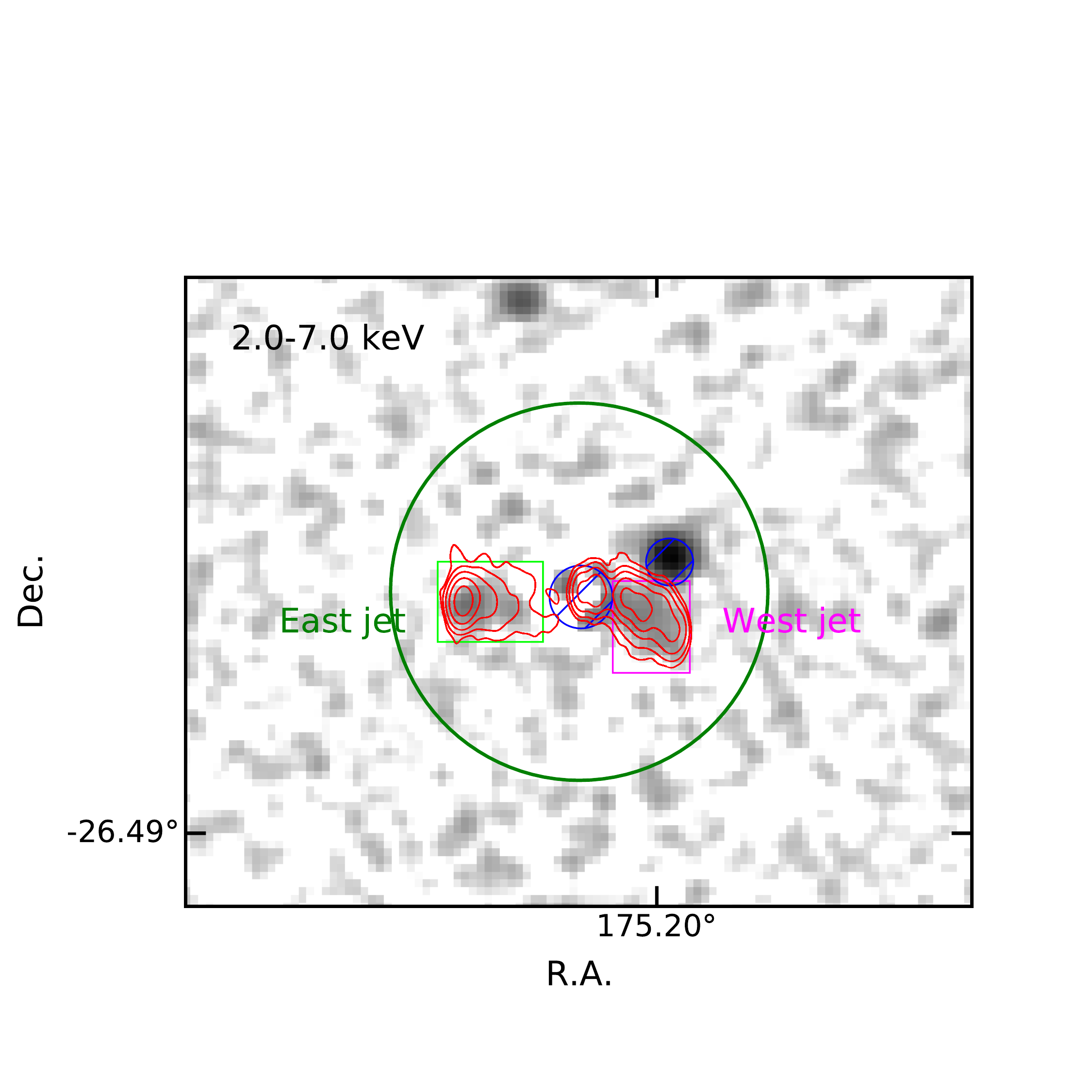}
\caption{Left: Background-subtracted soft band image of the Spiderweb Galaxy 
after AGN subtraction (with $n_d=4$).  The image has been smoothed with a 
Gaussian kernel with a sigma of 1 pixel, slightly degrading the effective resolution.  
The green and magenta boxes correspond to the regions used for the east and west jet spectral 
analysis, respectively.   Two AGN circular extraction regions (in blue) 
are removed (the AGN, with a radius of 2 arcsec, and a nearby AGN with a 
radius of 1.5 arcsec).  The large green circle is the region used for the 
spectral analysis of the isotropic diffuse emission, excluding both jet regions
and the AGNs.  
Red contours show radio emission observed in the 10 GHz band with the JVLA
\citep[][]{2022Carilli} at levels of $0.03$, $0.2$, $2$ and
$20$ mJy/beam.  }
\label{jet_diffuse}
\end{center}
\end{figure*}

In Figure \ref{jet_diffuse} we show the extraction regions used for our analysis of the diffuse 
emission, overplotted on the background-subtracted images after removing the central AGN 
(we used $n_d=4$ for display).
Consistently with the radio and X-ray combined analysis of the jets presented in 
\citet{2022Carilli} and \citet{2022Anderson}, we select two box regions for 
the east and west jet regions.  These regions cover all the radio emission 
observed at 10 GHz \citep[see][]{2022Carilli}, include most of the diffuse emission in the soft band
(see Figure \ref{jet_diffuse}, left) and almost the entire diffuse emission in the hard band 
(see Figure \ref{jet_diffuse}, right). 
The extraction region for the analysis of the diffuse emission outside the jet region is obtained
by selecting a circle centered on the Spiderweb Galaxy with a radius of 12 arcsec, corresponding to 
$\sim 100$ physical kpc at $z=2.156$.  From this circular region, 
we remove two boxes corresponding to the east and west jets, and two circular regions
corresponding to the central AGN (a circle with a radius of 2 arcsec) and 
to the source XID 7 (a circle with a radius of 1.5 arcsec) that is found 
to be a companion quasar (see Paper I).  
The boxy shape is chosen to follow at best the jet shape and to keep, at the same time, a simple
geometry.  The instrumental plus unresolved extragalactic X-ray background is sampled
from an annulus with inner and outer radius of 16 and 29.5 arcsec, respectively, 
which is the same used for the analysis of the nuclear emission.

In Figure \ref{jet_diffuse}, it is possible to appreciate that the diffuse emission in the soft band
is clearly visible outside both jet regions, out to a radius of 12$^{\prime\prime}$ from the nucleus, 
while the shape of the diffuse emission in the hard band is almost entirely elongated 
along the radio jets, with no emission elsewhere.  
On the basis of this preliminary imaging analysis, we argue that the emission along the jets is 
dominated by IC emission from the relativistic jet population, while the more isotropic
soft emission may be dominated by thermal emission from shocked gas. The 
diffuse emission along the jets is clearly harder, as expected in case of a power law, 
nonthermal emission, while the diffuse emission outside the jet regions has an hardness ratio 
$HR\sim -1$, being entirely soft, as expected in the case of thermal emission with 
a relatively low temperature, or a very steep power law.  
Incidentally, this result may provide another piece of evidence that inverse
Compton losses are important in high-z radio sources, and that this effect
is likely to be a strong contributor to the well-known correlation 
between ultra-steep spectrum and redshift for radio sources
\citep{1979Tielens,1979Blumenthal,2018Morabito}.
On the other hand, another plausible mechanism for a steep radio spectrum 
may be associated with an increased ambient density \citep{2006Klamer}.
Clearly, both mechanisms may be at play in the Spiderweb Galaxy.
A systematic exploration of high-z radio galaxies in a variety of 
environments would be key to understand the physics behind the 
correlation between redshift and 
their spectral properties. This relevant aspect is not explored in this work, 
but it constitutes a clear science case for future {\sl Chandra} programs.

While the nature of the diffuse emission is further investigated in our spectral analysis,
as a preliminary step, we obtain the photometry in the three different regions.  We 
find $207\pm 19$ and $228\pm 21$ net counts in the 0.5-7 keV band in the 
west and east jet regions, respectively, after subtracting the background.
In the circular
region with a radius of 12 arcsec, after removing the west jets, east jet, and AGN regions, we
measure $447\pm 53$ net counts in the 0.5-7 keV band. 
We note that the isotropic, diffuse component, where we expect to find thermal emission, 
overlaps with the west and east jet regions, while the AGN wings overlaps with 
all the diffuse components.  Therefore, 
we have to take into account two and three components to perform spectral analysis of the 
isotropic diffuse emission and the jet emission, respectively.  
While we already have established the shape and normalization 
of the AGN emission, we need to measure first the diffuse, isotropic emission 
(accounting for AGN contamination), and finally the emission in the two jet regions
(accounting for AGN contamination and isotropic emission).

\begin{table*}
\caption{Best-fit values for the reference spectral analysis of the 
diffuse emission.  We note that in this table, flux and luminosity values of the
isotropic (thermal) emission have not been corrected by the geometrical factor, but 
directly correspond to the best-fit values obtained from the extraction region.
Error bars refer to 1 $\sigma$ c.l. on 
a single parameter.}
\label{bestfit_table_diffuse}
\begin{center}
\begin{tabular}[width=0.5\textwidth]{lcccccc}
\hline
\hline
region   & $kT$ (mekal)  &   $F_S$  &  $F_H$  & $L_{0.5-2 keV}$ &  $L_{2-10 keV}$  \\
   & keV  &   erg s$^{-1}$ cm$^{-2}$ &  erg s$^{-1}$ cm$^{-2}$  & erg s$^{-1}$  &  erg s$^{-1}$ \\
\hline
   &  &   &   &   &   \\
isotropic   & $1.98_{-0.43}^{+0.70}$  &   $(2.5\pm0.6)\times 10^{-15}$  &   $(2.4\pm0.9)\times10^{-16}$  
 & $(7.7\pm 1.8)\times 10^{43}$ &  $(7.3\pm 1.7)\times 10^{43}$  \\
  &  &   &   &   &   \\
\hline
\hline
region   & $\Gamma$   &   $F_S$  &  $F_H$  & $L_{0.5-2 keV}$ &  $L_{2-10 keV}$  \\
         & -  &   erg s$^{-1}$ cm$^{-2}$ &  erg s$^{-1}$ cm$^{-2}$  & erg s$^{-1}$  &  erg s$^{-1}$ \\
\hline
   &  &   &   &   &   \\
west jet   & $1.90\pm 0.20$   &   $(1.32\pm 0.21)\times 10^{-15}$  &   $(1.78\pm 0.28)\times 10^{-15}$   & $(4.1\pm0.7)\times 10^{43}$ &  $(5.5\pm 0.9)\times 10^{43}$  \\
   &  &   &   &   &   \\
east jet   & $2.51\pm 0.21$   &   $(1.89\pm 0.25)\times 10^{-15}$  &   $(1.02\pm 0.13)\times 10^{-15}$   & $(1.18\pm 0.15)\times 10^{44}$ &  $(6.4\pm 0.8)\times 10^{43}$   \\
   &  &   &   &   &   \\
\hline
\end{tabular}
\end{center}
\end{table*}

\subsection{Spectral analysis of the diffuse emission: Isotropic component}

\begin{figure*}
\begin{center}
\includegraphics[width=0.49\textwidth]{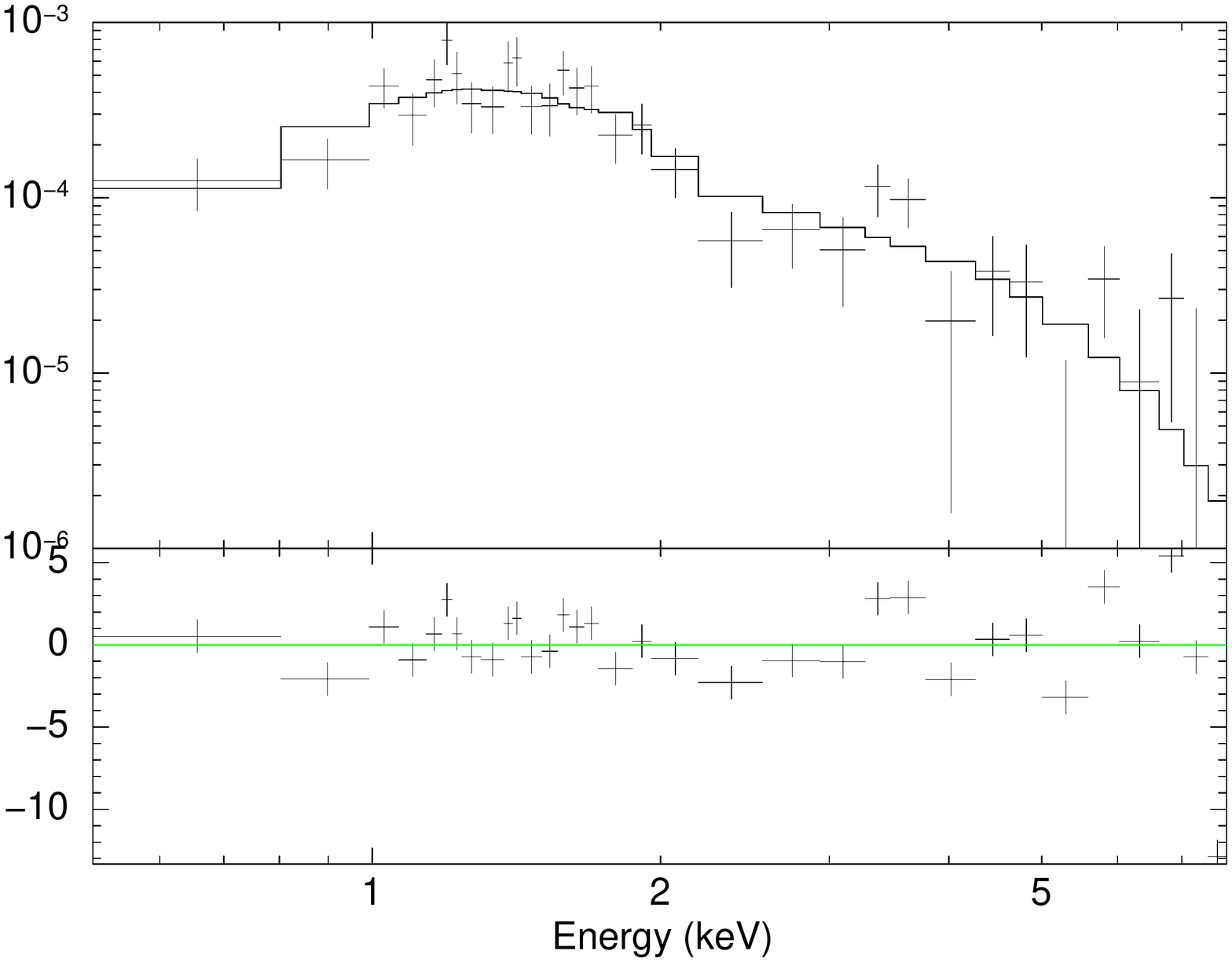}
\includegraphics[width=0.49\textwidth]{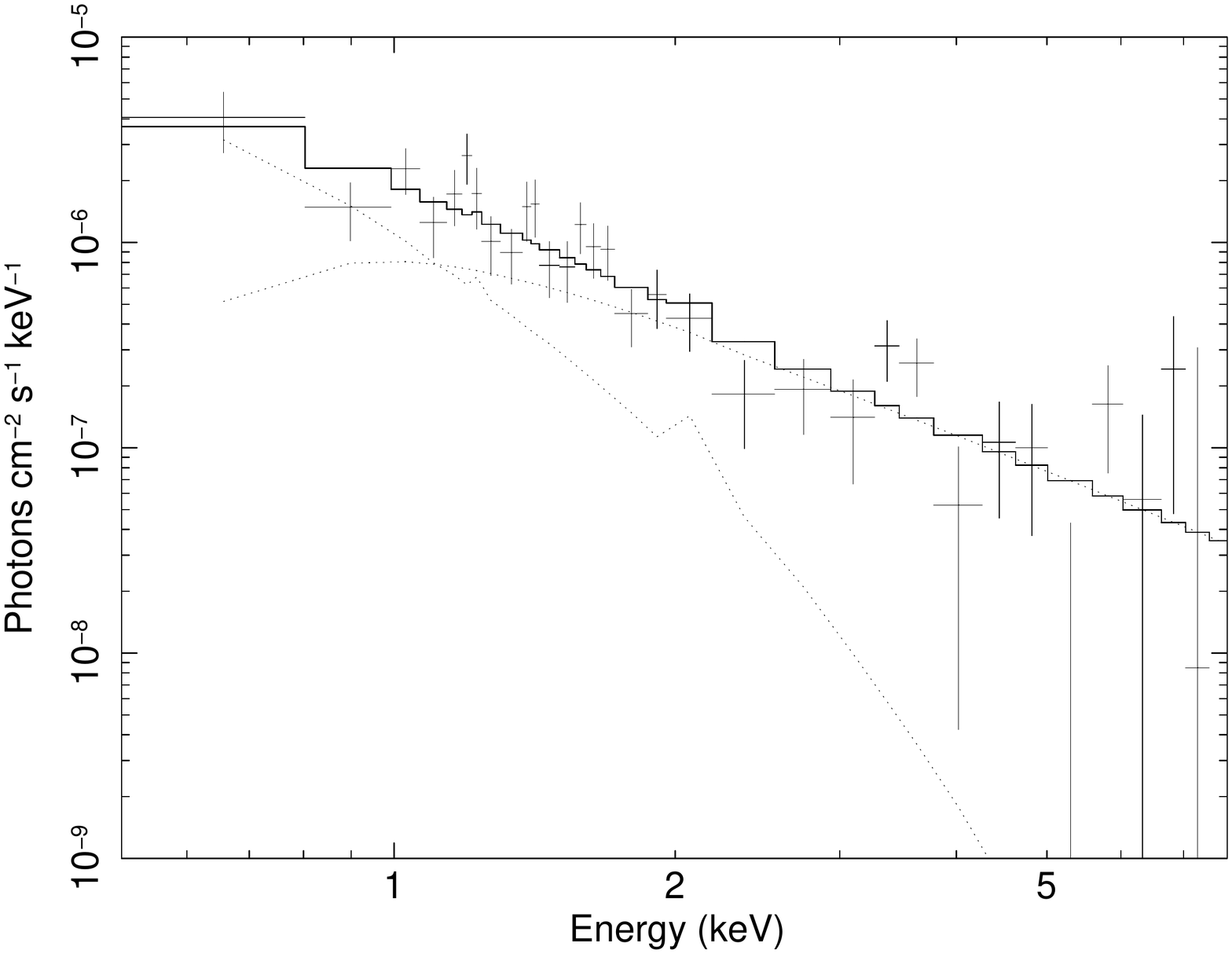}
\caption{Left: Spectrum (folded with the instrument spectral response) 
of the isotropic diffuse emission (after removing the central, AGN-dominated region, and the jet
regions) fitted with a thermal {\tt mekal} model plus
the AGN contamination.   Right: Unfolded spectrum with the two 
model components shown with dotted lines. 
The AGN contamination (absorbed power law) is dominant at energies $>1.5$ keV, while the thermal 
component is dominant below $1$ keV.
}
\label{diffuse_spectrum}
\end{center}
\end{figure*}

The spectral fits are performed with {\tt Xspec 12.11.1} \citep{1996Arnaud}
over the energy range 0.5–9.0 keV of the merged spectrum.  In fact, we do not use single
Obsid for the fits, since several observations would have
less than 10 net counts,  so that the spectral shape is completely undetermined 
in each separate Obsid. 
In this case, we use cumulative ARF and RMF files obtained 
by weighting single Obsid files by the corresponding exposure times.
Galactic absorption is described with the model {\tt tbabs}
and its value is fixed to $3.18\times 10^{20}$ cm$^{-2}$. 
We used Cash statistics \citep{1979cash} applied to the source plus background, which is preferable for 
low S/N spectra \citep[][]{1989Nousek}. 
We assume that the emission observed within 12 arcsec is approximately symmetric 
and centered on the Spiderweb Galaxy.  We fit the diffuse emission outside the jet 
regions, after removing the AGN as previously described, with a {\tt mekal} model 
and a {\tt power law} model.  

From the ray-tracing simulation, we find that 
$3.3$\% of the soft nuclear emission inside 2 arcsec is found in the extraction region. 
Therefore, we add this component modeling it with an absorbed power law with the parameters frozen
to the best-fit values found in the AGN extraction region, including the correction for the 
PSF effect that makes the AGN contamination slightly harder than the spectrum measured
within 2 arcsec.  This last correction implies that $\Gamma=1.85$ 
is a suitable expectation.  We also do not include the neutral 
iron line component, since it has been found to be negligible (see Section 5.1).  
The Galactic absorption is clearly the same for both components. Therefore, after accounting for 
a fixed component associated with the AGN contamination, we perform a fit of the 
isotropic diffuse emission, that, as thoroughly discussed in the imaging analysis 
(Section 4.2) is almost entirely included in the soft band.

The unabsorbed power law model has a best-fit slope of $\Gamma = 3.27\pm 0.43$
for a flux of $(2.7\pm 0.7) \times 10^{-15}$ erg s$^{-1}$ cm$^{-2}$ and $(4.9\pm 2.6)\times 10^{-16}$ erg s$^{-1}$ cm$^{-2}$
in the 0.5--2~keV and 2--10~keV bands.  If, instead, we assume a thermal model, we find that 
the fit improves with a $\Delta C \sim 10$.  We obtain a best-fit temperature of 
$1.98_{-0.43}^{+0.70}$ keV and metallicity of 
$Z=0.26_{-0.26}^{+1.35} Z_\odot$, or better, a $1\sigma$ upper limit $Z<1.6 Z_\odot$. 
\footnote{Based on our experience with lower redshift ICM, and considering the strong K-correction, 
we estimate that the signal should be higher by a factor $>3$ 
before obtaining a statistically significant measurement of the iron emission line complex. 
This would be of paramount relevance to investigate the 
chemical volution of the ICM well beyond its current limit at $z\sim 1$
\citep[see][]{2018Mernier,2020Liu}. This is clearly a relevant scientific goals
of future, high-angular
resolution X-ray missions, such as Lynx \citep{2018LynxTeam} 
and AXIS \citep{2019Mushotzky,2020Marchesi}.}.  We remark that the 1 $\sigma$ error bars 
(or upper limits) are obtained by varying a single parameter, 
keeping the other parameters frozen to their best-fit values.
The measured fluxes are $(2.5\pm 0.6) \times 10^{-15}$ erg s$^{-1}$ cm$^{-2}$ 
and $(2.4\pm 0.9)\times 10^{-16}$ erg s$^{-1}$ cm$^{-2}$ in the 0.5--2~keV and 2--10~keV bands, 
consistent with those found for the power law model.  The 0.5-10 keV rest-frame 
luminosity is $L_{0.5-10 {\tt keV}}=(1.50\pm 0.36)\times 10^{44}$ erg s$^{-1}$.  To correct for the 
excluded area (the jet regions and the AGN extraction region), 
we can assume an approximately constant surface brightness and apply a geometric
correction factor of 1.22, obtaining 
$L_{0.5-10 {\tt keV}}=(1.83\pm 0.44)\times (1+0.03\, n_d)\times 10^{44}$ erg s$^{-1}$, that, 
for our reference choice $n_d=4$, is equal to
$L_{0.5-10 {\tt keV}}=(2.0\pm 0.5)\times 10^{44}$ erg s$^{-1}$.
A more accurate study including a detailed analysis of the surface brightness 
distribution will be presented in a forthcoming paper (Lepore et al., in preparation).  

The spectrum fitted with a thermal component is shown in Figure \ref{diffuse_spectrum}. In the right
panel we show that the AGN contamination is dominant at energies $>1.5$ keV, while the thermal 
component is higher below 1 keV. We find that
the AGN contamination contributes 25\% and 92\% of the signal in the soft and hard
bands, respectively. 
From a statistical point of view, the thermal model is marginally better than 
a power law model ($\Delta C \sim 10$ with an additional degree of freedom).
However, the slope of the power law model is softer than the AGN emission at 3$\sigma$ c.l.
This implies that the diffuse emission cannot be simply ascribed to contamination from the 
AGN, and that it is unusually steep for nonthermal emission mechanisms.  
Therefore, adding these aspects to the statistical improvement obtained with the thermal model, 
we conclude that the diffuse almost isotropic emission detected within 12 arcsec around 
the Spiderweb Galaxy is very likely due to thermal emission from diffuse hot baryons. 
Unfortunately, as already mentioned, we cannot detect the 
emission line complex of the H-like and He-like iron ions because of the 
low S/N and for the unfortunate rest-frame position of the iron lines at $\sim 2.1-2.2$ keV.
Moreover, the large K-correction
combined with the loss of sensitivity below 1 keV, makes it unfeasible to 
use the iron L-shell complex to efficiently constrain metallicity and temperature.
However, the formal best-fit value for the metallicity $Z=0.26_{-0.26}^{1.35}\, Z_\odot$ 
leaves room for enrichment above solar abundance. 
Finally, we would like to point out that the presence of ICM 
is strongly corroborated by the successful detection 
of SZ signal in the same position and shape of the X-ray diffuse emission, providing
a completely independent and remarkably consistent measurement of the ICM (Di Mascolo et 
al., in preparation).

A final note on the  ICM temperature: we notice that
the best fit value of $1.98_{-0.43}^{+0.70}$ keV is somewhat lower than the $\sim 3$ keV 
component we assumed to account for 
the diffuse emission within a radius of 2 arcsec when fitting the AGN spectrum. In fact, 
one may expect the opposite, if any, with a lower temperature associated with 
denser, central ICM as in normal cool core.  However, we must bear in mind that, apart from
the unavoidable large uncertainties on its spectral shape, the diffuse component
overlapping the outshining AGN emission is necessarily a mix of thermal emission associated 
to the central ICM and nonthermal emission associated with the bases of the jets, and therefore 
it is actually expected to be harder. Needless to say, it is impossible to speculate about the 
share of thermal and nonthermal emission within a distance of 2 arcsec ($\sim 17$ physical kpc).

If we consider the best estimate for an evolved L-T relation, for a temperature of 2 keV, 
we obtain $L_X=1.0\times 10^{44}$ erg s$^{-1}$ and 
$L_X=0.7\times 10^{44}$ from \citet{2011Reichert} and 
\citet{2009Pratt}, respectively, where $L_X$ is the bolometric X-ray luminosity. 
In both cases we applied the phenomenologically estimated
evolution correction $E(z)^{-0.23}$ at $z=2.156$. 
Therefore, our estimate of $L_{0.5-10 {\tt keV}}=(2.0\pm 0.5)\times 10^{44}$ erg s$^{-1}$
(after including the geometrical correction and the core emission with $n_d=4$)
is about a factor of 2 above the expected average relation, but still in agreement
with it given the large intrinsic scatter observed particularly at low temperatures.
This result, if coupled to the forthcoming results from the SZ data, 
is consistent with the observation of a virialized halo, as we further discuss in Section 7.

\begin{figure*}
\begin{center}
\includegraphics[width=0.49\textwidth]{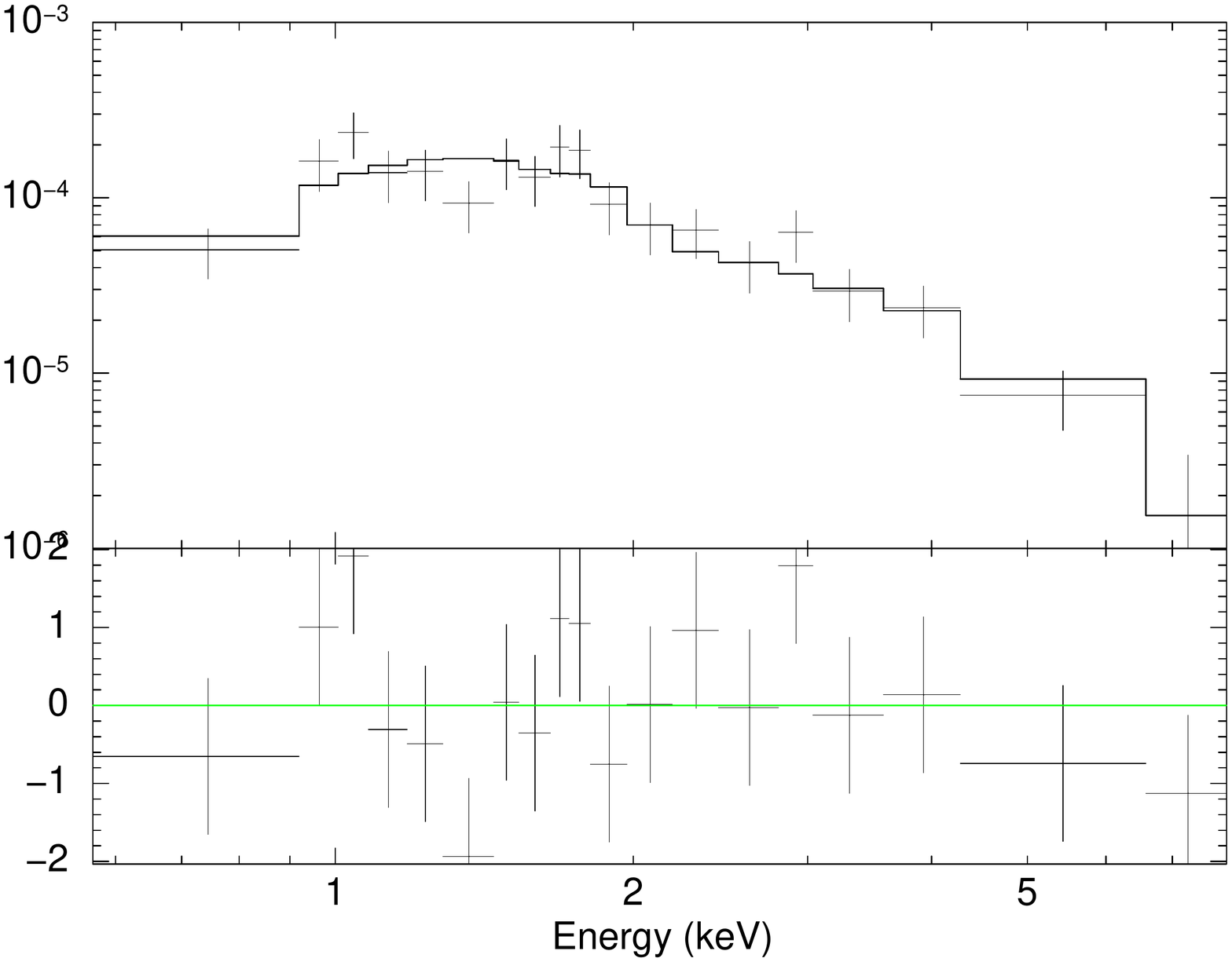}
\includegraphics[width=0.49\textwidth]{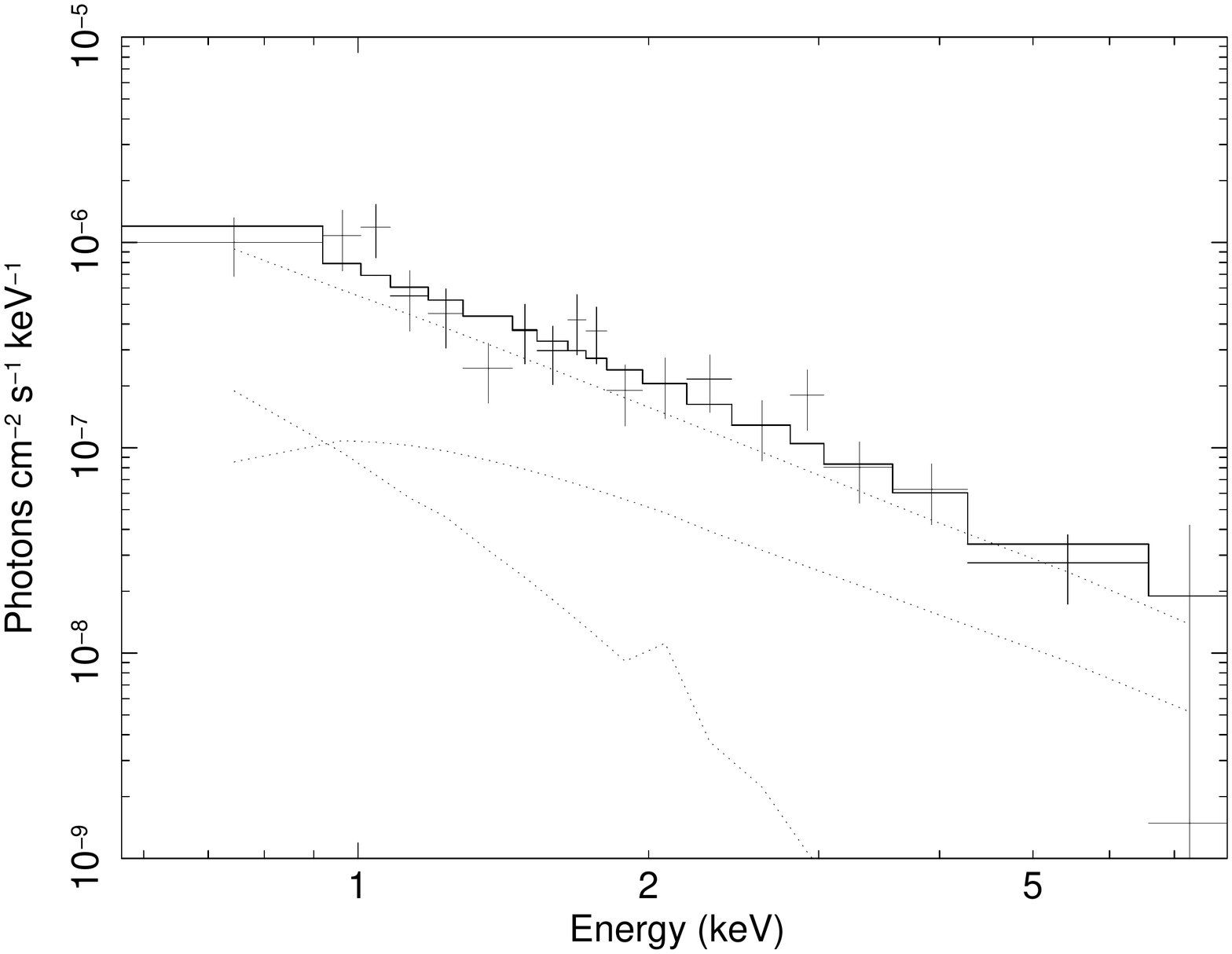}
\caption{Left: Spectrum (folded with the instrument spectral response) of the emission in the 
west jet region (including the AGN contamination and the residual thermal emission 
from the ICM) fitted with a power law.  Right: Unfolded spectrum with the three 
components in the model shown with dotted lines. The lowest dotted line 
corresponds to the thermal emission from the
overlapping ICM, while the line in the middle accounts for the AGN wings contamination.  The IC 
from the relativistic population in the west jet is shown by the unabsorbed power law that is 
dominating at any energy.}
\label{jet_spectrum}
\end{center}
\end{figure*}

\begin{figure*}
\begin{center}
\includegraphics[width=0.49\textwidth]{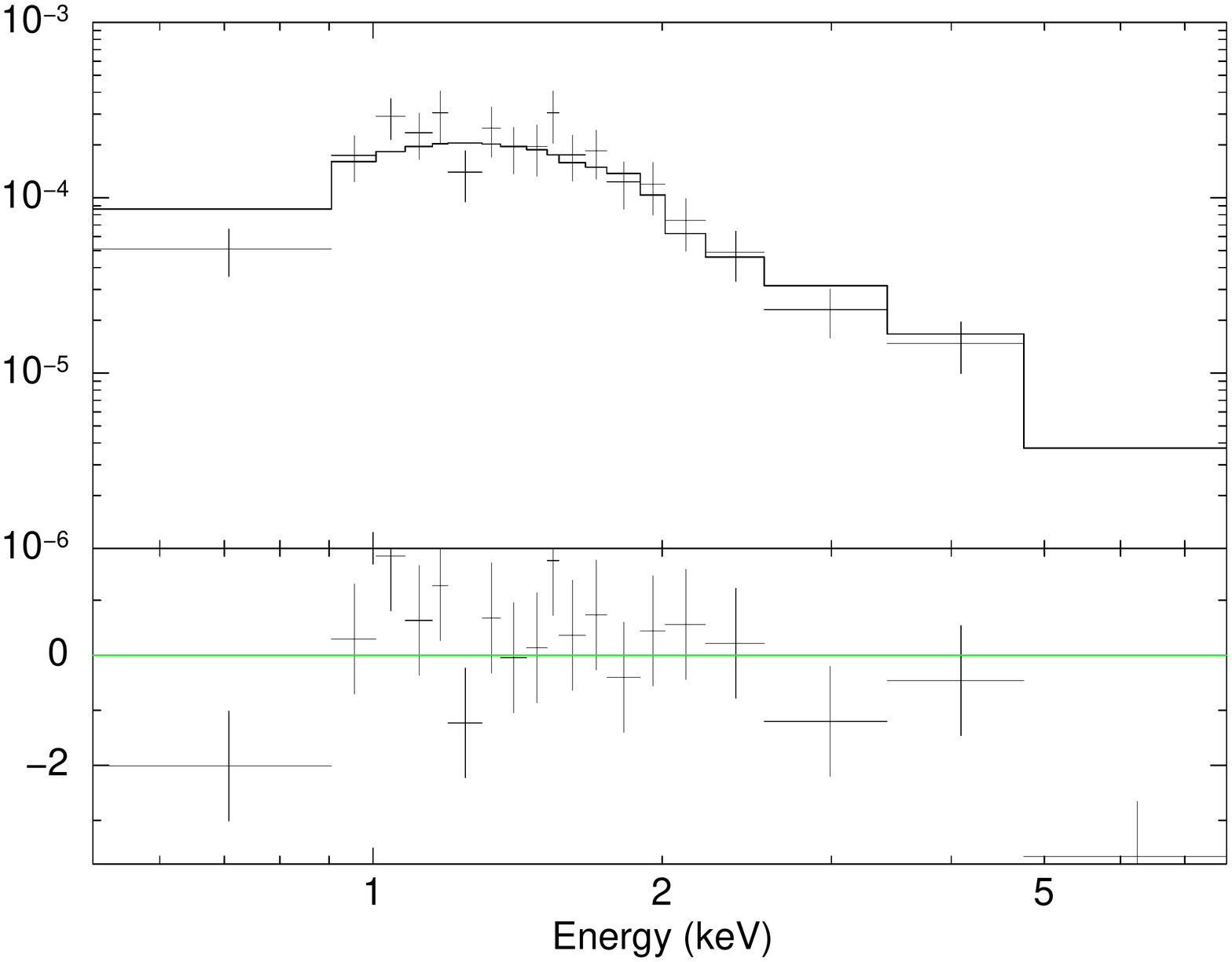}
\includegraphics[width=0.49\textwidth]{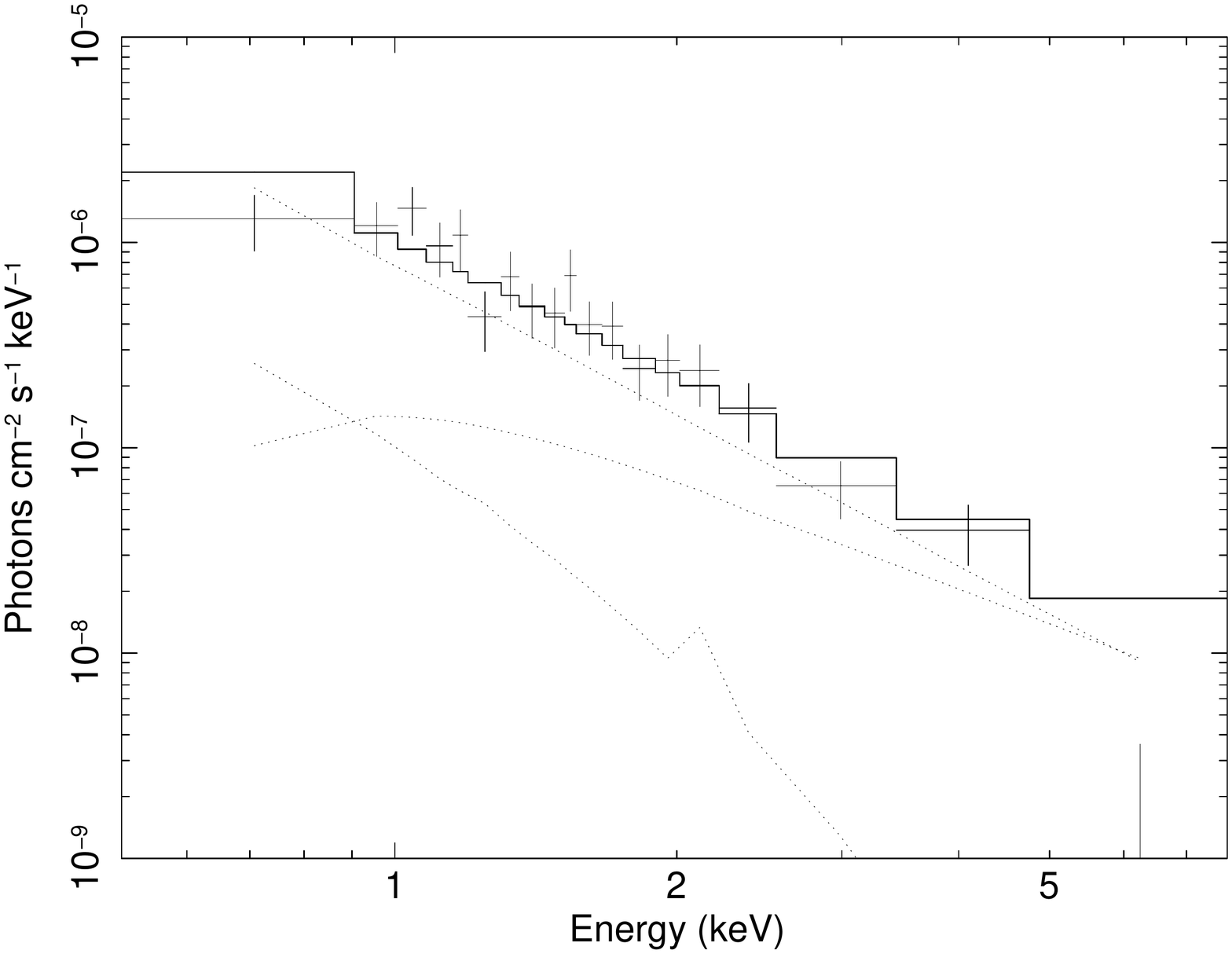}
\caption{Left: Spectrum (folded with the instrument spectral response) 
of the emission in the east jet region (including the AGN contamination and the residual 
thermal emission from the ICM) fitted with a 
power law.  Right: Unfolded spectrum with the three 
components in the model shown with dotted lines. The lowest dotted line 
corresponds to the thermal emission from the
overlapping ICM, while the line in the middle accounts for the AGN wings contamination.  The IC 
from the relativistic population in the east jet is shown by the unabsorbed power law that is 
dominating at any energy.
}
\label{counterjet_spectrum}
\end{center}
\end{figure*}

\subsection{Spectral analysis of the diffuse emission: Jet regions}

To fit the diffuse component in the west and east jet regions, we use a 
power law emission as a baseline model, being more appropriated for IC emission, 
as expected along radio jets.  On the other hand, we do not include any intrinsic
absorption, but only the Galactic absorption. 
As for the AGN contamination, we apply the same procedure described in Section 6.2.
In this case, however, we also need to include the presence of the thermal
diffuse emission that is expected to be present also in the jet regions. 
The uncertainty associated with the thermal component within the two box regions is, however, 
quite large,  since the surface brightness distribution is not known and it may be affected
by the jet itself.  
Therefore, we vary the contribution from the thermal component
when constraining the nonthermal emission in the jet regions to explore its effect on the 
best-fit parameters.

First we focus on the west jet region, which is a rectangle of $4.9\times 5.8$ arcsec$^2$
centered on RA=11:40:48.03,DEC=-26:29:10.89. The AGN contamination is obtained
by rescaling  the emission measured within 2 arcsec by the factor
$4.38\times 10^{-3}$.   The diffuse thermal component in this region is
obtained rescaling the emission measured in the diffuse extraction region 
by a factor of $7.7\times 10^{-2}$, which corresponds to assuming a constant surface 
brightness.  
The best-fit value for the slope of the IC emission is $\Gamma_{west} = 1.90\pm 0.20$. 
We investigate how the slope of the emission depends on our subtraction 
of the thermal component that we expect in that region. We find that the 
best-fit $\Gamma$ becomes slightly softer when reducing the normalization of the
thermal component, up to a value of $\Gamma = 2.00 \pm 0.20$ when the 
thermal component is ignored. The stability of the fit is not surprising, since in this region 
the fit is driven by the prominent emission in the hard band, and therefore is modestly
affected by the uncertainty in the soft band. If
we attempt to fit the emission with a thermal model, we obtain an unrealistic 
temperature of $9.3_{-2.5}^{+3.7}$ keV, and the quality of the fit  
decreases with $\Delta C_{stat} \sim 9$. 

We find fluxes of $(1.32\pm 0.21)\times 10^{-15}$ erg s$^{-1}$ cm$^{-2}$ 
and $(1.78\pm 0.28) \times 10^{-15}$ erg s$^{-1}$ cm$^{-2}$ 
in the 0.5-2 and 2-10 keV bands, respectively, after correcting for Galactic absorption.
These values correspond to luminosities of $(4.1\pm 0.7)\times 10^{43}$ 
erg s$^{-1}$ and $(5.5\pm 0.9)\times 10^{43}$ erg s$^{-1}$
in the 0.5-2 and 2-10 keV rest-frame bands, respectively.
The spectrum in the west jet region folded with the spectral response, 
and unfolded, including the three 
components used for the fit, is shown in Figure \ref{jet_spectrum}. We note that
the IC emission is more than 3 times larger than the AGN contamination 
across the entire energy range, while the thermal component is almost negligible.

We then focus on the east jet region, which is a rectangle of $7.0\times 5.1$ arcsec$^2$
centered on RA=11:40:48.76, DEC=-26:29:09.29. The AGN contamination is obtained
by rescaling  the emission measured within 2 arcsec by the factor
$ 5.79\times 10^{-3}$.   The diffuse thermal component in this region is
obtained rescaling the emission measured in the diffuse extraction region 
by a factor of $9.2\times 10^{-2}$, which corresponds again to a constant surface 
brightness.  
The best-fit value for the slope of the IC emission is $\Gamma_{east} = 2.51\pm 0.21$.
As in the previous case, we investigate how the slope of the emission depends on the 
thermal component that we expect in that region. We find that the 
best-fit $\Gamma$ becomes slowly softer when reducing the normalization of the
thermal component, up to a value of $\Gamma = 2.57 \pm 0.22$ when the 
thermal component is ignored. 
% REMOVE THIS TEST THAT IS JUST A CONSISTENCY CHECK AND 
% MAKES THE TEXT POTENTIALLY CONFUSING
% On the other hand, 
% the spectral slope gets harder up to $\Gamma =2.0$ when the thermal component is assumed
% to be 6 times larger than that obtained with geometrical scaling, which is 
% a factor way too large compared with the expectations. 
If we attempt to fit the emission with a thermal model, instead, we obtain a 
temperature of $4.2_{-0.7}^{+1.0}$ keV, and the quality of the fit significantly 
decreases with $\Delta C_{stat} \sim 20$.  Therefore, we conclude that also in this regions
the emission associated with the east jet is not consistent with being thermal, despite it
appears to be significantly softer than the emission found in the west jet region.
We thus conclude that also in the east jet region
the diffuse emission is dominated by nonthermal processes associated with IC from 
the relativistic electrons.  
In this regions we find fluxes of $(1.89\pm 0.25)\times 10^{-15}$ erg s$^{-1}$ cm$^{-2}$ 
and $(1.02\pm 0.13)\times 10^{-15}$ erg s$^{-1}$ cm$^{-2}$ in the 0.5-2 and 2-10 keV bands, 
respectively.  We find luminosity of $(1.18\pm 0.15)\times 10^{44}$ erg s$^{-1}$
and $(6.4\pm 0.8)\times 10^{43}$ ergs s$^{-1}$ cm$^{-2}$ in the 0.5--2 and 2--10 keV rest-frame 
bands, respectively.  The folded and unfolded spectra in the east jet region, 
including the three components used for the fit, are shown in Fig.~\ref{counterjet_spectrum}.

On a more quantitative ground, the difference in spectral slope
with respect to the west jet region is $\Delta \Gamma = 0.61\pm 0.29$, and therefore
is significant at 2$\sigma$ c.l.  This difference
can be interpreted in two ways.  The first possible explanation is the presence of an additional 
thermal component associated with some gas shocked by the east jet itself up to 
temperatures of few keV.  However, this would imply an amount of hot gas at least comparable if
not larger than that expected from the ICM within the east jet box, which seems 
unlikely.  A second explanation can be provided by the presence of an older 
population of relativistic electrons.  In fact, this is the scenario that
is discussed in much greater details in \citet{2022Carilli} and \citet{2022Anderson}
where the synergy between the {\sl Chandra} X-ray data and the JVLA radio data is
fully exploited.

\section{Nature of the isotropic diffuse emission and protocluster dynamical state}

As a first-cut analysis of the diffuse isotropic emission, we can consider a flat
ICM distribution at least in the core region.  This drastically simplified 
picture is consistent with a beta profile \citep{1978Cavaliere} with a large 
$\sim 100$ kpc core, and reflects our ignorance of the actual ICM distribution.  
On the other hand, since the X-ray emission depends on the square of the electron 
density, a homogeneous gas distribution would rather provide an upper limit to the 
true ICM mass.  Nevertheless, since we focus only on the innermost 100 kpc, 
the impact of the radial dependence of the ICM density is expected to be modest, 
and we expect an upper limit close to the actual mass value within the inner 12 arcsec.
A more accurate treatment of the distribution of the ICM, including the mild
decrease of the surface brightness with radius, and a possible
asymmetry, will be presented in a forthcoming paper (Lepore et al. in preparation).

Therefore, by assuming a constant gas density within to 12 arcsec, 
from the spectrum of the diffuse thermal emission 
we measure an average electron density of $n_e=(1.44\pm 0.23)\times 10^{-2}$ cm$^{-3}$,
for a total ICM mass of $M_{ICM}\leq (1.67\pm 0.25) \times 10^{12}M_\odot$ within a radius of physical 
100 kpc.  However, this value does not include the emission from the excised jet regions, 
plus the central 2 arcsec circle.  As we already discussed, we treated this aspect assuming a
simple geometrical correction applied to the surface brightness, corresponding to a factor 
of 1.22.  Since the electron density is proportional to the square root of the emission, this 
corresponds to a factor of $\sim 1.10$ for the density.  
Therefore, the geometrically corrected values are 
$n_e=(1.58\pm 0.24)\times 10^{-2}$ cm$^{-3}$ and $M_{ICM}\leq (1.84\pm 0.30) \times 10^{12}M_\odot$. 
A more conservative value would be a simple mean with the addition of a systematic uncertainty
bracketing the cases of no correction (corresponding to ICM-empty jet regions)
and geometrical correction (ICM-filled jet regions): 
$n_e=(1.51\pm 0.24 \pm 0.14)\times 10^{-2}$ cm$^{-3}$ and 
$M_{ICM}\leq (1.76\pm 0.30\pm 0.17) \times 10^{12}M_\odot$.  The error bars here
corresponds to the statistical errors on the normalization, and to the systematics 
associated with the geometrical correction.  

If we blindly apply the hydrostatic equilibrium relation of 
\citet[][see their Equation 7]{2006Vikhlinin} for a temperature
of $kT = 1.98^{+0.7}_{-0.4}$ keV out to 100 kpc (physical), we find a mass of 
$M(<100 \, {\tt kpc}) = (1.5^{+0.5}_{-0.3})\times 10^{13}\, M_\odot$ by assuming a roughly 
isothermal distribution (consistent with a total density decreasing approximately 
as $r^{-2}$  beyond 100 kpc and, therefore, a linearly increasing total mass). 
The implied upper limit to the
ICM mass fraction over the total mass is, therefore, $f_{ICM}\leq (0.12\pm 0.04)$.
Incidentally, this value is somewhat higher, and only marginal compatible, with the 
baryonic fraction measured at the group scale \citep{2021Eckert}, possibly
indicating that mechanical feedback has not yet depleted the ICM reservoir in the
core.  However, this is only a preliminary hint that must be verified after a more
careful modelization of the actual ICM distribution.

Applying instead the self-similar mass-temperature 
scaling relation at a redshift of $z=2.156$ (therefore, assuming that 
$M_{500}\propto T^\alpha E(z)^{-1}$ with $\alpha \sim 3/2$)
% well beyond the redshift range where this behaviour has been tested), 
calibrated on data, 
we find $M_{500} = (3.0\pm 1.1)\times 10^{13}\, M_\odot$ 
within a radius $r_{500} = (220\pm 30)$ kpc
% , adopting the best-fit parameters found in Table 5  of \citet{2006Vikhlinin} 
\citep[see also][that provides consistent results]{2004Ettori}.  
The hydrostatic-equilibrium mass value linearly extrapolated to $r_{500}$ provides
$M(<220 \, {\tt kpc}) = (3.2^{+1.1}_{-0.6})\times 10^{13}\, M_\odot$, in agreement
with the scaling relation.
We find consistent results also when scaling self-similarly with redshift the
$M-T$ relation observed in local groups by \citet{2015Lovisari}, obtaining
$M_{500}=(2.7\pm 1.3)\times 10^{13} M_\odot$.  
We note that the agreement toward a value $M_{500}=3.0 \times 10^{13} M_\odot$
essentially relies on the assumption of self similar scaling with redshift, which 
is approximately tested only up to $z\sim 1$.  If the dependence of $M_{500}$ 
on redshift at a given temperature is significantly milder than the self-similar 
behavior $M_{500}\propto (1+z)^{-3/2}$ for $z>1$, the extrapolated mass of the halo
can be more than a factor of two larger \citep[see, e.g.,][]{2019Bulbul}.

To interpret our findings we rely on the
cosmological hydrodynamic simulations used to predict the properties of proto-cluster 
regions at $z\sim 2$ \citep{2009Saro}. These simulations show evidences of 
ongoing assembly of a dominating central galaxy, with a pattern remarkably similar to 
that observed for the Spiderweb complex.  \citet{2009Saro} also find that the projected 
galaxy velocity dispersion, the observed level of star formation and the stellar mass 
of the dominant galaxy suggest that this region should trace the progenitor of a rich 
cluster, whose mass by $z=0$ would be $\simeq 10^{15}\, M_\odot$. 
It is found that such a structure 
at $z\sim 2$ should already contain a diffuse X-ray emitting atmosphere
of hot gas in hydrostatic equilibrium, which is chemically enriched at a level 
comparable to that of nearby galaxy clusters.  
To compare the predictions of \citet{2009Saro} with our analysis of the isotropic, thermal
diffuse emission in the Spiderweb galaxy, we rescale the values shown in their Table 
from 1.7 to 2 keV for the simulation C1, corresponding to a final mass by $z=0$ of 
$h^{-1} 10^{14}\, M_\odot$. We find that they predict a total luminosity $L_X$ which is 
lower by a factor of 2 with respect to the measured $L_X=(2.0\pm 0.5)\times 10^{44}$ erg s$^{-1}$, 
similarly to what we noticed in the comparison with the average $L_X-T$ relation. 
On the other hand, both the temperature and the luminosity found for the simulation C2, 
corresponding to a final mass by $z=0$ of $2\times 10^{15}\, h^{-1} M_\odot$, are in excess
with respect to what we observe.  From this simple comparison, we argue that the final mass 
of the Spiderweb complex may be few$\times 10^{14}M_\odot$ rather than $\geq 10^{15}M_\odot$, 
but clearly this is very hard to confirm, given the large dispersion expected in mass
accretion histories from simulations \citep[see][]{2013Chiang}. 
Remarkably, recent simulations
from the same group show that the SZ signal is 
more consistent with that of a progenitor of a few$\times 10^{14}\, M_\odot$ cluster
at $z=0$, in line with our X-ray analysis (Di Mascolo et al. in preparation).

When considering the dynamical mass estimates based on the velocity 
dispersion of the member galaxies, 
we note that the observed bimodality suggests the presence of two possibly virialized halos, 
whose virial masses are estimated to be
$9\times 10^{13}\, M_\odot$ and $3\times 10^{13}\, M_\odot$ in \citet{2000Pentericci},
and $17\times 10^{13}\, M_\odot$ and $6\times 10^{13}\, M_\odot$ in \citet{2004aKurk}. In \citet{2014Shimakawa} the dynamical
mass of the core is estimated to be $1.71 \times 10^{14}\, M_\odot$. In all cases, the 
mass of the largest halo is in excess if compared to our measurement. 
% , while the smaller halo mass estimates are consistent.  
We argue that the Spiderweb core is caught in a phase
of previrialization or very close to the first phase of virialization, making it 
unfeasible to perform a meaningful comparison of the dynamical mass from velocity dispersion to the
mass inferred from the hydrostatic equilibrium of the ICM. 
% Another possibility is that
We might instead observe an overlap of two recently virialized halos, 
each with its X-ray emitting ICM. 
The possible presence of multiple halos is further discussed in Di Mascolo et al. (in preparation)
based on SZ data analysis.

Yet another process that might affect the morphology in SZ and X-ray
bands is the cooling of the ICM.  The average cooling time in the inner
100 kpc of the Spiderweb hot halo is  $t_c =  (1.5\pm 0.4)$ Gyr. 
This is twice
shorter than the age of the Universe $t_{age}\sim 3$ Gyr at $z=2.156$, but still
longer than the dynamic time roughly estimated as $0.2\times t_{age}\sim 0.6$ Gyr. 
This implies that, even for a relatively well shaped dark matter
halo, the ICM distribution might be very sensitive to the history of
the assembly (especially for lower mass subhalos). This may help in explaining the 
marginal features of a somewhat irregular and off-centered distribution
of the thermal diffuse emission, that are not addressed in the present work, 
but will be thoroughly discussed in a forthcoming paper (Lepore et al., in preparation). 

At present, we conclude that the diffuse, isotropic emission 
observed around the Spiderweb Galaxy is due to an embryonic halo 
corresponding to a total mass $\sim 3.0\times 10^{13} \, M_\odot$.  
The proto-ICM appears to have an approximately isotropic distribution, 
showing no correlation with the direction of the jets, 
making it unlikely that direct feedback can provide a dominant
source of heating.  
Nevertheless, the coexistence of radio-mode feedback and a complex
dynamical status, makes it very hard to provide a unique and self-consistent modelization of the 
Spiderweb protocluster.  We are planning to provide a more accurate 
spatially resolved analysis of the X-ray data 
and combine them with SZ data, and to increase 
the numerical effort for a better comparison 
with simulations.  We also plan to add ALMA SZ data deeper than those currently available.
Furthermore, the next necessary step is to increase the sample of $z>2$ 
protoclusters with comparable properties, 
hopefully with an observational coverage similar in quality and depth 
to that of the archetypal Spiderweb.

\section{Conclusions\label{conclusions}}

A deep {\sl Chandra} X--ray observation of the field of the Spiderweb galaxy
(J1140-2629) at $z=2.156$ enabled us to investigate the properties
of the central AGN and of the diffuse emission in detail.  Our main conclusions are summarized as follows:

\begin{itemize}

\item we find that the Spiderweb galaxy hosts a mildly absorbed quasar, showing 
a modest yet significant spectral and flux variability on a timescale of $\sim 1$ year.

  \item We firmly identify significant extended emission out to a radius 
  of 12 arcsec (100 kpc), which appears to be made of a thermal and a nonthermal 
  component, and it is significantly contaminated by the wings of the strong AGN emission.
  
  \item After accurately accounting for all the three components plus the instrumental 
  background, we were able to construct an azimuthally averaged profile of the diffuse emission.
  
  \item We find that the emission in the jet regions (overlapping the radio 
  extended emission) is significantly harder than the rest  of the extended 
  emission, with $\Gamma \sim 2-2.5$, and consistent with IC 
  upscattering of the CMB photons by the relativistic electrons. 

\item  We find a roughly symmetric, diffuse emission in a circular region 
with a radius of $\sim 100$ kpc centered on the Spiderweb galaxy.  This emission is 
significantly softer than that in the jet regions, and it is consistent with
thermal emission from hot ICM with a temperature of 
$kT = 2.0_{-0.4}^{+0.7} $ keV, with an upper limit on the level of ICM enrichment of
$Z<1.6 Z_\odot$ at 1$\sigma$ c.l.

\item The diffuse emission, if assuming a constant density within 100 kpc, 
corresponds to an average electron density of $n_e=(1.51 \pm 0.24\pm 0.14)\times 10^{-2}$ cm$^{-3}$.
% $n_e=(1.44\pm 0.14)$ cm$^{-3}$,
If we make the simple assumption of an approximately flat ICM distribution, 
we obtain an upper limit to the ICM mass of $\leq (1.76\pm 0.30 \pm 0.17) \times 10^{12}M_\odot$ 
% $\sim (1.67\pm 0.27) \times 10^{12}M_\odot$ 
within the same radius
(error bars are $1\sigma$ statistical and systematic, respectively). 

\item The total rest-frame luminosity of the ICM, 
$L_{0.5-10 {\rm keV}}=(2.0 \pm 0.5) \times 10^{44}$ erg s$^{-1}$,
is about a factor of 2 higher than the central value of the
extrapolated $L-T$ relation for massive clusters, but still consistent within the scatter. 

\item If we apply hydrostatic equilibrium to the ICM, we estimate a total mass of 
$M(<100 \, {\rm kpc}) = (1.5^{+0.5}_{-0.3})\times 10^{13}\, M_\odot$. 
Applying the average scaling relation at a redshift of $z=2.156$, we estimate
a total mass of $M_{500} = (3.2^{+1.1}_{-0.6})\times 10^{13}\, M_\odot$ 
within a radius of $r_{500} = (220\pm 30)$ kpc.

\end{itemize}

We conclude that the Spiderweb galaxy is hosting a bright, mildly absorbed 
AGN showing limited but significant variability.  Significant diffuse emission
is detected within a radius of 12 arcsec, and it is shown to be dominated by 
IC scattering associated with the radio jets.  Outside the
jet regions, we also identify thermal emission showing that 
the protocluster is hosting a $\sim 100$ kpc core of 
hot, diffuse baryons that may represent the embryonic virialized halo 
of the forming cluster, and, based on numerical simulations, it is expected to
evolve into a massive halo with a mass a few times $10^{14}M_\odot$ by $z=0$.  
The origin and the detailed properties of the ICM
identified in the Spiderweb galaxy will be discussed in a paper presenting 
complementary SZ results (Di Mascolo et al. in preparation) and in another paper further 
investigating the spatial properties of the X-ray diffuse emission (Lepore et al. in preparation).  
The main results of the multiwavelength survey of the Spiderweb galaxy field 
and the corresponding data products can be found on the 
project webpage\footnote{\url{http://www.arcetri.inaf.it/spiderweb/ .}}.

Finally, our results demonstrate the central role of a sharp Chandra-like 
PSF for detailed analysis of the ICM properties of high-z protoclusters. They also pave the way for 
future X--ray studies of distant proto-clusters to be carried out with 
the advent of the next generation of X--ray satellites, such as Lynx \citep{2018LynxTeam}, 
AXIS \citep{2019Mushotzky,2020Marchesi}, and, to some extent, the survey and time domain
mission STAR-X
\footnote{The Survey and Time-domain Astrophysical Research Explorer, or STAR-X, is 
targeted for launch in 2028. The mission has been recently selected in the Phase A of 
the Medium Explorer category.} \citep[][]{2022Zhang}.

\begin{acknowledgements}
This work was carried out during the ongoing COVID-19 pandemic.
The authors would like to acknowledge the health workers
all over the world for their role in fighting in the frontline of this
crisis.  We thank the
anonymous referee for detailed comments and positive criticism that helped
improving the quality of the paper. 
S.B., P.T., E.R. and R.G. acknowledge financial contribution from the agreement 
ASI-INAF n.2017-14-H.0. M.N. acknowledges INAF-1.05.01.86.20.
L.D.M., M.P., and A.S. are supported by the ERC-StG 'ClustersXCosmo"'
grant agreement 716762. A.S. is also supported by the FARE-MIUR grant 'ClustersXEuclid' 
R165SBKTMA, and by INFN InDark Grant.
S.B. acknowledges partial financial support from the Indark INFN Grant.
HD acknowledges financial support from AEI-MCINN under the grant with reference PID2019-105776GB-I00/DOI:10.13039/501100011033, 
and from the ACIISI, Consejeria de Economia, Conocimiento y Empleo del Gobierno de Canarias and the 
European Regional Development Fund (ERDF) under grant with reference PROID2020010107.
We thank Hans Moritz G\"unther for help with the use of the MARX software. We thank
Malgorzata Sobolewska for assistance during the Chandra observations.
\end{acknowledgements}

\bibliographystyle{aa}
\bibliography{references_Spiderweb}

\end{document}